\documentclass{lmcs} 
\pdfoutput=1
\usepackage[utf8]{inputenc}

\usepackage{lastpage}
\lmcsdoi{21}{4}{21}
\lmcsheading{}{\pageref{LastPage}}{}{}%
{Jul.~11,~2024}{Nov.~10,~2025}{}

\keywords{logics of programs; relational properties; semantic completeness; relational Hoare logic; relative completeness; inductive assertion method}

\usepackage{latexsym}
\usepackage{amssymb}
\usepackage{hyperref}
\usepackage{color} 
\usepackage{listings} 
\usepackage{mathpartir} 
\usepackage{stmaryrd} 
\usepackage{relsize}
\usepackage{tikz}
\usetikzlibrary{automata,positioning,arrows,arrows.meta,decorations.pathmorphing}
\usepackage{thmtools} 
\usepackage{thm-restate}
\usepackage{arydshln} 

\definecolor{bleudefrance}{rgb}{0.19, 0.55, 0.91} 

\newcommand{\dt}[1]{\textbf{\emph{#1}}} 

\newcommand{\sep}{\mathbin{\mid}} 
\newcommand{\Sep}{\mathbin{\:\mid\:}} 
\newcommand{\smSep}{\mbox{\tiny$|$}}

\def\rightharpoonupfill{$\mathsurround=0pt \mathord- \mkern-6mu
  \cleaders\hbox{$\mkern-2mu \mathord- \mkern-2mu$}\hfill
  \mkern-6mu \mathord\rightharpoonup$}

\def\leftharpoonupfill{$\mathsurround=0pt \mathord\leftharpoonup \mkern-6mu
  \cleaders\hbox{$\mkern-2mu \mathord- \mkern-2mu$}\hfill
  \mkern-6mu \mathord-$}
\def\overleftharpoonup#1{\vbox{\ialign{##\crcr
      \leftharpoonupfill\crcr\noalign{\kern-1pt\nointerlineskip}
      $\hfil\displaystyle{#1}\hfil$\crcr}}}

\def\overrightharpoonup#1{\vbox{\ialign{##\crcr
      \rightharpoonupfill\crcr\noalign{\kern-1pt\nointerlineskip}
      $\hfil\displaystyle{#1}\hfil$\crcr}}}

\newcommand{\means}[1]{\llbracket\, #1 \,\rrbracket}

\newcommand{\eqdef}{\mathrel{\,\hat{=}\,}}

\newcommand{\mkt}[1]{\llparenthesis\, #1 \,\rrparenthesis}

\renewcommand{\neg}{\mbox{\relsize{-1}$\lnot$}}

\newcommand{\union}{\cup}

\newcommand{\proves}{\vdash}
\newcommand{\aftqua}{.\:}
\newcommand{\all}[2]{\forall #1 \aftqua #2}
\newcommand{\some}[2]{\exists #1 \aftqua #2}

\newcommand{\allSet}[2]{\dot{\forall} #1 \aftqua #2}
\newcommand{\someSet}[2]{\dot{\exists} #1 \aftqua #2}

\renewcommand{\mod}{\mathbin{\mathsf{mod}}}
\newcommand{\KE}{\mathit{KE}} 
\newcommand{\KB}{\mathit{KB}} 
\newcommand{\Hyp}{\mathconst{Hyp}} %

\newcommand{\nat}{\mathbb{N}} 
\newcommand{\Z}{\mathbb{Z}} 

\newcommand{\imp}{\Rightarrow}

\renewcommand{\iff}{\Leftrightarrow}

\newcommand{\ok}{\mathconst{ok}}
\newcommand{\okf}{\mathconst{okf}}
\newcommand{\okfd}{\mathconst{okfd}}

\newcommand{\enab}{\mathconst{enab}}
\newcommand{\totalIf}{\mathconst{total\mbox{-}if}}

\newcommand{\lab}{\mathconst{lab}}
\newcommand{\labs}{\mathconst{labs}}
\newcommand{\sub}{\mathconst{sub}}

\newcommand{\fsuc}{\mathconst{fsuc}}
\newcommand{\ghost}{\mathconst{ghost}}
\newcommand{\erase}{\mathconst{erase}}

\newcommand{\addPC}{\mathconst{add}^{pc}}
\newcommand{\map}{\mathconst{map}}
\newcommand{\indep}{\mathconst{indep}} 
\newcommand{\nfax}{\mathconst{nfax}} 
\newcommand{\aprod}{\prod}
\newcommand{\dom}{\mathconst{dom}}  

\newcommand{\encode}{\tilde} 
\newcommand{\proj}[1]{{#1}^{\mbox{\tiny\ensuremath{\downarrow}}}}

\newcommand{\POST}{\mathconst{post}} 
\newcommand{\WP}{\mathconst{wp}}

\newcommand{\Var}{\mathconst{Var}} 

\newcommand{\aut}{\mathconst{aut}}

\newcommand{\init}{\mathit{init}}
\newcommand{\fin}{\mathit{fin}}
\newcommand{\trans}{\mapsto} 
\newcommand{\tranSeg}[1]{\stackrel{#1}{\longmapsto}}
 
\newcommand{\biTrans}{\Mapsto} 
\newcommand{\biTranSeg}[1]{\stackrel{#1}{\Mapsto}}

\newcommand{\ctrans}{\rightarrowtriangle} 
\newcommand{\config}[2]{\langle #1,\: #2\rangle}

\newcommand{\kateq}{\mathrel{\simeq}} 

\newcommand{\relKAT}{\mathfrak{L}} 
\newcommand{\anyKAT}{\mathfrak{M}} 

\newcommand{\mathconst}[1]{\mbox{\upshape\textsf{#1}}} 

\newcommand{\rn}[1]{\textsc{\relsize{-1}#1}} 

\newcommand{\updateSym}{\mbox{\relsize{-1}$\,\mapsto\,$}}
\newcommand{\update}[3]{#1[#2\updateSym#3]} 
\newcommand{\scol}{\mathord{:}} 

\newcommand{\leftex}[1]{ \raisebox{.25ex}{\relsize{-1}$\langle\hspace*{-2.0pt}[$} #1 \raisebox{.25ex}{\relsize{-1}$\langle\hspace*{-2.1pt}]$} }
\newcommand{\rightex}[1]{ \raisebox{.25ex}{\relsize{-1}$[\hspace*{-2.1pt}\rangle$} #1 \raisebox{.25ex}{\relsize{-1}$]\hspace*{-2.0pt}\rangle$} }
\newcommand{\leftF}[1]{\leftex{#1}}
\newcommand{\rightF}[1]{\rightex{#1}}
\newcommand{\bothF}[1]{%
  \raisebox{.25ex}{\relsize{-1}$\langle\hspace*{-2.1pt}[$}
  #1 
  \raisebox{.25ex}{\relsize{-1}$]\hspace*{-2.1pt}\rangle$ } 
}

\newcommand{\codeleft}[1]{{^\bullet\!#1}} 
\newcommand{\coderight}[1]{#1^\bullet} 

\newcommand{\hint}[1]{{\qquad\mbox{#1}}} 
\newcommand{\Agr}{\ensuremath{\mathbb{A}}}

\newcommand{\keyw}[1]{\ensuremath{\mathsf{#1}}}

\newcommand{\pp}[1]{\ensuremath{#1{\relsize{-1}++}}}

\newcommand{\skipc}{\keyw{skip}}
\newcommand{\ifc}[3]{\keyw{if}\ {#1}\ \keyw{then}\ {#2}\ \keyw{else}\ {#3}\ \keyw{fi}}

\newcommand{\havc}[1]{\keyw{hav}\ {#1}} 
\newcommand{\lhavc}[2]{\keyw{hav}^{#1}\ {#2}} 

\newcommand{\lskipc}[1]{\keyw{skip}^{#1}}
\newcommand{\lassg}[3]{#2:=^{#1}#3}
\newcommand{\assg}[2]{#1:=#2}

\newcommand{\ifgc}[1]{\keyw{if}\ {#1}\ \keyw{fi}}
\newcommand{\dogc}[1]{\keyw{do}\ {#1}\ \keyw{od}}
\newcommand{\lifgc}[2]{\keyw{if}^{#1}\ {#2}\ \keyw{fi}}
\newcommand{\ldogc}[2]{\keyw{do}^{#1}\ {#2}\ \keyw{od}}
\newcommand{\gcsep}{\talloblong} 
\newcommand{\gcto}{\mathrel{\shortrightarrow}} 

\renewcommand{\P}{\mathcal{P}}  
\renewcommand{\S}{\mathcal{S}} 
\newcommand{\Lrel}{\mathcal{L}} 
\newcommand{\Q}{\mathcal{Q}} 
 
\newcommand{\R}{\mathcal{R}} 
\newcommand{\I}{\mathcal{I}} 
\newcommand{\T}{\mathcal{T}} 

\newcommand{\subst}[3]{{#1}^{#2}_{#3}}

\definecolor{light-gray}{gray}{0.88}
\definecolor{dark-gray}{gray}{0.25}
\newcommand{\graybox}[1]{\colorbox{light-gray}{#1}} 

\newcommand{\specSym}{\leadsto}
\newcommand{\rspecSym}{\ensuremath{\mathrel{\mbox{\footnotesize$\raisebox{-.07ex}{$\thickapprox$}\hspace*{-1.2ex}>$}}}}
\newcommand{\spec}[2]{#1\specSym #2} 
\newcommand{\rspec}[2]{#1\rspecSym #2} 

\newcommand{\aespecSym}{\ensuremath{\mathrel{\mbox{\footnotesize$\stackrel{\exists}{\raisebox{-.07ex}{$\thickapprox$}\hspace*{-1.2ex}>}$}}}}

\newcommand{\aespec}[2]{#1\aespecSym #2}

\newcommand{\especSym}{\ensuremath{\mathrel{\mbox{\footnotesize$\stackrel{\exists}{\leadsto}$}}}}
\newcommand{\espec}[2]{#1\especSym #2}

\newcommand{\norm}[3]{#1 \mathbin{/} #2 \hookrightarrow  #3}
\newcommand{\normSmall}[3]{#1 \mathbin{/} #2 \hookrightarrow #3}

\newcommand{\spc}{\mathord{!}}
\newcommand{\tpc}{\mathord{?}}

\newcommand{\quant}[4]{(#1\,#2 \::\: #3 \::\: #4)}

\newenvironment{ditemize}{
\begin{list}{$\bullet$}{%
\setlength{\itemsep}{0pt}\setlength{\rightmargin}{0pt}%
\setlength{\leftmargin}{.6em}\setlength{\parsep}{0ex}}}{
\end{list}}

\theoremstyle{plain} 

\def\opcit{{\em op.~cit.}} 

\begin{document}

\title[Alignment complete relational Hoare logics for some and all]{Alignment complete relational Hoare logics\newline for some and all}

\thanks{Nagasamudram and Naumann were partially supported by NSF grants CNS 1718713 and CNS 2426414.
Banerjee's research
was based on work supported by the NSF, while working at the
Foundation.  Any opinions, findings, and conclusions or
recommendations expressed in this article are those of the authors and
do not necessarily reflect the views of the NSF}

\author[R.~Nagasamudram]{Ramana Nagasamudram\lmcsorcid{0000-0003-2779-2071}}[a]
\author[A.~Banerjee]{Anindya Banerjee\lmcsorcid{0000-0001-9979-1292}}[b]
\author[D.~Naumann]{David A. Naumann\lmcsorcid{0000-0002-7634-6150}}[a]

\address{Stevens Institute of Technology, USA}	

\address{Dartmouth College, USA}	


\begin{abstract} 
In relational verification, judicious alignment of computational steps
facilitates proof of relations between programs using simple relational
assertions. Relational Hoare logics (RHL) provide compositional rules that
embody various alignments of executions.  Seemingly more flexible alignments can
be expressed in terms of product automata based on program transition
relations. A single degenerate alignment rule (sequential composition), atop a
complete Hoare logic, comprises a RHL for $\forall\forall$ properties that is
complete in the sense of Cook. The notion of alignment
completeness was previously proposed as an additional measure, and some
rules were shown to be alignment complete with respect to a few ad hoc forms of
alignment automata. This paper proves alignment completeness with respect to a
general class of $\forall\forall$ alignment automata, for a RHL comprised of
standard rules together with a rule of semantics-preserving rewrites based on
Kleene algebra with tests.  A new logic for $\forall\exists$ properties is
introduced and shown to be sound and alignment complete for a new general class of automata.  The
$\forall\forall$ and $\forall\exists$ automata are shown to be semantically
complete.  Thus both logics are complete in the sense of Cook. 
The paper includes discussion of why alignment is not the only important principle for relational reasoning and proposes entailment completeness
as further means to evaluate RHLs.
\end{abstract}

\maketitle

\section{Introduction}\label{sec:intro}

A ubiquitous problem in programming is reasoning about relational properties, such as equivalence between two programs in the sense that from any given input they produce the same output.  Relational properties are also of interest for a single program. For example, a basic notion in security is noninterference: any two executions with the same public inputs should result in the same public outputs, even if the secret inputs differ~\cite{Sabelfeld:Myers:JSAC}.  
We write $c\sep c':\rspec{\R}{\S}$ to say program $c$ relates to program $c'$ in the sense that 
for any pair of initial states related by $\R$, and terminated executions of $c$ and $c'$ from those states,
the final states are related by $\S$.
For example, letting $\Agr low$ say two states agree on the values of low-security variables, $c\sep c:\rspec{\Agr low}{\Agr low}$ specifies termination-insensitive noninterference for $c$.

Another kind of relational property is written $c\sep c': \aespec{\R}{\S}$ and deals with nondeterminacy.  It says that from any $\R$-related pair of initial states, and for any terminated run of $c$ from the left state, there is a terminated run of $c'$ from the right state such that $\S$ relates the final states.  For example, a standard notion of refinement is obtained by taking $\R$ and $\S$ to be the identity.
For data refinement~\cite{deRdataref}, let $\R$ connect two different data representations and let $\S$ be $\R$.
For brevity we refer to the two kinds of relational properties as $\forall\forall$ and $\forall\exists$. 

Often a pair of executions are similar, perhaps because the related programs are similar, or even the same as in the case of noninterference.  So reasoning often relies on the \emph{alignment} of similar steps ---think of a programmer viewing two versions of a similar program, side by side on screen.  Having chosen to align a pair of points in control flow, one can consider a relational assertion that holds whenever execution reaches those points. 

Most work on formal reasoning about relational properties has focused on $\forall\forall$ properties which capture many requirements for deterministic programs.
There are two main approaches.
One is deductive systems inspired by Hoare logic (\dt{HL}), which we call \emph{relational Hoare logics} (\dt{RHL}s)~\cite{Francez83,Benton:popl04}.
The other approach is to reason about program semantics in the form of an automaton (i.e., transition system),
adapting the inductive assertion method (\dt{IAM})~\cite{Floyd67} to use relational assertions at aligned pairs of control points.  
By representing the two programs as a product automaton one can easily express which control points are meant to be aligned.
Alignments can be conditioned on the programs' data, with such conditions expressed using relational assertions.  
In case the program transition relation(s) can be expressed in a solvable fragment of first order logic, it is even possible to automatically infer alignment conditions and relational assertions~\cite{ShemerGSV19,UnnoTerauchiKoskinen21}, 
using proof search in the constraint language instead of proof search in a logic of programs.

Deductive systems are important for several reasons.
Because a deductive system applies to ordinary program syntax, it caters for human interaction which is essential for verification of programs involving dynamically allocated data and other features for which program semantics and/or specifications are not easily expressible in solvable fragments.
Deductive proofs can also serve as independently checkable certificates to represent proofs that may be found by other means.  

Finally, deductive rules can embody principles beyond IAM.
The rule of conjunction decomposes a goal with a conjunctive postcondition into simpler subgoals which may be proved by IAM or deductively.  
Rules for procedure call and linking account for procedure modular verification, which in the IAM setting is deployed through use of specs as procedure summaries.  
These principles apply in both unary and relational reasoning, but there are further principles for relational reasoning. 
A well known relational principle is vertical ---i.e., transitive--- composition, which is sound for $\forall\exists$ but not $\forall\forall$.
This makes $\forall\exists$ logic useful even for deterministic programs~\cite{hawblitzelklr13,DosualdoFD2022}.

Aside from soundness, the fundamental criterion for a deductive system is completeness: any true correctness judgment should be provable.  For program logics one seeks Cook's \emph{relative completeness}~\cite{Cook78,AptOld3,WinskelBook}, 
factoring out completeness and expressiveness of the assertion logic.
The problem we address in this paper is that Cook completeness is unsatisfactory for RHLs.

Consider this sound rule: from $c\sep\skipc: \rspec{\R}{\Q}$ and $\skipc\sep c':\rspec{\Q}{\S}$ infer
$c\sep c':\rspec{\R}{\S}$.  
The rule shows how deductive rules can embody alignments,
in this case a \emph{sequential alignment} wherein the final state of $c$ is aligned with the initial state of $c'$.  
Under mild assumptions about encoding a pair of states as a single state, and thereby treating a relation as a \emph{unary} assertion, the premises of this rule can be expressed as 
partial correctness judgments in HL~\cite{Francez83,BartheDArgenioRezk}.
The sequential alignment rule, together with a complete set of unary rules, provides a Cook complete relational 
logic~\cite{Francez83,BartheDArgenioRezk,Beringer11}.
Yet it is well known that sequential alignment is unsatisfactory. 
Consider proving $c\sep c:\rspec{\I}{\I}$ where $\I$ is a conjunction of agreements, one for each variable of $c$.  
Surely $c$ is equivalent to itself as expressed by $\rspec{\I}{\I}$.
By aligning $c$ with itself step by step, the only intermediate assertion we need is $\I$ ---whereas for reasoning about the sequential alignment, the intermediate assertion $\Q$ may have to ``remember everything''~\cite{Francez83} about $c$. 
In particular this may require finding strong invariants for loops.
Good alignment is essential for relational verification because it enables the use of relatively simple assertions~\cite{ShemerGSV19}.
As an example how different alignments can be formalized, here is a standard rule of RHL:
from $c\sep c':\rspec{\R}{\Q}$ and $d\sep d':\rspec{\Q}{\S}$ infer $c;d\sep c';d':\rspec{\R}{\S}$.
Another standard rule aligns loop iterations (see \rn{rAlgnDo} in \autoref{fig:derivedRHL}).
To be clear, sequential alignment has important uses, including cases where the two programs have different structure or simply differ in the order of atomic operations like assignments.

The notion of \emph{alignment completeness} has been proposed as 
an additional criterion for RHLs~\cite{NagasamudramN21}.
Here's the idea: For any valid alignment-based proof there should be a proof of the same judgment in the RHL \emph{using essentially the same assertions}.
The approach of~\cite{NagasamudramN21} uses automata to represent alignments and IAM proofs 
(as in e.g.~\cite{ShemerGSV19,ChurchillP0A19,UnnoTerauchiKoskinen21,ItzhakySV24}).
In~\cite{NagasamudramN21}, alignment completeness results are only given for $\forall\forall$ properties and a few specialized kinds of automata
for which there is a straightforward connection with specific proof rules.
For example, for lockstep alignments of two programs with the same control structure one gets alignment completeness for the rules of Benton~\cite{Benton:popl04} (our rules 
\rn{rAsgnAsgn}, \rn{rSeq}, \rn{rAlgnIf}, \rn{rAlgnDo}, \rn{rConseq}
in Figures~\ref{fig:RHL} and~\ref{fig:derivedRHL}).

In a nutshell, we make two contributions. First, we answer the challenge in~\cite{NagasamudramN21} to obtain alignment completeness for 
a general class of $\forall\forall$ alignment automata.
Second, we introduce a new logic for $\forall\exists$ properties and prove its
soundness and alignment completeness.
These achievements overcome several challenges.

\underline{Challenge:} Find an interesting/useful general class of alignment automata. A key requirement is that the automaton be \emph{adequate} in the sense of covering all pairs of executions.  Adequacy has been worked out in several independent works (with terminology like ``fair scheduler''), for $\forall\forall$ properties, but there is no standard theory, and there is little relevant work on $\forall\exists$.
Our solutions are in sections~\ref{sec:aut} and~\ref{sec:eacomplete}.  Alignment conditions and the choice of existential witnesses are specified by state relations as is done in many practical works on relational verification, like those cited above.
For $\forall\exists$, our automata formulation (in \autoref{sec:eacomplete})
shows that to prove a $\forall\exists$-spec $\aespec{\R}{\S}$, instead of explicitly constructing a positive witness of the existential~\cite{BaumeisterCBFS21,LamportS21}
one can filter out the non-witnesses on the right, so what's left satisfies the corresponding $\forall\forall$-spec $\rspec{\R}{\S}$.  All for some!

\underline{Challenge:} Find a good set of RHL rules.  In this paper we confine attention to simple imperative programs but, even so, a large number of rules can be found in the $\forall\forall$ RHL literature, 
and there is little prior work on $\forall\exists$ RHLs.
For $\forall\forall$, several rules seem to be needed to account for the many possible program structures one may wish to align, e.g., relating a loop to a sequence of loops.  The key is to include some form of semantics-preserving rewriting, so programs can be rewritten to have more similar control structure for which fewer rules are needed.

\underline{Challenge:} Find a sensible notion of equivalence for use in rewriting.
Prior works that appeal to rewriting in a deductive logic use ad hoc sets of rewrite rules
(e.g.,~\cite[Appendix~D]{BNN16},\cite{BartheGHS17,BNNN19}), which raises a fresh question about completeness.  Another question is whether equivalence should be a distinct judgment as opposed to a special case of the main relational judgment, and if so, should it involve preconditions.

\underline{Challenge:} For alignment completeness, find a systematic way to obtain deductive proofs from automata-based ones.

Our answer to the latter two challenges is a key technical result:
We show that every program can be rewritten, using laws of Kleene algebra with tests (\dt{KAT})~\cite{Kozen97}, into \emph{automaton normal form} which mimics its transition system representation (\autoref{sec:KATnf}).  
As noted in the related work \autoref{sec:related}, 
this is not too surprising
but the specific form is new and plays a crucial role in proving the main results.

\underline{Technical contributions}:
\begin{itemize}
\item We show that RHL+, a logic featuring a KAT-based rewrite rule (\autoref{fig:RHL}), is alignment complete for $\forall\forall$ proofs based on a general class of alignment automata.
\item We introduce a new general class of $\forall\exists$ alignment automata.
\item We introduce ERHL+ and show it is sound and alignment complete for
  proofs of $\forall\exists$ properties.
\item We show that both classes of automata are sound and semantically complete.
  Together with alignment completeness, this yields Cook completeness for both logics.
\end{itemize}

As a conceptual contribution, we show that for capturing IAM-style proofs,
these core rules of RHL for imperative programs suffice:
(a) one-sided rules for primitive commands,
(b) same-structure rules for sequence, conditional, and (conditionally aligned) loop,
and (c) rewriting a program to an unconditionally equivalent one.
With these rules, many convenient rules found in the literature are derivable.  

Although the IAM is clearly fundamental and widely used in relational verification
(see~\autoref{sec:related}), it is not the only useful reasoning principle.
Even for unary reasoning about simple imperative programs, it is common 
to augment the core complete set of rules (i.e., syntax-directed plus Consequence) 
with, for example, the rule of Conjunction which facilitates modular proofs~\cite{AptOld3}.  
For relational reasoning there are other natural principles such as transitivity
which are not derivable from the core rules we use to obtain alignment completeness.
As future work we highlight some reasoning challenges that motivate additional rules.
This raises the question what are good criteria for RHLs in addition to alignment completeness, a question for which we propose a possible answer dubbed entailment completeness. 

\paragraph{Outline.}
Section~\ref{sec:overview} provides an overview.
Section~\ref{sec:logics} presents the $\forall\forall$ logic RHL+.
Section~\ref{sec:aut} develops alignment products and their verification conditions for $\forall\forall$ properties.
Section~\ref{sec:KATnf} develops the normal form.
Section~\ref{sec:autUnary} gives the unary Floyd completeness result sketched in \autoref{sec:overview}, illuminating some ingredients of the proof of alignment completeness of RHL+ which appears in Section~\ref{sec:acomplete}.  
Section~\ref{sec:ERHL} introduces the $\forall\exists$ logic ERHL+.
Section~\ref{sec:eacomplete} 
extends alignment products with filtering conditions
and shows alignment completeness of ERHL+.
Sections~\ref{sec:unaryRevisited} and~\ref{sec:unaryRevisitedE} revisit 
Cook completeness both based on alignment completeness and based on sequential alignment and unary logics.
Related work is discussed in \autoref{sec:related}
and future work in \autoref{sec:future}.
Section~\ref{sec:concl} concludes.
Some details are in the appendix.

This article is meant to be self-contained and we include considerable details 
which lengthen the text but may shorten the reading.
There is a longer document with additional technical details~\cite{BNN23v5}.

\section{Overview}\label{sec:overview}

Floyd's formulation of the IAM is based on annotation of control points,
representing programs as automata.
We use Dijkstra's guarded command syntax but with labelled control points.
This somewhat silly command $c0$ will be a running example.  
Here $\mod$ means remainder and the superscripts $1\ldots 5$ are labels.
\[ \graybox{$c0:$} \quad
        \lassg{1}{x}{y} ;
        \keyw{do}^2 x > 0 \gcto 
                            \keyw{if}^3 \: x\mod 2 = 0 \gcto \lassg{4}{x}{x-1} 
                            \gcsep x\mod 2 \neq 0 \gcto \lassg{5}{x}{x-2} 
                             ~ \keyw{fi} ~ \keyw{od}
\]
We aim to prove $c0\sep c0:\rspec{\Agr y}{\Agr x}$, 
which says any two runs of $c0$, from initial stores that agree on $y$, result in final stores that agree on $x$.
(This expresses that the final value of $x$ depends only on the initial value of $y$.)
Pairs of runs can be represented by executions of a product automaton whose control is successively the same control point on left and right.  An IAM-style proof annotates each
control point
$(n,n)$ of the product with the relation $\Agr x$, except for $(1,1)$ which is annotated by the precondition $\Agr y$.
A deductive proof formulates such lockstep alignment 
using rules like \rn{rSeq} (\autoref{fig:RHL}) and \rn{rAsgnAsgn} (\autoref{fig:derivedRHL}).
By contrast, a sequential product automaton would run the left execution to completion and then the right, annotated with more complicated assertions than $\Agr x$.  Such a proof can be presented using the sequential alignment rule mentioned in \autoref{sec:intro} (cf.\ \rn{rLRseq} in \autoref{fig:derivedRHL}).

The mentioned forms of automata can be made precise using conditions on control points to designate when left-side, right-side, or joint steps should be taken.  Later we consider products where the alignment conditions can also depend on the stores.

\paragraph{Rewriting facilitates expression of alignment.}

A core element of our logic RHL+ is a KAT-based rule for rewriting of programs.  
It says that to show
$c \sep c' : \rspec{\P}{\Q}$, it suffices to show $d \sep d' : \rspec{\P}{\Q}$
provided $d$ and $d'$ are equivalent to $c$ and $c'$ respectively.  A minimal amount of
equivalence-preserving rewriting using simple laws derived from KAT can open up opportunities for better alignments.   
KAT equality is sufficient for our purposes and it is well suited to serve as an auxiliary judgment in a program logic because it can be presented by a deductive system (equational logic) and even as a decidable premise of the rewrite rule as we note in passing later (\autoref{rem:KAT}).

As an example we consider a variation on the loop tiling
optimization example of Barthe et al~\cite{BartheCK13}.  Both $c1$ and $c1'$ shown
below compute the sum of all integers from $0$ to $n \times m - 1$ for some
$n, m > 0$.  For this discussion we can omit labels on commands.
\[
\begin{array}{ll}
  \graybox{$c1:$} & \dogc{i<n\times m \gcto s:=s+i;\ \pp{i}} \\
  \graybox{$c1':$} & \dogc{i<n \gcto j:=0;\
                 \dogc{j<m \gcto (s:=s+i\times m+j;\ \pp{j})};\ \pp{i}}
\end{array}
\]
We aim to show equivalence under a precondition:
\begin{equation}\label{eq:c12}
c1\sep c1' : \rspec{\Agr i \land \leftex{i}=0 \land \Agr s \land \leftex{s}=0 \land \Agr n \land \Agr m}{\Agr s}
\end{equation}
(Some notation: $\leftex{i}$ is the value of $i$ in the left state,
and $\rightex{i}$ is its value on the right, and $\Agr i$ is an abbreviation for  $\leftex{i}=\rightex{i}$.)
A naive alignment would require lining up the inner loop in $c1'$ with the body
of the loop in $c1$.  But this would require us to summarize the effect of the inner loop in $c1'$ which is an avoidable complication.
Instead we start by rewriting the two programs to the following.
\[
\hspace*{-3pt}\begin{array}{ll@{\;}l}
\graybox{$c2:$}\!\! & \keyw{do} & i < n \times m \gcto s:=s+i; \pp{i};
       \dogc{i < n \times m\land i \mod m\neq0 \gcto s:=s+i; \pp{i}}\ \keyw{od} \\[1ex]
\graybox{$c2':$}\!\! & \keyw{do} & i < n \gcto j:=0;\  
       \ifc{j<m}{s:=i\times m+j;\ \pp{j}}{\skipc}; \\
       & & \phantom{i<n\gcto} \dogc{j<m \gcto (s:=s+i\times m+j;\ \pp{j})};\ \pp{i};\ \keyw{od}
\end{array}
\]
We obtain $c2'$ by unrolling the inner
loop in $c1'$ once, and $c2$ by using the fact that $(\dogc{b\gcto c})$ and
$(\dogc{b\gcto c; \dogc{b\land b_0\gcto c}})$ are equivalent for any choice of
$b, b_0$ and $c$.  
Unconditional equivalences like these are purely about control structure,
and encompassed by KAT's ``propositional'' view of programs~\cite{Kozen97}.
Now, by the rewriting rule of RHL+ it suffices to show $c2$ and
$c2'$ satisfy the spec in (\ref{eq:c12}).
We can establish this by aligning the outer and inner
loops in lockstep.  The outer loops admit 
$\leftex{i} = \rightex{i \times m} \land \Agr s$ 
as relational invariant and this is sufficient to establish the postrelation.
The inner loops admit 
$\leftex{i} = \rightex{i \times m + j} \land \Agr s$ 
as relational invariant.

\paragraph{Nondeterminacy and $\forall\exists$ judgments.}

While $\forall\forall$ specs capture a wide class of requirements, they
do not express all interesting properties of nondeterministic programs.  
Consider the havoc command $\lhavc{}{x}$ which sets $x$ to any integer.
The $\forall\forall$ judgment
$\lhavc{}{x}\sep\lhavc{}{x} : \rspec{\mathit{true}}{\Agr x}$ does not
hold.
However, for every execution of $\lhavc{}{x}$, there exists an execution of
$\lhavc{}{x}$ such that $\Agr x$.
This property is captured by the $\forall\exists$ judgment
$\lhavc{}{x}\sep\lhavc{}{x} : \aespec{\mathit{true}}{\Agr x}$.  We establish this
judgment by picking an alignment and, in addition, a way of filtering out
executions of the right program that violate the post-relation.  One proof uses  the sequential
alignment that performs $\lhavc{}{x}$ on the left followed by $\lhavc{}{x}$ on the right.
Filtering is achieved by \emph{assuming} the relation $\Agr x$ after the
$\lhavc{}{x}$ on the right. 
In other words, to establish $\Agr x$, we only consider executions of
$\lhavc{}{x}$ on the right which match the left side value.

Filtering conditions are sound provided they permit all executions of the left
program and at least one execution of the right program.  In this example, the
filtering condition $\Agr x$ is justified because every possible value of
$x$ that can be produced by $\lhavc{}{x}$ on the left
can also be produced by $\lhavc{}{x}$ on the right.  An alignment automaton that enforces the
filter effectively reduces the $\forall\exists$ to a $\forall\forall$ property
that can be proved by IAM.
As another example, consider the invalid judgment
\[ \lhavc{}{x} \Sep \lhavc{}{x};x:=2\times x \::\: \aespec{\mathit{true}}{\Agr x} \qquad\mbox{(invalid)} \] 
It is not sound to assume $\Agr x$ because odd values of $x$ on the left cannot be matched on the right, considering that $x$ is an integer variable.
In our $\forall\exists$ logic ERHL+, such ``assumptions'' are expressed by postconditions in $\aespecSym$ specs.
We do not use verification exotica like assume statements or nonstandard program semantics.  

\paragraph{Possibilistic Noninterference.}

\begingroup 
\newcommand{\vhigh}{\ensuremath{\mathit{high}}}
\newcommand{\vlow}{\ensuremath{\mathit{low}}}

For a more involved example with nondeterminism, consider the following program
adapted from~\cite{UnnoTerauchiKoskinen21}. 
It illustrates how alignment facilitates use of simple conditions for filtering.
\label{pg:c3}
\[
\begin{array}{ll@{\;}l}
  \graybox{$c3:$}
  & \keyw{if}^1 \vhigh \neq 0 \gcto & \lhavc{2}{x}; \\
  & & \!\!\!\begin{array}[t]{l}
              \keyw{if}^3\ x \geq \vlow \gcto \lskipc{4}
              \gcsep\ x < \vlow \gcto \ldogc{5}{\mathit{true}\gcto\lskipc{6}}\ \keyw{fi}
            \end{array} \\
  & \gcsep \ \ \vhigh = 0 \gcto & \lassg{7}{x}{\vlow};\ \lhavc{8}{b}; \\
  & & \ldogc{9}{b \neq 0 \gcto \lassg{10}{x}{x+1};\ \lhavc{11}{b}}\ \keyw{fi}
\end{array}
\]
This satisfies a possibilistic noninterference property: from low-equivalent
states, for any terminated execution of $c3$, there exists an execution of $c3$
such that the final values of $x$ in both states are in agreement.  
This is expressed by the $\forall\exists$ judgment $c3\sep c3 : \aespec{\Agr\vlow}{\Agr x}$.

We verify this property of $c3$ by choosing a convenient alignment and a
collection of filtering conditions associated with aligned points.
The alignment is as follows.
From a pair of initial states where both runs take the same branch of the conditional at label $1$, consider the executions in lockstep.  Otherwise, consider the left run up to termination, then reason about the right run, i.e., reason in terms of a left-first sequential alignment.

Suppose $\vhigh\neq 0$ in both initial states, so we reason in terms of a
lockstep alignment: the points labelled (2,2) are aligned, as are (3,3).  At (3,3) we assume $\Agr x$, to filter out executions where the aligned havocs disagree.
With this filtering, it's easy to prove the post-relation $\Agr x$ holds, because the inner conditional takes the same branch on both sides.  Notice that the diverging loop at label
5 does not falsify the noninterference property since we only consider
pairs of executions in which $c3$ has terminated on the left.  The case where $\vhigh=0$ in both initial states is similar.

When $\leftex{\vhigh}\neq0$ and $\rightex{\vhigh}=0$ holds in the initial states, we reason in terms of a left-first sequential alignment.
Given that the left run of $c3$ is terminated, we have $\leftex{x}\geq\leftex{\vlow}$ preceding the point we start to reason about the right run.  This run first sets $x$ on the right to $\vlow$ and then the loop at label 9 increments it some nondeterministically chosen number of times.  We filter the right execution so that we maintain $\rightex{b}\geq0 \land \rightex{b}=\leftex{x}-\rightex{x}$ as an invariant of this loop.  Intuitively, we only permit executions on the right in which $\lhavc{8}{b}$ sets $b$ to the difference between $x$ on the two sides and $\lhavc{11}{b}$ decrements $b$ by $1$ to maintain the invariant.  Since the $\forall\exists$ judgment concerns existence of right executions, we must prove termination of the loop at label 9.  We do so using the value of $b$ as a variant of the loop.  By the chosen filtering conditions, this quantity decreases in each iteration.

Left-first sequential alignment is also used for the case when $\leftex{\vhigh} = 0$ and $\rightex{\vhigh} \neq 0$.  The key idea
here is to resolve $\lhavc{2}{x}$ on the right so that $\rightex{x} \geq \rightex{\vlow}$ holds, so we only consider right executions that don't diverge due to the loop at label 5.  This is again done by filtering right executions so we maintain agreement on $x$.

\endgroup

\paragraph{On heuristics}

The preceding proof sketch 
reflects some straightforward heuristics that are also applicable to $\forall\forall$ properties.
To reason about executions of two programs with similar control structure, consider lockstep alignment following the control structure.  For programs with different control structure, consider aligning the executions sequentially.  
In addition, for $\forall\exists$ judgments, align the left ($\forall$) execution before the right, so that filter conditions for right side nondeterminacy can refer to what happened on the left.

\paragraph{Data dependent alignment}

Prior works showed the need for alignments to be conditioned on program state, 
which can be realized in some forms of product automata~\cite{ShemerGSV19,ChurchillP0A19}. 
In a deductive proof of the $c3$ example, the case distinctions of the standard 4-way if-rule of RHLs directly enable the use of different alignments in different cases as we show later (\autoref{eg:c3ded}).  A deductive proof can also use the disjunction rule to introduce different cases in which different alignments can be used.
For loops, a number of published RHLs offer only restricted alignment patterns, but Beringer~\cite{Beringer11} formulated a loop rule that features auxiliary relations to designate the conditions under which iterations are aligned together or proceed on just one side.  Our logics use this rule.  Our results show that these ingredients suffice for deductive proofs to use any alignments expressed using automata.

Alignment conditions in product automata like those cited above are local in the sense of referring to the current (pair of) program states.  It is conceivable to formulate alignment notions or other relational reasoning globally, in terms of entire program executions (see~\autoref{sec:related}), but this is beyond the assertional methods on which we focus in this paper.

\paragraph{Alignment completeness.}

The examples show the value of good alignment for both $\forall\forall$ and $\forall\exists$ reasoning.
They also show that a deductive system must cater for relating both similar and different programs,
to express various alignments. For $\forall\exists$, filtering must be expressed somehow, 
while disallowing unjustified assumptions.
Our main results show that for a minimal set of rules it suffices to have one-sided rules
for primitives (e.g., \rn{rAsgnSkip} and \rn{rSkipAsgn} in \autoref{fig:RHL})
and same-structure rules for control structures (e.g., \rn{rSeq} in \autoref{fig:RHL}).
Known examples show the need for alignment conditioned on 
data~\cite{ShemerGSV19,ChurchillP0A19,BNNN19}; our main results show that it suffices to have conditional alignment in the loop rule.  
A key rule allows rewriting the programs into equivalent ones with different control structure.

In a human guided deductive proof one should rewrite the programs as little as possible, just enough to express a desired alignment as in the examples above.  
(A proof search procedure might select rewrites based on a similarity measure between programs to be related, but that is beyond the scope of this paper.)
Rewriting can also be used to derive convenient general rules for common cases (\autoref{fig:derivedRHL}).
Indeed, a few such derived rules suffice to capture the alignments described above for the $c3$ example.  
However, for the theoretical purpose of proving alignment completeness
we go to the opposite extreme, rewriting the program into a normal form like a fetch-execute loop.\footnote{Please note that we are using the term ``normal form'' for a syntactic notion (like A-normal form used in compilation~\cite{FlanaganSDF93}),
not in the sense of term rewriting systems~\cite{BaaderNipkow99}.}
By ``compiling'' programs we make it possible to obtain a deductive proof from an IAM-style proof in the form of an aligned and annotated product automaton.
Section~\ref{sec:future} revisits minimal rewriting in connection with alignment completeness as a criterion for RHLs.

Our approach to proving alignment completeness 
can be explained most simply by considering unary correctness.
Suppose we have an IAM proof that $c0$ satisfies an ordinary pre-post spec, so there is an inductive annotation of its control points.
Instrument the program with a fresh variable $pc$ serving as program counter. We write $\spc n$ to abbreviate $pc:=n$
and $\tpc n$ to abbreviate the expression $pc=n$
(to be precise, $\spc n$ is $\lassg{i}{pc}{n}$ for some arbitrary $i$).
The result looks 
as follows, using 6 as a final control point, which in IAM is annotated with the postcondition.
\label{page:c0} 
\[ 
\begin{array}{l}
\graybox{$c0^+:$} \quad
 \spc 1; \lassg{1}{x}{y} ;
        \spc 2; \keyw{do}^2 x > 0 \gcto 
                            \spc 3; \!\!\!\begin{array}[t]{l}
                            \keyw{if}^3  x\mod 2 = 0 \gcto \spc 4;\lassg{4}{x}{x-1} \\
                            \gcsep x\mod 2 \neq 0 \gcto \spc 5;\lassg{5}{x}{x-2} 
                             ~ \keyw{fi} ~ ; \spc 2 ~ \keyw{od} 
                             \,;\; \spc 6
                             \end{array}
                             \end{array}
 \]
Using simple transformations, this can be rewritten into the equivalent form $\spc 1; d0$ where 
\label{page:d0page}
\[ \begin{array}[t]{lllll}
\graybox{$d0:$}\quad \keyw{do} 
   & \tpc 1 \gcto x:=y; \spc 2 
      & \gcsep \tpc 3 \land x\mod 2=0 \gcto \spc 4 
      &  \gcsep \tpc 4 \gcto \lassg{}{x}{x-1}; \spc 2 \\
   &  \gcsep \tpc 2 \land x > 0 \gcto \spc 3 
      & \gcsep \tpc 3 \land x\mod 2\neq 0 \gcto \spc 5 
      & \gcsep \tpc 5 \gcto \lassg{}{x}{x-2}; \spc 2 \\
   &  \gcsep \tpc 2 \land x \ngtr 0 \gcto \spc 6  & & &  \keyw{od}
      \end{array}
\]
(When considering programs of this form, with explicit $pc$, the labels no longer matter and we omit them.)
For the loop invariant we use\footnote{We use the notation $\quant{operator\;}{variables}{range}{term}$ of
Dijkstra and Scholten~\cite{DijkstraScholten}.}
$1 \leq pc \leq 6 \land 
\quant{\land}{i}{0\leq i\leq 6}{ pc=i \imp an(i)} 
$ 
where $an(i)$ is the given annotation at control point $i$.
To use the loop rule (\autoref{fig:HLplus}) there is a premise for each guarded command in $d0$, 
and \emph{each of these corresponds closely to one of the VCs in the IAM proof}.
It turns out that this enables us to prove each premise  
using assignment, sequence, and consequence rules.
Thus we get that $\spc 1; d0$ satisfies the spec.  Then, since 
$\spc 1;d0$ is equivalent to $c0^+$, we have by the \rn{Rewrite} rule that $c0^+$ satisfies the spec. But $pc$ is fresh so the rule for ghost variables lets us erase the instrumentation.
Erasing produces extraneous skips; these are removed by another application of \rn{Rewrite},
completing the proof that $c0$ satisfies the spec.
All assertions used in this proof are essentially boolean combinations of 
the assertions in the given IAM proof.

This argument works for any IAM proof of a unary spec, owing to our normal
form \autoref{thm:normEquiv} that says any program, instrumented with $pc$, is equivalent to its corresponding automaton normal form.
The general result that one can obtain a deductive proof from an IAM proof is called Floyd completeness; we prove it for HL+ introduced later
(\autoref{thm:FloydComplete}).

The same approach works to prove alignment
completeness for a relational logic.  
Suppose we are given an IAM-style proof using an alignment product.
Compile each of the programs to its normal form. Instantiate the relational loop rule
(\rn{rDo} in \autoref{fig:RHL}, \rn{eDo} in \autoref{fig:ERHL}) using the
automaton's alignment conditions.  The side conditions of the loop rule
follow from assumed adequacy conditions of the automaton-based proof.  The
premises of the loop rule correspond to the verification conditions of the
automaton.

There are a number of technical challenges to work out these ideas in detail.
Many of the details in \autoref{sec:aut} for $\forall\forall$ alignment completeness 
are also used for the $\forall\exists$ alignment completeness 
result where there are additional complications about the 
filtering conditions and well-founded right-side alignment.

\section{The programming language and its \texorpdfstring{\protect{$\forall\forall$}}{∀∀} relational Hoare logic}\label{sec:logics}

This section formalizes the programming language and presents RHL+, the $\forall\forall$ relational logic.
Given that we will use KAT extensively, one might carry out the entire formal development using a KAT-like notation for programs (as is nicely done in \cite{OHearn2019} for example).
We make a different design decision, using a more conventional notation for programs.  This may facilitate comparison with other logics and it enables direct application of the logics to examples.  
It does require a little extra work for translating programs to KAT terms, which is largely confined to \autoref{sec:KATnf}.

\paragraph{Guarded command language}  

The labelled guarded command syntax is defined as follows, where 
$x$ ranges over a countable set $\Var$ of integer variables, $e$ ranges over integer and boolean expressions, $n$ ranges over integer literals.
\[ \begin{array}[t]{lcl}
c & ::= & \lskipc{n} \:\mid\: \lassg{n}{x}{e} \:\mid\: \lhavc{n}{x} \:\mid\: c;c 
       \:\mid\: \lifgc{n}{gcs} \:\mid\: \ldogc{n}{gcs} \\
gcs & ::= & e \gcto c \:\mid\: e \gcto c \gcsep gcs 
\end{array} \]
We omit details about expressions except to note that boolean expressions are given from some primitives $bprim$ and the logical operators are written $\land,\lor,\neg$.
A \dt{guarded command} has the form $e \gcto c$ where $e$ is a boolean expression.
The category $gcs$ is essentially non-empty lists of guarded commands and we sometimes treat it as such. The havoc command $\lhavc{n}{x}$ nondeterministically assigns any integer to $x$;
it can model inputs, randomization, and unknowns.  
The usual imperative control structures are special cases:
$\ifgc{e\gcto c\gcsep \neg e\gcto d}$ 
and $\dogc{e\gcto c}$. 

Later we define a predicate, $\ok$, on commands that says their labels are unique and positive.  Labels play an important role in some results, for which the $\ok$ condition is needed. But many definitions and results do not involve labels or require $\ok$; for those definitions and results we omit labels, meaning that any labels are allowed.

Programs act on \dt{variable stores}, i.e., total functions $\Var\to\Z$.
Later we consider automata with arbitrary sets of stores that need not be variable stores,
and the term ``state'' will refer to a store together with a control point.  
In this section we say simply ``store'', meaning variable store.
We assume expressions $e$ are always defined.
We write $\means{e}(s)$ for the value of expression $e$ in store $s$.  
In case $e$ is a boolean expression, the value is in $\{true,\mathit{false}\}$; otherwise 
it is in $\Z$.
We use standard big-step semantics and 
write \graybox{$\means{c}\, s\, t$} to express that from initial store $s$ the command $c$ can terminate with final store $t$.  
(See \autoref{fig:denot} in \autoref{sec:additional}.)
We call $\means{c}$ the \dt{denotation} of $c$.

The standard semantics for guarded commands considers $\lifgc{}{gcs}$ to fail if none of its guards is enabled~\cite{AptOld3}.
Modeling failure would clutter the semantics without shedding any light,
so we disallow such ifs, as follows.
Define \graybox{$\enab(gcs)$} as the disjunction of the guards.
For example, $\enab(x>0\gcto y:=1 \gcsep z < 1\gcto y:=2)$ is the expression $x>0 \lor z<1$.

\begin{defi}\label{def:lang}
A command $c$ is \dt{well formed} if 
(a) $c$ is typable in the sense that guards are boolean expressions and both integer and boolean operators are used sensibly, considering that all variables have type int; and
(b) $c$ satisfies the \dt{$\totalIf$} condition which says:
for every subprogram $\lifgc{}{gcs}$ of $c$, $\means{\enab(gcs)}=\means{true}$.
In the sequel we say \dt{command} to mean well formed command.\footnote{It is not difficult to enforce $\totalIf$ syntactically.
One way is given in \cite[Remark A.1]{BNN23v5}.} 
We refer to the language of well formed commands as the \dt{guarded command language}, (\dt{GCL}).
\end{defi}

\paragraph{Program logic.} 

Our relational logics are not based on unary HL, but for expository purposes we occasionally touch on HL.
The partial correctness judgment is written \graybox{$c:\spec{P}{Q}$} rather than the conventional $\{P\}c\{Q\}$.
To focus on what's interesting 
we treat assertions as shallowly embedded:\footnote{This approach is  
   popular~\cite{NipkowCSL02,OHearn2019,Pierce:SF2} because it factors
   out the issue of expressiveness~\cite{Cook78,AptOld3,WinskelBook}.}
so $P$ and $Q$ are sets of stores.
But we use formula notation for clarity, e.g., $P\land Q$ means their intersection.
A boolean expression $e$ in a formula stands for $\{s\mid\means{e}(s)=true\}$,
which we sometimes write as $\means{e}$. 
For a set $P$ of variable stores, we use substitution notation $\subst{P}{x}{e}$
with the standard meaning: $s$ is in $\subst{P}{x}{e}$ iff the updated store 
$\update{s}{x}{\means{e}(s)}$ is in $P$.
(We write $\update{s}{x}{v}$ for the store like $s$ but with $x$ mapped to $v$.)
Quantifiers, as operators on store sets, are defined by
$\graybox{$\allSet{x}{P}$} \eqdef \{ s \mid \all{v\in\Z}{s\in \subst{P}{x}{v}} \}$
and 
$\graybox{$\someSet{x}{P}$} \eqdef \{ s \mid \some{v\in\Z}{s\in \subst{P}{x}{v}} \}$.
Define \graybox{$\indep(x,P)$} to mean that $P=\someSet{x}{P}$,
that is, $P$ is independent of $x$.
Occasionally we restrict attention to assertions $P$ that are \dt{finitely supported}, meaning that $P$ is independent of all but finitely many variables.
We use $\models$ to indicate a \dt{valid} correctness judgment, defined as follows:
\begin{equation}\label{eq:valid}
\graybox{$\models c: \spec{P}{Q}$}
\eqdef
\mbox{For all $s,t$ such that 
$\means{c}\, s\, t$,
if $s\in P$ then $t\in Q$.} 
\end{equation}

\begin{figure}[t]
\begin{footnotesize}
\begin{mathpar}
\inferrule[Rewrite]{
    c: \spec{P}{Q} \\ c\kateq d
}{
    d: \spec{P}{Q} }

\inferrule[Ghost]{
    c: \spec{P}{Q} \\ \ghost(x,c) \\ \indep(x,P)\\ \indep(x,Q) 
}{
    \erase(x,c) : \spec{P}{Q}
}

\inferrule[Do]{
   c: \spec{e\land P}{P} \mbox{ for every $e\gcto c$ in $gcs$}
}{
    \dogc{gcs}: \spec{P}{P\land \neg \enab(gcs) }
}


\inferrule[Asgn]{}{ \lassg{}{x}{e} : \spec{\subst{P}{x}{e}}{P} }

\inferrule[Hav]{}{ 
\havc{x}:\spec{(\allSet{x}{P})}{P}
}

\inferrule[Skip]{}{ 
\skipc:\spec{P}{P} 
}

\inferrule[Seq]
{
c:\spec{P}{R} \\ d:\spec{R}{Q}
}{
c;d : \spec{P}{Q} 
}

\inferrule[If]
{
c: \spec{e\land P}{Q} \mbox{ for every $e\gcto c$ in $gcs$}
}{
    \ifgc{gcs}: \spec{P}{Q}
}

\inferrule[Conseq]
{
P\imp R \\ 
c : \spec{R}{S} \\
S\imp Q
}{
c : \spec{P}{Q} 
}

\inferrule[False]{}{ 
c:\spec{\mathit{false}}{P} 
}

\end{mathpar}
\end{footnotesize}
\vspace*{-3ex}
\caption{Proof rules of HL+.}
\label{fig:HLplus}
\end{figure}
The proof system we call \dt{HL+} comprises the standard rules of HL plus a rule for elimination of ghost\footnote{Also known as auxiliary variables, see~\cite{AptOld3}.} variables~\cite{FilliatreGP16} and a rule for rewriting programs
(\autoref{fig:HLplus}).
Define \graybox{$\ghost(x,c)$} to mean that variable $x$ occurs in $c$ only in assignments to $x$
and havocs of $x$.  
Define \graybox{$\erase(x,c)$} to be $c$ with every assignment to $x$, and havoc of $x$, replaced by $\skipc$.

Because we are using shallow embedding for assertions, HL+ does not need to be accompanied by a formal system for proving entailments between assertions.
But the logic also uses command equality $c\kateq d$, in rule \rn{Rewrite}.  
For this we rely on KAT as formalized in \autoref{sec:KATnf}.
For soundness of rule \rn{rewrite} we just need that the relation $\kateq$ implies equal denotations.

\paragraph{Relational specs and proof rules.}

For relational pre- and post-conditions we use (binary) relations on stores,
by shallow embedding just like for unary predicates.  
To express that a unary predicate holds in the left store of a pair,
we write \graybox{$\leftF{P}$} for the set of pairs $(s,t)$ where $s\in P$.
Similarly, $\rightF{P}$ is the set of $(s,t)$ where $t\in P$.
Combined with our coercion of boolean expressions $e$ to predicates,
$\leftF{e}$ says that $e$ is true in the left store.
Notation: 
\graybox{$\bothF{P\sep Q}$} abbreviates  $\leftF{P}\land\rightF{Q}$.
Recall from \autoref{sec:overview} that we write $\leftex{e}$ (resp.\ $\rightex{e}$) for the value of expression $e$ in the left (resp. right) state, in formulas like 
$\leftex{e}\leq\rightex{e'}$ which describes the set of $(s,t)$ such that 
$\means{e}(s) \leq \means{e'}(t)$.
As an abbreviation, define \(\graybox{$\Agr e$} \eqdef \leftex{e}=\rightex{e}\).

For relation $\R$ on stores we write 
$\subst{\R}{x\smSep}{e\smSep}$ for substitution of $e$ for $x$ in the left store.
Similarly $\subst{\R}{\smSep x}{\smSep e}$ substitutes on the right
and $\subst{\R}{x\smSep x'}{e\smSep e'}$ does both.
We write $\someSet{x\smSep x'}{\R}$ for quantification over $x$ on the left side and $x'$ on the right. 
Specifically, $\graybox{$\someSet{x\smSep x'}{\R}$} 
\eqdef\{(s,s')\mid \some{v,v'}{ (s,s')\in \subst{\R}{x\smSep x'}{v\smSep v'} } \}$.
Define \graybox{$\indep(x|x',\R)$} iff $\R = \someSet{x|x'}{\R}$.
We also need one-sided quantifier forms, and note that 
$(\allSet{x\smSep x'}{\R}) = \allSet{x\smSep}{(\allSet{\smSep x'}{\R})}
= \allSet{\smSep x'}{(\allSet{x\smSep}{\R})}$.
For store relations $\R,\S$,
a relational spec is written $\rspec{\R}{\S}$.
We use $\models$ to indicate valid judgment,
and define \graybox{$\models c\sep c': \rspec{\R}{\S}$} $\eqdef$
\[ 
\begin{array}{l}
\mbox{For all $s,s',t,t'$ such that 
$\means{c}\, s\, t$ and $\means{c'}\, s'\, t'$,}
\\
\mbox{if $(s,s')\in \R$ then $(t,t')\in \S$.}
\end{array}
\]
We write $\forall\exists$ specs as $\aespec{\R}{\S}$ and 
define the valid judgments by 
\graybox{$\models c\sep c': \aespec{\R}{\S}$} $\eqdef$
\begin{equation}\label{eq:aespec}
\begin{array}{l}
\mbox{For all $s,s',t$ such that $(s,s')\in \R$ and $\means{c}\, s\, t$,} 
\\ 
\mbox{there is $t'$ such that $\means{c'}\, s'\, t'$ and $(t,t')\in \S$.}
\end{array}
\end{equation}

\begin{figure}[t]
\begin{footnotesize}
\begin{mathpar}
\mprset{sep=1.4em}
\inferrule[rSkip]{}{
\skipc \sep \skipc : \rspec{\R}{\R}
}

\inferrule[rAsgnSkip]{}{
\lassg{}{x}{e} \sep \skipc  : \rspec{\subst{\R}{x|}{e|}}{\R}
}

\inferrule[rSkipAsgn]{}{
\skipc\sep \lassg{}{x}{e} : \rspec{\subst{\R}{|x}{|e}}{\R}
}

\inferrule[rHavSkip]{}{ 
\havc{x}\sep\skipc: \rspec{(\allSet{x\smSep}{\P})}{\P}
}

\inferrule[rSkipHav]{}{ 
\skipc\sep\havc{x}: \rspec{(\allSet{\smSep x}{\P})}{\P}
}

\inferrule[rSeq]{
  c\sep c' : \rspec{\R}{\S} \\
  d\sep d' : \rspec{\S}{\T}
}{
  c;d \Sep c';d' : \rspec{\R}{\T}
}

\inferrule[rIf]{
  c\sep c' : \rspec{\R\land \leftF{e}\land\rightF{e'}}{\S} 
\quad\mbox{for all $e\gcto c$ in $gcs$ and $e'\gcto c'$ in $gcs'$}
}{
  \ifgc{gcs} \Sep \ifgc{gcs'} : 
     \rspec{\R}{\S}
}

\inferrule[rDo]{
  {\begin{array}{l}
  c\sep \skipc : \rspec{\Q\land \leftF{e}\land\Lrel }{\Q}
\quad\mbox{for all $e\gcto c$ in $gcs$} 
\\
  \skipc\sep c' : \rspec{\Q\land \rightF{e'}\land\R }{\Q}
\quad\mbox{for all $e'\gcto c'$ in $gcs'$} 
\\
  c\sep c' : \rspec{\Q\land \leftF{e}\land\rightF{e'} \land \neg\Lrel \land \neg\R}{\Q}
\quad\mbox{for all $e\gcto c$ in $gcs$ and $e'\gcto c'$ in $gcs'$} 
\\
\Q\imp (\leftex{\enab(gcs)} = \rightex{\enab(gcs')}  
          \lor (\Lrel \land \leftF{\enab(gcs)})
          \lor (\R \land \rightF{\enab(gcs')}))
  \end{array}}
}{ 
  \dogc{gcs} \Sep \dogc{gcs'} : \rspec{\Q}{\Q\land \neg\leftF{\enab(gcs)}\land\neg\rightF{\enab(gcs')}}
}

\inferrule[rRewrite]{ c\sep c': \rspec{\R}{\S} \\ c\kateq d \\  c'\kateq d' }
{ d\sep d': \rspec{\R}{\S}  }

\inferrule[rConseq]{
  \P\imp \R \\
  c\sep c' : \rspec{\R}{\S} \\
  \S \imp \Q 
}{
  c\sep c' : \rspec{\P}{\Q} \\
}

\inferrule[rDisj]{
  c\sep c' : \rspec{\Q}{\S} \\
  c\sep c' : \rspec{\R}{\S} \\
}{
  c\sep c' : \rspec{\Q\lor\R}{\S} \\
}

\inferrule[rFalse]{}{
  c\sep c' : \rspec{\mathit{false}}{\R} 
}

\inferrule[rGhost]{
c\sep c': \rspec{\R}{\S} \\ \ghost(x,c) \\ \ghost(x',c') \\ 
     \indep(x|x',\R) \\ \indep(x|x',\S) }
{   \erase(x,c)\sep \erase(x',c') : \rspec{\R}{\S}  }

\end{mathpar}
\end{footnotesize}
\vspace*{-3ex}
\caption{The rules of RHL+.}
\label{fig:RHL}
\end{figure}

\begin{figure}[t]
\begin{footnotesize}
\begin{mathpar}
\inferrule[rLRseq]{
c\sep \skipc : \rspec{\P}{\Q} \\
\skipc\sep c' : \rspec{\Q}{\R} 
}{
c\sep c' : \rspec{\P}{\R}
}

\inferrule[rRLseq]{
\skipc\sep c' : \rspec{\P}{\Q}  \\
c\sep \skipc : \rspec{\Q}{\R}
}{
c\sep c' : \rspec{\P}{\R}
}

\inferrule[rAsgnAsgn]{}{
\lassg{}{x}{e} \sep \lassg{}{x'}{e'} : \rspec{\subst{\R}{x|x'}{e|e'}}{\R}
}

\inferrule[rHavHav]{}{
\havc{x}\sep \havc{x'} : \rspec{(\allSet{x\smSep x'}{\P})}{\P}
}

\inferrule[rAlgnIf]{
  c \sep c' : \rspec{\R\land \leftF{e}}{\S} \\
  d \sep d' : \rspec{\R\land\leftF{\neg e}}{\S} \\ 
  \R\imp \leftF{e}=\rightF{e'} 
}{
  \lifgc{}{e\gcto c \gcsep \neg e\gcto d} \Sep 
  \lifgc{}{e'\gcto c' \gcsep \neg e'\gcto d'} :\rspec{\R}{\S} 
}

\inferrule[rAlgnDo]{
  c \sep c' : \rspec{\R\land \leftF{e}\land\rightF{e'}}{\R}  \\
  \R\imp \leftF{e}=\rightF{e'}  
}{
  \ldogc{}{e\gcto c} \Sep 
  \ldogc{}{e'\gcto c'} :\rspec{\R}{\R\land\neg\leftF{e}\land\neg\rightF{e'}} 
}

\inferrule[rSeqSkip]{
  c \sep\skipc : \rspec{\R}{\Q} \\
  d \sep \skipc : \rspec{\Q}{\S} 
}{
  c;d \sep \skipc : \rspec{\R}{\S}
}

\inferrule[rIfSkip]{
  c\sep\skipc : \rspec{\R\land \leftF{e}}{\S} 
\quad\mbox{for all $e\gcto c$ in $gcs$}
}{
  \lifgc{}{gcs} \Sep \skipc : \rspec{\R}{\S}
}

\inferrule[rDoSkip]{ 
   c\sep \skipc : \rspec{\Q\land \leftF{e} }{\Q}
       \quad\mbox{for all $e\gcto c$ in $gcs$} 
}{
   \dogc{gcs} \Sep \skipc : \rspec{\Q}{\Q\land \neg\leftF{\enab(gcs)}}
}

\inferrule[rDisjN]{
  c\sep d : \rspec{\R_i}{\S} \mbox{ for all $i\in X$} \\
 \mbox{$X$ finite}
}{
  c\sep d : \rspec{\quant{\lor}{i}{i\in X}{\R_i}}{\S} \\
}

\end{mathpar}
\end{footnotesize}
\vspace*{-3ex}
\caption{Some derived rules of RHL+.}
\label{fig:derivedRHL}
\end{figure}

\autoref{fig:RHL} gives the proof rules of the $\forall\forall$ logic RHL+.
The $\forall\exists$ logic appears in \autoref{sec:eacomplete};
until then we focus on $\forall\forall$.
We are not aware of a prior RHL that has been formulated for GCL,
but the rules are straightforward adaptations of rules found in prior work.
In particular, \rn{rIf} generalizes the standard 4-way if-rule:
instead of $2\times 2$ cases there are $i\times j$ where $i$ and $j$ are the number of branches left and right. 

Rules such as the same-branch-if rule popular in RHLs~\cite{Benton:popl04,Yang07relsep} 
are easily derived, see \rn{rAlgnIf} and \rn{rAlgnDo} in 
\autoref{fig:derivedRHL}.  It is also easy to derive 
rules relating different control structures, in particular the ``one sided'' rules like \rn{rSeqSkip} and \rn{rSeqIf} 
in \autoref{fig:derivedRHL}.  These are useful for reasoning by sequential alignment which is embodied in rules \rn{rLRseq} and \rn{rRLseq}.

Rule \rn{rDoSkip} is derived as follows.  
Observe that $\skipc\kateq\dogc{\mathit{false}\gcto\skipc}$,
so we get the conclusion of \rn{rDoSkip} by \rn{rRewrite} from 
$\dogc{gcs} \sep \dogc{\mathit{false}\gcto\skipc}
: \rspec{\Q}{\Q\land\neg\rightF{\enab(gcs)}}$.
This we prove using \rn{rDo} with $\Lrel,\R:=\mathit{true},\mathit{false}$.
The left-only premises have the form 
$c\sep\skipc : \rspec{\Q\land \leftF{e}\land\mathit{true}}{\Q}$.
They follow from the premises of \rn{rDoSkip} using \rn{rConseq} to 
add the conjunct $\mathit{true}$.\footnote{By shallow embedding, \rn{rConseq} is 
never needed to manipulate equivalent relations, but we mention it for clarity.} 
The right-only and joint premises are proved by \rn{rFalse} and \rn{rConseq}.
The side condition of this instantiation of \rn{rDo} simplifies to  
$\Q\imp(\leftex{\enab(gcs)}=\rightex{\mathit{false}})\lor\leftF{\enab(gcs)}$
and the consequent simplifies to true.\footnote{The side condition in \rn{rDo} uses notation
$\leftex{\enab(gcs)} = \rightex{\enab(gcs')}$ for a store relation that some readers may prefer to write as $\leftex{\enab(gcs)} \iff \rightex{\enab(gcs')}$.  Similarly for the side conditions of \rn{rAlignIf} and \rn{rAlignDo}.}
Proofs of the other rules in \autoref{fig:derivedRHL} are similar.

Unlike some RHLs in the literature, RHL+ does not make use of unary correctness judgments.  
However, unary correctness can be encoded as relational correctness, in the sense that 
validity of $c:\spec{P}{Q}$ 
is equivalent to validity of the judgment
$c\sep\skipc: \rspec{\leftF{P}}{\leftF{Q}}$
and also 
$\skipc\sep c: \rspec{\rightF{P}}{\rightF{Q}}$.
We return to this in \autoref{sec:unaryRevisited}.

\begin{exa}\label{exa:majorize}
Consider these programs adapted from~\cite{NagasamudramN21} (and in turn from~\cite{Francez83}). 
\begin{small}
\[
\begin{array}{ll}
\graybox{$c4:$}
& \assg{y}{x}; \assg{z}{24}; \assg{w}{0}; 
\\
& \dogc{
   y\neq 4 \gcto
      \ifgc{ w \mod 2 = 0 \gcto \assg{z}{z*y}; \assg{y}{y-1} 
           \gcsep w\mod 2 \neq 0 \gcto \skipc}
      ; \assg{w}{w+1} }
\\[.3ex]
\graybox{$c5:$}
& \assg{y}{x}; \assg{z}{16}; \assg{w}{0}; 
\\
&    \dogc{y\neq 4 \gcto 
          \ifgc{w \mod 3 = 0 \gcto \assg{z}{z*2}; \assg{y}{y-1}
              \gcsep w \mod 3 \neq 0 \gcto \skipc}
    ; \assg{w}{w+1} }
\end{array}
\]
\end{small}
For input $x$ with $x\geq 4$ these compute factorial and exponent respectively.  Without reasoning about those functions, however, one can show that factorial majorizes exponent for arguments at least 4.  Formally:
$c4 \sep c5 : \rspec{\Agr x\land \leftex{x} \geq 4}{\leftex{z}>\rightex{z}}$.
The reader may enjoy to use the proof rules to prove the spec, 
using rule \rn{rDo} with loop alignment conditions $\Lrel := \leftF{ w\mod 2\neq 0 }$ and $\R := \rightF{  w'\mod 3\neq 0 }$.  
As a loop invariant, try $\Agr y\land \leftex{y}\geq 4 \land \leftex{z}>\rightex{z}
\land \rightex{z}>0$.

Another approach is to rewrite the loops, unfolding them 2 (resp. 3) times.
This is not possible, however, if we replace literals 2 and 3 by 
an input variable $v$ with precondition $\leftF{v\geq 1}\land\rightF{v\geq 1}$. 
Of course the example is contrived, but such data-dependent alignment patterns arise in settings such as equivalence checking for optimizing compilers.
\qed
\end{exa}

\section{Alignment automata, adequacy, and verification conditions}\label{sec:aut}

This section lays groundwork for alignment completeness, defining alignment automata 
for $\forall\forall$ properties and the verification conditions for program alignment automata. 
The definitions are adapted to $\forall\exists$ properties in \autoref{sec:filt}.

\subsection{Automata, alignment automata and adequacy}

We adapt a number of technical definitions from~\cite{NagasamudramN21},
where automata are formulated in a way that can represent program semantics using, in essence, a finite control flow graph.

\begin{defi}\label{def:automaton}
An \dt{automaton} is a tuple 
\( (Ctrl,Sto,\init,\fin,\trans) \)
where $Sto$ is a set (called the data stores),
$Ctrl$ is a finite set (the control points)
that contains distinct elements $\init$ and $\fin$,
and ${\trans} \subseteq (Ctrl\times Sto)\times(Ctrl\times Sto)$ is the transition relation.\footnote{The symbol $\trans$ here has nothing to do with 
the store update notation, e.g., $\update{s}{x}{v}$,  where the square brackets
should prevent any confusion.}
We require 
$(n,s)\trans (m,t)$ to imply $n\neq \fin$ and $n\neq m$ 
and call these the \dt{finality} and \dt{non-stuttering} conditions respectively.
Absence of stuttering loses no generality and facilitates definitions involving product automata.
A \dt{state} of the automaton is an element of $Ctrl\times Sto$.
\end{defi}

We use the term ``alignment product'' informally, in reference to various constructions in the literature.
We define a particular construction that we call alignment automaton.

\begin{defi}\label{def:alignProd}
Let $A = (Ctrl,Sto,\init,\fin,\trans)$ and 
$A' = (Ctrl',Sto',\init',\fin',\trans')$ be automata.
Let $L$, $R$, and $J$ be subsets of $(Ctrl\times Ctrl')\times (Sto\times Sto')$
that are \dt{live for} $A,A'$, meaning that:
\( \begin{array}[t]{l} 
\all{ ((n,n'),(s,s'))\in L }{(n,s)\in\dom(\trans) } \\
\all{ ((n,n'),(s,s'))\in R }{(n',s')\in\dom(\trans') } \\
\all{ ((n,n'),(s,s'))\in J }{(n,s)\in\dom(\trans) \land (n',s')\in\dom(\trans')}
\end{array}\). \\
The \dt{alignment automaton} \graybox{$\aprod(A,A',L,R,J)$} is 
the automaton 
\[ ((Ctrl\times Ctrl'), (Sto\times Sto'), (\init,\init'), (\fin,\fin'), \biTrans) \]
where \graybox{$\biTrans$} is defined by:
$((n,n'),(s,s')) \biTrans ((m,m'),(t,t'))$ iff one of these conditions holds:
\begin{description}
\item[LO]\quad
$((n,n'),(s,s'))\in L$ and
$(n,s)\trans(m,t)$ and $(n',s')=(m',t')$
\item[RO]\quad
$((n,n'),(s,s'))\in R$ and 
$(n,s)=(m,t)$ and $(n',s')\trans'(m',t')$
\item[JO]\quad
$((n,n'),(s,s'))\in J$ 
and $(n,s)\trans(m,t)$ and $(n',s')\trans'(m',t')$
\end{description}
\end{defi}

Notice that the states of $\aprod(A,A',L,R,J)$ are 
$((Ctrl\times Ctrl')\times (Sto\times Sto'))$.
So the \dt{alignment conditions} $L$, $R$, and $J$ are sets of alignment automaton states. 
We write $[n|n']$ for the set of states where control is at $(n,n')$, i.e.,
\( \graybox{$[n|n']$} \eqdef \{ ((i,i'),(s,s')) \mid n=i \land n'=i'\} \).
For example $[\fin|\fin']$ is the set of terminated states.  
Let $\graybox{$[n|*]$} \eqdef \{ ((i,i'),(s,s')) \mid n=i \}$.

Roughly speaking, taking $L$ and $R$ to be false and $J$ true, we obtain an automaton that runs $A$ and $A'$ in lockstep.
Taking $L,R,J$ all true, we obtain a nondeterministic alignment automaton that represents very many alignments.  
Taking $L,R,J$ all false, we obtain an alignment automaton that represents no alignments whatsoever.  
Taking $J$ to be false, $L$ to be $[*|\init']$, and $R$ to be $[\fin|*]$, we obtain an automaton that runs only $A$, unless it terminates, in which case it proceeds to run $A'$ ---the sequential alignment.\footnote{To be precise, 
one must ensure that $L,R,J$ are live (per \autoref{def:alignProd}).
Later we construct automata from programs, and for such automata every state has a successor except when control is final (\autoref{lem:autLive}).  
Hence, for programs, an arbitrary triple $(L,R,J)$ can be made live very simply, as
$(L\setminus[\fin|*],R\setminus[*|\fin'],J\setminus[\fin|\fin'])$
where $\setminus$ is set subtraction.}

\begin{defi}\label{def:adequacy}
Consider an alignment automaton $\aprod(A,A',L,R,J)$
and relation $\P\subseteq Sto\times Sto'$.
The alignment automaton  is \dt{$\P$-adequate} provided for
all $(s,s')\in\P$ and $t,t'$ with
$(\init,s)\trans^*(\fin,t)$ and $(\init',s')\trans'^*(\fin',t')$,
we have
$((\init,\init'),(s,s'))\biTrans^* ((\fin,\fin'),(t,t'))$.
\end{defi}

An alignment automaton may be adequate for reasons that are specific to the 
underlying automata, for example it may not cover all traces but still cover all outcomes.
We focus on alignment automata that are adequate in the sense that they cover 
all traces, which can be ensured as follows.
%
Given a relation $\P\subseteq Sto\times Sto'$, an alignment automaton is
\dt{manifestly $\P$-adequate} provided 
that $L \lor R \lor J \lor [\fin|\fin']$ is \dt{$\P$-invariant}.
This means $L\lor R \lor J \lor [\fin|\fin']$ holds at every state reachable from some 
$((\init,\init'),(s,s'))$ such that $(s,s')\in\P$.

\begin{lem}\label{lem:manifest}
\upshape
If alignment automaton $\aprod(A,A',L,R,J)$ is manifestly $\P$-adequate 
then it is $\P$-adequate (in the sense of \autoref{def:adequacy}).
\end{lem}

\subsection{Correctness of automata, and the inductive assertion method}

Generalizing slightly from \autoref{sec:logics},
we consider specs $\spec{P}{Q}$ where $P$ and $Q$ are sets of automaton stores,
not necessarily variable stores.
Satisfaction of a spec by an automaton is written 
\graybox{$A \models \spec{P}{Q}$} and defined to mean:
For all $s,t$ such that $(\init,s)\trans^*(\fin,t)$,
if $s\in P$ then $t\in Q$.

Let $A$ and $A'$ be automata with store sets $Sto$ and $Sto'$ respectively.
Generalizing slightly from \autoref{sec:logics},
we consider relational specs $\rspec{\R}{\S}$
where $\R$ and $\S$ are relations from $Sto$ to $Sto'$.
Satisfaction of the spec by the pair $A,A'$ is written 
\graybox{$A,A'\models\rspec{\R}{\S}$} and defined 
to mean, for any $s,s',t,t'$:
\begin{quote}
\mbox{If $(\init,s)\trans^*(\fin,t)$ and $(\init',s')\trans'^*(\fin',t')$
and $(s,s')\in\R$ then $(t,t')\in\S$.}
\end{quote}

A store relation $\Q$ for $A,A'$ can be seen as a predicate on the stores
of an alignment automaton $\aprod(A,A',L,R,J)$ because the latter are pairs of stores.
Hence, for relational spec $\rspec{\Q}{\S}$, 
the unary spec $\spec{\Q}{\S}$ makes sense 
for $\aprod(A,A',L,R,J)$.

\begin{lem}[adequacy semantically sound and complete]\label{prop:adequate-sound}
\upshape
Suppose that $\aprod(A,A',L,R,J)$ is $\Q$-adequate.
Then $\aprod(A,A',L,R,J)\models \spec{\Q}{\S}$
if and only if 
$A,A'\models\rspec{\Q}{\S}$.
\end{lem}


Given automaton $A$ and spec $\spec{P}{Q}$,
an \dt{annotation} is a function $an$ from control points to store predicates 
such that $P\imp an(\init)$ and $an(\fin)\imp Q$.
The requirement $\init\neq\fin$ in \autoref{def:automaton} ensures that annotations exist for any spec.\footnote{In Floyd's formulation, an annotation only needs to be defined on a subset of control points that cut every loop in the control flow graph.  Such an annotation can always be extended to one for all control points.}  
We lift $an$ to a function $\hat{an}$ that yields states:
$\graybox{$\hat{an}(n)$} = \{ (m,s) \mid s\in an(n), m\in Ctrl \}$.
Put differently:
\begin{equation}\label{eq:hatALT}
(m,s)\in\hat{an}(n) \quad\mbox{iff}\quad s\in an(n) \qquad\mbox{(for all $m,n,s$)} 
\end{equation}
For each pair $(n,m)$ of control points
there is a \dt{verification condition} (\dt{VC}):
\begin{equation}\label{eq:VC}
\POST(\tranSeg{n,m})(\hat{an}(n)) \subseteq \hat{an}(m) 
\end{equation}
Here $\POST$ gives the direct image (i.e., strongest postcondition) of a relation,\footnote{Defined for any set $X$ and relation $R$ 
   by $t\in\graybox{$\POST(R)(X)$}$ iff $\some{s}{(s,t)\in R \land s\in X}$.}
and \graybox{$\tranSeg{n,m}$} is the \dt{fixed-control transition relation} restricted to starting control point $n$ and ending   $m$, i.e.,
\begin{equation}\label{eq:tranSeg}
\mbox{
$(i,s)\tranSeg{n,m}(j,t) \; $ iff $\; i=n$, $j=m$, and $(n,s)\trans(m,t)$
}
\end{equation}
It is well known that (\ref{eq:VC}) is equivalent to \( \hat{an}(n) \subseteq \WP(\tranSeg{n,m})(\hat{an}(m)) \)
using the universal preimage operator $\WP$.\footnote{Defined for any set $X$ and relation $R$ by $s\in\graybox{$\WP(R)(X)$}$ iff $\all{t}{(s,t)\in R \imp t\in X}$.}

The VC (\ref{eq:VC}) says that for every transition from control point $n$ and store $s\in an(n)$,
if the step goes to control point $m$ with store $t$, then $t$ is in $an(m)$.
Annotation $an$ is \dt{valid} if the VC is  true for every pair $(n,m)$ of control points.
In most automata, including those we derive from programs, some pairs $(n,m)$ have no transitions,
in other words $\tranSeg{n,m}$ is empty.
In that case the VC (\ref{eq:VC}) is true regardless of $an(n)$ and $an(m)$.

In case the stores of $A$ are variable stores,
a set $S$ of $A$-states is \dt{finitely supported} provided that 
for each control point $n$ the set of stores $\{t \mid (n,t)\in S \}$ is finitely supported.  Similarly for states of an alignment product.
We say $A$ is finitely supported if its transition relation acts on finitely many variables.\footnote{This can be formalized as follows: $\dom(\trans)$ is finitely supported and
 for any $x$ outside the support of $\dom(\trans)$ and any 
 states $(n,s)$ and $(m,t)$, if 
 $(n,s)\trans(m,t)$ then 
 $(n,\update{s}{x}{i})\trans(m,\update{t}{x}{i})$ 
 for all $i\in\nat$.}
An annotation $an$ is finitely supported provided $an(n)$ is, for all control points $n$.

\begin{lem}[semantic soundness and completeness of IAM~\cite{Floyd67}]\label{prop:IAM}
\upshape
There is a valid annotation of $A$ for $\spec{P}{Q}$ iff
$A\models \spec{P}{Q}$.
Moreover, in case $A$ acts on variable stores and $P,Q$ are finitely supported,
$A\models \spec{P}{Q}$ implies there is a finitely supported valid annotation.
\end{lem}

\begin{cor}[soundness of alignment automata]\label{cor:relIAM}
\upshape
Suppose that $\aprod(A,A',L,R,J)$ is $\Q$-adequate and 
$an$ is an annotation of $\aprod(A,A',L,R,J)$ for $\spec{\Q}{\S}$.
If $an$ is valid then $A,A'\models \rspec{\Q}{\S}$.
\end{cor}

Together, \autoref{cor:relIAM} and \autoref{lem:manifest} 
provide a method to verify
$A,A'\models \rspec{\Q}{\S}$:
Find alignment conditions $L,R,J$ and annotation $an$ of $\aprod(A,A',L,R,J)$
such that the annotation is valid and 
$\breve{an}(i,j)\imp L \lor R \lor J \lor [\fin|\fin']$ for every $i,j$.
This uses the abbreviation 
\( \graybox{$\breve{an}(n,n')$} \eqdef \hat{an}(n,n')\land [n|n'] \).
This implication ensures manifest adequacy.


\begin{cor}[semantic completeness of alignment automata]\label{cor:relIAMcomplete}
\upshape
Suppose $A,A'\models \rspec{\S}{\T}$.
Then there are $L,R,J$ and a valid annotation $an$ of $\aprod(A,A',L,R,J)$ for $\spec{\S}{\T}$ 
such that $\breve{an}(i,j)\imp L \lor R \lor J \lor [\fin|\fin']$ for every $i,j$.
Moreover, 
if $A$ and $A'$ act on variable stores, and $A,A',\S,\T$ are finitely supported,
then so are $L$, $R$, $J$, and $an$.
\end{cor}

\subsection{Automata from programs and their VCs} 

\begin{figure}[t] 
\begin{footnotesize}
\begin{mathpar}

\inferrule{}{ \config{\lhavc{n}{x}}{s} \ctrans \config{\lskipc{-n}}{\update{s}{x}{v}}}

\inferrule{}{ \config{\lskipc{n};c}{s} \ctrans \config{c}{s} }

\inferrule{ e\gcto c \mbox{ is in } gcs \\ \means{e}(s) = \mathit{true} }
{ \config{ \lifgc{n}{gcs} }{s} \ctrans \config{c}{s} }

\inferrule{ \enab(gcs)(s) = \mathit{false} }
{ \config{\ldogc{n}{gcs} }{s} \ctrans  \config{\lskipc{-n}}{s}
}

\inferrule{}{ \config{\lassg{n}{x}{e}}{s} \ctrans \config{\lskipc{-n}}{\update{s}{x}{\means{e}(s)}} }

\inferrule{ e\gcto c \mbox{ is in } gcs \\ \means{e}(s) = \mathit{true} }
{ \config{ \ldogc{n}{gcs} }{s}  \ctrans 
  \config{ c;\ldogc{n}{gcs} }{s} 
}

\inferrule{
\config{c}{s} \ctrans \config{d}{t} }
{ \config{c;b}{s} \ctrans \config{d;b}{t} }

\end{mathpar}
\end{footnotesize}
\vspace*{-3ex}
\caption{Transition semantics (with $n$ and $v$ ranging over $\Z$).}
\label{fig:progtrans}
\end{figure}

Labels on commands serve as basis for defining the automaton for a program,
to which end we make the following definitions (adapted from~\cite{NagasamudramN21}). 
Write \graybox{$\ok(c)$} to say no label in $c$ occurs more than once and all labels are positive.
Write $\lab(c)$ for the label of $c$ 
and $\labs(c)$ for the set of labels that occur in $c$.
(The only non-obvious case is $\lab(c;d) \eqdef \lab(c)$,
see \autoref{fig:lab}.) 
For small-step semantics,
we write $\config{c}{s} \ctrans \config{d}{t}$ if command $c$ with store $s$ transitions to continuation command $d$ and store $t$ (see \autoref{fig:progtrans}).

Negative labels are used in the small-step semantics
in a way that facilitates defining the automaton of a program.
In a configuration reached from an $\ok$ command, 
the only negative labels are those introduced by the transition for assignment
and the transition for termination of a loop.
For every $c,s$, either $\config{c}{s}$ has a successor via $\ctrans$
or $c$ is $\lskipc{n}$ for some $n\in\Z$.  (This relies on $\totalIf$ of Def.~\ref{def:lang}.)
When a negative label is introduced, the configuration is either terminated 
or has the form $\config{\lskipc{-n};c}{s}$
in which case the next transition is 
$\config{\lskipc{-n};c}{s} \ctrans \config{c}{s}$.

\begin{lem}
\upshape
For any $c,s,t$, we have 
$\means{c}\, s\, t$
iff 
$\config{c}{s} \ctrans^* \config{\lskipc{n}}{t}$
for some $n$.
\end{lem}

\begin{figure}[t]
\begin{footnotesize}
\[
\begin{array}{l@{\;}c@{\;}l}
\fsuc(n,c;d,f)                 & \eqdef & \fsuc(n,c,\lab(d)) \mbox{ , if $n\in\labs(c)$} \\
                               & \eqdef & \fsuc(n,d,f) \mbox{ , otherwise} \\ 
\fsuc(n,\lifgc{n}{gcs}, f) &\eqdef & f \\
\fsuc(m,\lifgc{n}{gcs}, f) &\eqdef & \fsuc(m,c,f) \mbox{ , if $e\gcto c \in gcs$ and $m\in\labs(c)$} \\
\fsuc(n,\ldogc{n}{gcs}, f) &\eqdef & f \\
\fsuc(m,\ldogc{n}{gcs}, f) &\eqdef & \fsuc(m,c,n) \mbox{ , if $e\gcto c \in gcs$ and $m\in\labs(c)$} 
\\
\multicolumn{3}{l}{ 
\fsuc(n,\lskipc{n},f)           \;\eqdef\; 
\fsuc(n,\lassg{n}{x}{e},f)      \;\eqdef\; 
\fsuc(n,\lhavc{n}{x},f)         \;\eqdef\; f } \\
\end{array}
\]
\end{footnotesize}
\vspace*{-3ex}
\caption{Following successor \graybox{$\fsuc(n,c,f)$},
assuming $\ok(c)$, $n\in\labs(c)$, and $f\notin\labs(c)$.
}
\label{fig:fsuc}
\end{figure}

Write \graybox{$\sub(n,c)$} for the sub-command of $c$ with label $n$, if $n$ is in $\labs(c)$.
Let $c$ and $\fin$ be such that $\ok(c)$ and $\fin\notin\labs(c)$.
We write $\fsuc(n,c,\fin)$ for the \dt{following successor} of $n$ in the control flow graph of $c$, i.e., the control successor of the subprogram at $n$,
in the sense made precise in \autoref{fig:fsuc}.
Note that $\fin$ serves as a final or exit label.
The key case in the definition is for loops: the following successor is the control point after termination of the loop.
For the running example, we have $\fsuc(2,c0,6) = 6$ and $\fsuc(4,c0,6)=2$.
Define \graybox{$\okf(c,f)$} (``ok, fresh'') to abbreviate the conjunction
of $\ok(c)$ and $f\notin\labs(c)$.

If $\okf(c,f)$ then we define the \dt{automaton of $c$ for $f$}, written \graybox{$\aut(c, f)$}, with control set 
$\labs(c)\union \{ f \}$ and transitions in accord 
with the small-step semantics.

\begin{defi}\label{def:aut}
Suppose $\okf(c,f)$. 
The \dt{automaton of $c$ for $f$}, written \graybox{$\aut(c, f)$}, is
\( (\labs(c)\union \{ f \}, (\Var\to\Z), \lab(c), f, \trans) \) 
where 
$(n,s)\trans (m,t)$ iff either
\begin{small}
\begin{ditemize}
\item $
\some{d}{ \config{\sub(n,c)}{s}\ctrans\config{d}{t} 
   \land \lab(d) > 0 \land m = \lab(d) }$, or 
\item 
$\some{d}{ \config{\sub(n,c)}{s}\ctrans\config{d}{t} \land \lab(d) < 0 \land m = \fsuc(n,c,f) }$, or 
\item $\sub(n,c)=\lskipc{n} \land m = \fsuc(n,c,f) \land t = s$
\end{ditemize}
\end{small}
\end{defi}
The first two cases use the semantics of \autoref{fig:progtrans}
for a sub-command on its own.  
The second case uses $\fsuc$ for a sub-command that takes a terminating step 
(either assignment, havoc, or loop).
The third case handles skip, which on its own would be stuck 
but which should take a step when it occurs as part of a sequence.\footnote{For example, with $\ctrans$ we have
$\config{\lskipc{n};\lskipc{m}}{s} \ctrans \config{\lskipc{m}}{s}$
but $\lskipc{n}$ by itself has no transitions and the first two cases 
do not apply.  Owing to the third case we have $(n,s)\trans(m,s)\trans(f,s)$.
This also shows that the automaton steps are not in exact correspondence with 
those via $\ctrans$ but this does not matter.
}

\begin{exa} 
The automaton for the command $c3$ on page~\pageref{pg:c3} in \autoref{sec:overview},
using 12 as final label, is depicted in \autoref{fig:c3aut}. 
One can see an alignment automaton for $c3$ in \autoref{fig:filteredProductEx}, 
ignoring the dashed boxes in the figure.
\qed
\end{exa}

\begingroup
\newcommand{\vhigh}{\ensuremath{\mathit{high}}}
\newcommand{\vlow}{\ensuremath{\mathit{low}}}

\begin{figure}[t]
\begin{footnotesize}
\begin{tikzpicture}
  [-{Latex[length=2mm]},every state/.style={fill=gray!10},initial text=$ $,auto,
  node distance=2.5cm,
  line width=0.1mm,
  scale=0.9,
  transform shape]

  \node[state] (n1) {$1$};
  \node[state, right of=n1, yshift=5em] (n2) {$2$};
  \node[state, right of=n2] (n3) {$3$};
  \node[state, right of=n3, yshift=2.5em] (n4) {$4$};
  \node[state, right of=n3, yshift=-2.5em] (n5) {$5$};
  \node[state, right of=n5] (n6) {$6$};
  \node[state, right of=n6, yshift=-2.5em] (f) {$f$};
  \node[state, right of=n1, yshift=-5em] (n7) {$7$};
  \node[state, right of=n7] (n8) {$8$};
  \node[state, right of=n8] (n9) {$9$};
  \node[state, below of=n9, yshift=2em, xshift=-4em] (n10) {$10$};
  \node[state, below of=n9, yshift=2em, xshift=4em] (n11) {$11$};

  \draw
  (n1) edge[bend left=35] node[above, xshift=-1em] {$\vhigh\neq0$} (n2)
  (n1) edge[bend right=35] node[below, xshift=-1em] {$\vhigh=0$} (n7)
  (n2) edge node[above] {$\havc{x}$} (n3)

  (n3) edge[bend left] node[above,xshift=-.5em] {$x\geq\vlow$} (n4)
  (n3) edge[bend right] node[below,xshift=-.5em] {$x<\vlow$} (n5)

  (n4) edge[bend left] node[above] {$\skipc$} (f)

  (n5) edge[bend left] node[above] {$\mathit{true}$} (n6)
  (n6) edge[bend left] node[below] {$\skipc$} (n5)
  (n5) edge[bend right=45] node[above] {$\mathit{false}$} (f)

  (n7) edge node[above] {$x:=\vlow$} (n8)
  (n8) edge node[above] {$\havc{b}$} (n9)

  (n9) edge[bend right] node[left] {$b\neq0$} (n10)
  (n10) edge[bend right] node[below] {$x:=x+1$} (n11)
  (n11) edge[bend right] node[right] {$\havc{b}$} (n9)

  (n9) edge[bend right] node[below] {$b=0$} (f)
  ;
\end{tikzpicture}
\end{footnotesize}
\vspace*{-2ex}
\caption{Automata for program $c3$ (in \autoref{sec:overview}): $\aut(c3,f)$ where $f=12$.}\label{fig:c3aut}
\end{figure}
\endgroup

\begin{lem}\label{lem:autLive}
\upshape
The only stuck states of $\aut(c, f)$ are terminated ones, i.e.,  where the control is $f$.
\end{lem}

\begin{lem}[automaton consistency]\label{lem:autConsistent}
\upshape
Suppose $\okf(c,f)$ and  let $n=\lab(c)$.
For any $s,t$ we have 
$\means{c}\, s\, t $ iff $(n,s)\trans^*(f,t)$ in $\aut(c,f)$.
Hence 
$\models c:\spec{P}{Q}$ iff $\aut(c,f)\models\spec{P}{Q}$
for any $P,Q$.
Also $ \models c\sep c':\rspec{\Q}{\S}$ iff $\aut(c,f),\aut(c',f')\models\rspec{\Q}{\S}$
for any $\Q,\S$ and 
$\okf(c',f')$. 
The same holds for the $\aespecSym$ judgment.
\end{lem}

By inspection of the transition semantics in \autoref{fig:progtrans},
there are six kinds of transitions
for $\ctrans$ and also for the automaton relation $\trans$ derived from it.
(There is a seventh rule for $\ctrans$ that says a transition can occur for the first command in a sequence, but that is used together with one of the other six, 
and it is not relevant to \autoref{def:aut}.)
Thus there are six kinds of verification conditions, which can be derived from 
the semantic definitions.

\begin{figure}[t]
\begin{footnotesize}
\begin{tabular}{lll}
if $\sub(n,c)$ is\ldots & and $m$ is\ldots & then the VC for $(n,m)$ is equivalent to\ldots \\\hline

$\lskipc{n}$ &
$\fsuc(n,c,f)$ & $an(n)\imp an(m)$ 
\\

$\lassg{n}{x}{e}$ &
$\fsuc(n,c,f)$ & $an(n)\imp \subst{an(m)}{x}{e}$ 
\\

$\lhavc{n}{x}$ &
$\fsuc(n,c,f)$ & $an(n)\imp \allSet{x}{an(m)}$ 
\\

$\lifgc{n}{gcs}$ & $\lab(d)$ where $e\gcto d$ is in $gcs$
& $an(n)\land e \imp an(m)$ 
\\

$\ldogc{n}{gcs}$ & $\lab(d)$ where $e\gcto d$ is in $gcs$
& $an(n)\land e \imp an(m)$ 
\\

$\ldogc{n}{gcs}$ & $\fsuc(n,c,f)$ & $an(n)\land \neg \enab(gcs) \imp an(m)$
\\[.5ex]
\multicolumn{3}{l}{
In all other cases, there are no transitions from $n$ to $m$ so the VC is $true$ by definition.}
\end{tabular}
\end{footnotesize}
\vspace*{-1ex}
\caption{VCs for the automaton $aut(c,f)$ of $\ok$ program $c$ and annotation $an$.}
\label{fig:VC}
\end{figure}

\begin{lem}[VCs for programs]\label{lem:VCprog}
\upshape
Consider $c$ and $f$ such that $\ok(c,f)$, and 
let $an$ be an annotation of $\aut(c,f)$.
For each pair $n,m$ of labels, the VC of equation (\ref{eq:VC}) 
can be expressed as in \autoref{fig:VC}.
\end{lem}
\begin{proof}
We give two cases. The other cases are similar.

\smallskip

\textbf{Case} $\lskipc{n}$ with $m = \fsuc(n,c,f)$.
To show: the VC is equivalent to $an(n)\imp an(m)$.
The VC of (\ref{eq:VC}) is the first line of the following calculation.
\[\begin{array}{lll}
    & \POST(\tranSeg{n,m})(\hat{an}(n)) \subseteq \hat{an}(m) \\
\iff & \hint{def of $\POST$} \\
    & \all{i,s,j,t}{ (i,s)\in\hat{an}(n) \land (i,s)\tranSeg{n,m}(j,t)
     \; \imp \; (j,t)\in \hat{an}(m) } \\
\iff & \hint{def $\tranSeg{n,m}$ and one-point rule of predicate calculus} \\
    & \all{s,t}{ (n,s)\in\hat{an}(n) \land (n,s)\trans(m,t)
     \imp (m,t)\in \hat{an}(m) } \\
\iff & \hint{def (\ref{eq:hatALT})} \\
    & \all{s,t}{ s\in an(n) \land (n,s)\trans(m,t) \imp t\in an(m) } \\
\iff & \hint{def $\trans$ for the case of skip with successor $m$ (\autoref{def:aut})} \\
    & \all{s,t}{ s\in an(n) \land s = t \imp t\in an(m) } \\
\iff & \hint{one-point rule} \\
    & \all{s}{ s\in an(n) \imp s\in an(m) } 
\end{array}\]
The last line is equivalent to $an(n)\subseteq an(m)$ which we write as $an(n)\imp an(m)$.

\medskip
\textbf{Case}
$\lhavc{n}{x}$ with $m=\fsuc(n,c,f)$. 
To show: the VC is equivalent to $an(n)\imp \allSet{x}{an(m)}$.
As in the previous case, the VC is equivalent to the first line of this calculation:
\begin{samepage}
\[\begin{array}{lll}
    &  \all{s,t}{ s\in an(n) \land (n,s)\trans(m,t) \imp t\in an(m) } \\
\iff & \hint{using def $\trans$ for case  $\lhavc{n}{x}$ with successor $m$}\\
    & \all{s,v}{ s\in an(n) \imp \update{s}{x}{v}\in an(m) } \\
\iff & \hint{using def of substitution} \\
    &  \all{s,v}{ s\in an(n) \imp s \in \subst{an(m)}{x}{v} } \\
\iff &\hint{by predicate calculus} \\
    & \all{s}{ s\in an(n) \imp \all{v}{s \in \subst{an(m)}{x}{v} }} \\
\iff & \hint{by def} \\
     & \all{s}{ s\in an(n) \imp s\in \allSet{x}{an(m)} } 
\end{array}\]
which we write as $an(n) \imp \allSet{x}{an(m)}$.
\end{samepage}
\end{proof}

\subsection{Relational VCs}\label{sec:RVC}

Consider an alignment automaton $\aprod(A,A',L,R,J)$ where the underlying automata $A$ and $A'$ are obtained from programs by \autoref{def:aut}.
An annotation of $\aprod(A,A',L,R,J)$ thus maps pairs of control points of $A,A'$ to
relations on variable stores.  
Verification conditions are associated with tuples $((n,n'),(m,m'))$
that represent edges in the control flow graph of $\aprod(A,A',L,R,J)$,
i.e., VCs are given by (\ref{eq:VC}) instantiated with $n:=(n,n')$ and $m:=(m,m')$.

For a given pair $((n,n'),(m,m'))$ of alignment automaton control points,
the transitions go only via LO, or only via RO, or only via JO in \autoref{def:alignProd}.
If $n=m$, i.e., control does not change on the left, the transitions must be via RO
because the non-stuttering condition for automata (\autoref{def:automaton})
ensures there is no unary transition where control does not change.
Similarly, if $n'=m'$ the transitions are via LO.
If $n\neq m $ and $n'\neq m'$ the transitions only go via JO.

\begin{figure}[t]
\begin{footnotesize}
\begin{tabular}{lll}
if $\sub(n,c)$ is\ldots     & 
and $m$ is\ldots &
then the VC for $((n,n'),(m,n'))$ is equivalent to $\ldots$ 
\\\hline
$\lskipc{n}$ & $\fsuc(n,c,f)$ & $L \land \breve{an}(n,n')\imp \hat{an}(m,n')$ 
\\
$\lassg{n}{x}{e}$ & $\fsuc(n,c,f)$ & $L\land \breve{an}(n,n')\imp \subst{\hat{an}(m,n')}{x\smSep}{e\smSep}$ 
\\
$\lhavc{n}{x}$ & $\fsuc(n,c,f)$ & $L\land \breve{an}(n,n')\imp \allSet{x\smSep}{\hat{an}(m,n')}$ 
\\
$\lifgc{n}{gcs}$ & $\lab(d)$ where $e\gcto d$ is in $gcs$ 
& $L\land \breve{an}(n,n')\land \leftF{e} \imp \hat{an}(m,n')$ 
\\ 
$\ldogc{n}{gcs}$ & $\lab(d)$ where $e\gcto d$ is in $gcs$ 
& $L\land \breve{an}(n,n')\land \leftF{e} \imp \hat{an}(m,n')$ 
\\ 
$\ldogc{n}{     gcs}$ & $\fsuc(n,c,f)$ 
& $L\land \breve{an}(n,n')\land \neg\leftF{\enab(gcs)} \imp \hat{an}(m,n')$ 
\\
\multicolumn{3}{l}{
In all other cases, there are no transitions from $(n,n')$ to $(m,n')$ so the VC is $true$ by definition.}
\end{tabular}
\end{footnotesize}
\vspace*{-1ex}
\caption{The left-only VCs for annotation $an$ of $\aprod(\aut(c,f),\aut(c',f'),L,R,J)$.}
\label{fig:RVClo}
\end{figure}

\setlength{\dashlinedash}{.5ex}
\setlength{\dashlinegap}{1ex}

\begin{figure}[t]
\begin{footnotesize}
\begin{tabular}{lll}
if\hspace*{-.5ex} $\begin{array}[t]{l} 
    \sub(n,c) \\
    \sub(n',c') 
    \end{array}$
    \hspace*{-1ex}are\ldots     & 
and $\begin{array}[t]{l} m \\ m' \end{array}$
    \hspace*{-1ex}are\ldots &
then the VC for $((n,n'),(m,m'))$ is equiv.\ to$\ldots$ \\\hline

$\lskipc{n}$ &
$\fsuc(n,c,f)$ & $J \land \breve{an}(n,n')\imp \hat{an}(m,m')$ 
\\
$\lskipc{n'}$ &
$\fsuc(n',c',f')$ & 
\\\hdashline

$\lassg{n}{x}{e}$ &
$\fsuc(n,c,f)$ & $J \land \breve{an}(n,n')\imp \subst{\hat{an}(m,m')}{x|x'}{e|e'}$ 
\\
$\lassg{n'}{x'}{e'}$ &
$\fsuc(n',c',f')$ & 
\\\hdashline

$\lhavc{n}{x}$ &
$\fsuc(n,c,f)$ & $J \land \breve{an}(n,n') \imp \allSet{x\smSep x'}{\hat{an}(m,m')}$ 
\\
$\lhavc{n'}{x'}$ &
$\fsuc(n',c',f')$ & 
\\\hdashline

$\lifgc{n}{gcs}$ & $\lab(d)$ where $e\gcto d$ is in $gcs$ 
& $J \land \breve{an}(n,n')\land \leftF{e} \land \rightF{e'} \imp \hat{an}(m,m')$ 
\\
$\lifgc{n'}{gcs'}$ & $\lab(d')$ where $e'\gcto d'$ is in $gcs'$
& 
\\\hdashline

$\ldogc{n}{gcs}$ & $\lab(d)$ where $e\gcto d$ is in $gcs$ 
& $J \land \breve{an}(n,n')\land \leftF{e} \land \rightF{e'} \imp \hat{an}(m,m')$ 
\\
$\ldogc{n'}{gcs'}$ & $\lab(d')$ where $e'\gcto d'$ is in $gcs'$
& 
\\\hdashline

$\ldogc{n}{gcs}$ & $\fsuc(n,c,f)$  
& $J \land \breve{an}(n,n')\land \neg\leftF{\enab(gcs)} \land \neg\rightF{\enab(gcs')} $ 
\\
$\ldogc{n'}{gcs'}$ & $\fsuc(n',c',f')$ 
                   & $\imp \hat{an}(m,m')$ 

\\\hdashline

$\ldogc{n}{gcs}$ & $\lab(d)$ where $e\gcto d$ is in $gcs$ 
& $J \land \breve{an}(n,n')\land \leftF{e} \land \neg\rightF{\enab(gcs')} \imp \hat{an}(m,m')$ 
\\
$\ldogc{n'}{gcs'}$ & $\fsuc(n',c',f')$ & 
\\\hdashline

$\ldogc{n}{gcs}$ & $\fsuc(n,c,f)$ 
& $J \land \breve{an}(n,n')\land \neg\leftF{\enab(gcs)} \land \rightF{e'} \imp \hat{an}(m,m')$ 
\\
$\ldogc{n'}{gcs'}$ & $\lab(d)$ where $e\gcto d$ is in $gcs$ 
\\\hdashline

$\lassg{n}{x}{e}$ & $\fsuc(n,c,f)$ & $J \land \breve{an}(n,n')\imp \subst{\hat{an}(m,m')}{x|}{e|}$  \\
$\lskipc{n'}$ & $\fsuc(n',c',f')$ & 
\\\hdashline


$\lassg{n}{x}{e}$ & $\fsuc(n,c,f)$ 
& $J \land \breve{an}(n,n')\land \rightF{e'} \imp \subst{\hat{an}(m,m')}{x|}{e|}$ 
\\
$\lifgc{n'}{gcs'}$ & $\lab(d')$ where $e'\gcto d'$ is in $gcs'$
\\\hdashline

$\lassg{n}{x}{e}$ & $\fsuc(n,c,f)$ & $J\land \breve{an}(n,n')\land \rightF{e'} \imp \subst{\hat{an}(n,m')}{x|}{e|}$  \\
$\ldogc{n'}{gcs'}$ & $\lab(d')$ where $e'\gcto d'$ is in $gcs'$ \\\hdashline

$\lassg{n}{x}{e}$ & $\fsuc(n,c,f)$ & $J\land \breve{an}(n,n')\land \neg\rightF{e'} \imp \subst{\hat{an}(n,m')}{x|}{e|}$  \\
$\ldogc{n'}{gcs'}$ & $\fsuc(n',c',f')$ & 

\\[.5ex]
\multicolumn{3}{l}{Omitted: the other 24 cases with nontrivial VCs.}
\end{tabular}

\end{footnotesize}
\vspace*{-1ex}
\caption{Selected joint VCs for 
annotation $an$ of 
$\aprod(\aut(c,f),\aut(c',f'),L,R,J)$.}
\label{fig:RVCjo}
\end{figure}

\autoref{fig:RVClo} gives the six VCs for transitions that go by the LO condition.  
The VCs for RO are symmetric (and omitted).
There are 36 combinations for JO transitions of an alignment automaton; some of their VCs are in \autoref{fig:RVCjo}.
By contrast with \autoref{fig:VC} 
and \autoref{lem:VCprog} we do not eliminate lift (hat) notation 
in \autoref{fig:RVClo} and \autoref{fig:RVCjo},
because the alignment conditions $L$, $R$, $J$ are sets of states, not sets of stores.
We return to this later, using the $pc$  variable to encode the control part of the state.
In \autoref{fig:RVClo} and \autoref{fig:RVCjo}, the substitution notation is lifted from store relations to state relations and likewise for $\leftF{e}$ and $\rightF{e}$.

\begin{lem}[VCs for program alignment automata] 
\upshape
Consider programs $c$ and $c'$, with alignment automaton
$\aprod(\aut(c,f),\aut(c',f'),L,R,J)$.
Let $an$ be an annotation.
For each pair $(n,n'),(m,m')$ of alignment automaton control points, the VC 
can be expressed 
as in \autoref{fig:RVCjo}, \autoref{fig:RVClo}, and in the omitted RO cases that are symmetric to \autoref{fig:RVClo}.
\end{lem}
\begin{proof}
First, for any $(n,n')$ and $(m,m')$ the VC is
equivalent to the first line of this calculation:
\[\begin{array}{lll}
     & \POST(\biTranSeg{(n,n'),(m,m')})(\hat{an}(n,n')) \subseteq \hat{an}(m,m') \\
\iff & \hint{definitions} \\
     & \all{i,i',s,s',j,j',t,t'}{
     \begin{array}[t]{l}
     ((i,i'),(s,s'))\in \hat{an}(n,n') \land 
     ((i,i'),(s,s')) \\ 
      \biTranSeg{(n,n'),(m,m')} ((j,j'),(t,t')) \\
      \imp ((j,j'),(t,t')) \in\hat{an}(m,m') 
     \end{array} } \\
\iff & \hint{using def (\ref{eq:hatALT}) of hat} \\ 
     & \all{i,i',s,s',j,j',t,t'}{
    \begin{array}[t]{l}
     (s,s')\in an(n,n') \land 
     ((i,i'),(s,s'))\biTranSeg{(n,n'),(m,n')} ((j,j'),(t,t')) \\
      \imp (t,t') \in an(m,m') 
     \end{array} } \\
\iff & \hint{using def (\ref{eq:tranSeg})} \\
     & \all{i,i',s,s',j,j',t,t'}{
    \begin{array}[t]{l}
     (s,s')\in an(n,n') \land
     (i,i')=(n,n') \land (j,j')=(m,m') \\ 
     \land \: ((i,i'),(s,s'))\biTrans ((j,j'),(t,t')) 
      \imp (t,t') \in an(m,m') 
     \end{array} } \\
\iff & \hint{predicate calculus (the one-point rule)} \\
     & \all{s,s',t,t'}{
     \begin{array}[t]{l}
     (s,s')\in an(n,n') \land
     ((n,n'),(s,s'))\biTrans ((m,m'),(t,t')) 
\\
      \imp (t,t') \in an(m,m') 
     \end{array} 
} 
\hfill \stepcounter{equation}\label{eq:VCx}(\theequation) 
\end{array}
\]
From here we proceed by distinguishing the three cases discussed preceeding this lemma in Sec.~\ref{sec:RVC}, depending on whether $n=m$, $n'=m'$, or neither equality holds.

In case $n'=m'$, the LO case applies and unfolding the definition of $\biTrans$ in (\ref{eq:VCx}) we get 
\begin{equation}\label{eq:VCy}
   \all{s,s',t}{
    \begin{array}[t]{l}
     (s,s')\in an(n,n') \land ((n,n'),(s,s'))\in L \land (n,s)\trans (m,t)  \\
      \imp (t,s') \in an(m,n') 
     \end{array} 
} 
\end{equation}
In case $n=m$, the RO case applies, which is similar.
In case $n\neq m$ and $n'\neq m'$, the JO case applies and unfolding $\biTrans$ in (\ref{eq:VCx}) we get 
\[ \all{s,s',t,t'}{
    \begin{array}[t]{l}
     (s,s')\in an(n,n') \land ((n,n'),(s,s'))\in J \land (n,s)\trans (m,t) \land (n',s')\trans (m',t') \\
     \imp (t,t') \in an(m,m') 
     \end{array} 
} \]
From here, we must consider for each LO case the six possible commands $\sub(n,c)$ and successors $m$,
for each RO case the possible commands $\sub(n',c')$ and successors $m'$, 
and for each JO all 36 combinations. 

Here is one case: LO where $\sub(n,c)$ is $\lassg{n}{x}{e}$,
$m = \fsuc(n,c,f)$, so the VC is for $((n,n'),(m,n'))$.
By definition of $\trans$, (\ref{eq:VCy}) is equivalent to 
\[   \all{s,s',t}{
    \begin{array}[t]{l}
     (s,s')\in an(n,n') \land ((n,n'),(s,s'))\in L \land t = \update{s}{x}{\means{e}(s)} \imp (t,s') \in an(m,n') 
     \end{array} 
} \]
which equivales (by def of semantic substitution and one-point rule) 
\[   \all{s,s'}{
    \begin{array}[t]{l}
     (s,s')\in an(n,n') \land ((n,n'),(s,s'))\in L \imp (s,s') \in \subst{an(m,n')}{x|}{e|} 
     \end{array} 
} \]
which, using def (\ref{eq:hatALT}) of lift and def of the control predicate $[n|n']$, is equivalent to
\[ 
\all{i,i',s,s'}{
    \begin{array}[t]{l}
     ((i,i'),(s,s'))\in \hat{an}(n,n') \land ((i,i'),(s,s'))\in[n|n'] 
     \land ((i,i'),(s,s'))\in L \imp \\ 
     ((i,i'),(s,s')) \in \subst{\hat{an}(m,n')}{x|}{e|}
     \end{array} 
} \]
Now that the condition is phrased uniformly in terms of state sets, 
we can write it as 
$L\land \hat{an}(n,n')\land [n|n'] \imp \subst{\hat{an}(m,n')}{x}{e}$,
and even more succinctly as 
$L \land \breve{an}(n,n') \imp \subst{\hat{an}(m,n')}{x|}{e|}$ using the $\breve{an}$ abbreviation.
This concludes the proof for the left assignment case in \autoref{fig:RVClo}.
(Note: It is for the sake of this last step that we did not eagerly simplify using the one-point rule with $(i,i')=(n,n')$.   
For the quantifier-free notation to make sense, everything needs to be a predicate of the same type.)

All the other cases are proved similarly.
\end{proof}

\paragraph{Encoding relational VCs in terms of store relations.}

The unary VCs are given in terms of store predicates (\autoref{fig:VC}).
For relational VCs, although the annotation comprises store relations,
the relational VCs involve alignment conditions ($L,R,J$) which are sets of 
alignment automaton states.  
So the relational VCs are given in terms of states, using $\hat{an}$ 
and $\breve{an}$ in Figs.~\ref{fig:RVClo} and~\ref{fig:RVCjo}.  
For use in RHL+ proofs about programs in automaton normal form,
we will encode the VCs in terms of relations on stores that use 
a variable $pc$ (for ``program counter'') to encode control information.  
(The normal form is sketched in \autoref{sec:overview} and developed in \autoref{sec:KATnf}.)

We often use the following abbreviations for assignments and tests of 
the chosen $pc$ variable.
\begin{equation}\label{eq:pcabbrev}
\mbox{
\graybox{$\spc n$} for $pc:=n$
\qquad 
\graybox{$\tpc n$} for $pc=n$ 
\qquad  for literal $n\in\nat$
}
\end{equation}
To be precise, we define $\spc n$ to be $\lassg{0}{pc}{n}$, using $0$ arbitrarily as the label. Our uses of $\spc n$ will be in contexts where labels are irrelevant, 
in commands that are not required to be $\ok$.
Note that the notations $\spc n$ and $\tpc n$ depend, implicitly, on the choice of the variable $pc$.
\begin{defi} 
Let $R$ be a set of states of a program alignment automaton.
Let $pc$ be a variable such that $\indep(pc|pc,R)$.
Define the \dt{$pc$-encoded $R$} to be a relation on stores as follows:
\( \graybox{$\encode{R}$} \eqdef \{ (s,s') \mid  ((s(pc),s'(pc)),(s,s'))\in R  \} \).
\end{defi}

\begin{lem}\label{lem:state-to-store}
\upshape
Let $R$ be set of alignment automaton states such that $\indep(pc|pc,R)$.
If $(s,s') \in \encode{R}\land\bothF{\tpc n\sep\tpc n'} $
then $((n,n'),(s,s')) \in R$.
\end{lem}
\begin{proof}
Suppose $(s,s')\in  \encode{R}\land\bothF{\tpc n\sep\tpc n'}$.
Then $s(pc)=n$ and $s'(pc)=n'$
by definition of $\bothF{\tpc n\sep\tpc n'}$,
hence $((n,n'),(s,s'))\in R$ by definition of $\encode{R}$.
\end{proof}

The encoding makes it possible to derive $pc$-encoded forms of the verification conditions in terms of store relations, enabling their use in deductive proofs.

\begin{figure}[t]
\begin{footnotesize}
\begin{tabular}{lll}
if $\sub(n,c)$ is\ldots     & 
and $m$ is\ldots &
then the encoded VC for $((n,n'),(m,n'))$ is$\ldots$ 
\\\hline
$\lskipc{n}$ & $\fsuc(n,c,f)$ & $\encode{L}\land\bothF{\tpc n\sep\tpc n'} \land an(n,n')\imp an(m,n')$ 
\\
$\lassg{n}{x}{e}$ & $\fsuc(n,c,f)$ & $\encode{L}\land\bothF{\tpc n\sep\tpc n'}\land an(n,n')\imp \subst{an(m,n')}{x\smSep}{e\smSep}$ 
\\
$\lhavc{n}{x}$ & $\fsuc(n,c,f)$ & $\encode{L}\land\bothF{\tpc n\sep\tpc n'}\land an(n,n')\imp \allSet{x\smSep }{an(m,n')}$ 
\\
$\lifgc{n}{gcs}$ & $\lab(d)$ where $e\gcto d$ is in $gcs$ 
& $\encode{L}\land\bothF{\tpc n\sep\tpc n'}\land an(n,n')\land \leftF{e} \imp an(m,n')$ 
\\ 
$\ldogc{n}{gcs}$ & $\lab(d)$ where $e\gcto d$ is in $gcs$ 
& $\encode{L}\land\bothF{\tpc n\sep\tpc n'}\land an(n,n')\land \leftF{e} \imp an(m,n')$ 
\\ 
$\ldogc{n}{     gcs}$ & $\fsuc(n,c,f)$ 
& $\encode{L}\land\bothF{\tpc n\sep\tpc n'}\land an(n,n')\land \neg\leftF{\enab(gcs)} \imp an(m,n')$ 
\end{tabular}
\end{footnotesize}
\vspace*{-1ex}
\caption{The $pc$-encoded left-only VCs for $an$ and 
$\aprod(\aut(c,f),\aut(c',f'),L,R,J)$.}
\label{fig:RVClo-encoded}
\end{figure}

\begin{figure}[t]
\begin{footnotesize}
\begin{tabular}{lll}
if $\begin{array}[t]{l} 
    \sub(n,c) \\
    \sub(n',c') 
    \end{array}$
    \!\!\! are\ldots \!\!\!     & 
and $\begin{array}[t]{l} m \\ m' \end{array}$ are\ldots &
then the encoded VC for $((n,n'),(m,m'))$ is$\ldots$ \\\hline

$\lskipc{n}$ &
$\fsuc(n,c,f)$ & $\encode{J}\land\bothF{\tpc n\sep\tpc n'} \land an(n,n')\imp an(m,m')$ 
\\
$\lskipc{n'}$ &
$\fsuc(n',c',f')$ & 
\\\hdashline

$\lassg{n}{x}{e}$ &
$\fsuc(n,c,f)$ & $\encode{J}\land\bothF{\tpc n\sep\tpc n'} \land an(n,n')\imp \subst{an(m,m')}{x|x'}{e|e'}$ 
\\
$\lassg{n'}{x'}{e'}$ &
$\fsuc(n',c',f')$ & 
\\\hdashline

$\lifgc{n}{gcs}$ & $\lab(d)$ where $e\gcto d$ is in $gcs$ 
& $\encode{J} \land\bothF{\tpc n\sep\tpc n'}\land an(n,n')\land \leftF{e} \land \rightF{e'} \imp an(m,m')$ 
\\
$\lifgc{n'}{gcs'}$ & $\lab(d')$ where $e'\gcto d'$ is in $gcs'$
& 
\end{tabular}
\end{footnotesize}
\vspace*{-1ex}
\caption{Selected $pc$-encoded joint VCs for $an$ and $\aprod(\aut(c,f),\aut(c',f'),L,R,J)$.}
\label{fig:RVCjo-encoded}
\end{figure}

\begin{lem}[pc-encoded VCs for alignment automata]\label{lem:liftRVC}
\upshape
Let $an$ be an annotation of an alignment automaton 
$\aprod(\aut(c,f),\aut(c',f'),L,R,J)$ for commands $c,c'$.
Suppose $pc$ is a fresh variable in the sense that 
it does not occur in $c$ or $c'$, and all of $L,R,J$ and any $an(i,j)$ are independent from
$pc$ on both sides.
Then each left-only relational VC of \autoref{fig:RVClo} implies
the corresponding condition on store relations in \autoref{fig:RVClo-encoded}.
Each joint relational VC in
\autoref{fig:RVCjo} implies a corresponding condition (see~\autoref{fig:RVCjo-encoded}).
Similarly for right-only VCs.
\end{lem}
\begin{proof}
Consider the left-only skip case.
The primary VC is $L \land \breve{an}(n,n')\imp \hat{an}(m,n')$
(first row of \autoref{fig:RVClo}),
which abbreviates 
\begin{equation}\label{eq:primVC}
 L \land [n|n'] \land \hat{an}(n,n')\imp \hat{an}(m,n') 
 \end{equation}
We show this implication entails 
the first line in \autoref{fig:RVClo-encoded}, i.e.,
\( \encode{L}\land\bothF{\tpc n\sep\tpc n'} \land an(n,n')\imp an(m,n') \).
To prove the latter implication, observe for any $(s,s')$
\begin{small}
\[\begin{array}{lll}
    & (s,s')\in \encode{L}\land (s,s')\in\bothF{\tpc n\sep\tpc n'} \land (s,s')\in an(n,n') \\
\imp 
    & ((n,n'),(s,s'))\in L \land (s,s')\in an(n,n') & \mbox{Lemma~\ref{lem:state-to-store}} \\
\iff
    & ((n,n'),(s,s'))\in L \land((n,n'),(s,s'))\in[n|n']\land (s,s')\in an(n,n')  & \mbox{def $[n|n']$} \\
\iff 
    & ((n,n'),(s,s'))\in L \land((n,n'),(s,s'))\in[n|n']\land ((n,n'),(s,s'))\in \hat{an}(n,n') & \mbox{def (\ref{eq:hatALT})} \\
\imp 
    & ((n,n'),(s,s'))\in \hat{an}(m,n') & \mbox{primary VC (\ref{eq:primVC}) } \\
\iff 
    & (s,s')\in an(m,n') & \mbox{def (\ref{eq:hatALT})} \\
    \end{array}
\]
\end{small}
The argument is essentially the same for all cases in Figs.~\ref{fig:RVClo}
and~\ref{fig:RVCjo}.
\end{proof}

\section{Automaton normal form reduction and KAT}\label{sec:KATnf}

\subsection{Command equivalence and KAT}

\newcommand\kK{\mathbb{K}}
\newcommand\kB{\mathbb{B}}
\newcommand\kdot{\mathbin{;}}
\newcommand\kstar{*}
\newcommand\kplus{+}
\newcommand\kneg{\neg}
\newcommand\kNeg[1]{\overline{#1}}
\newcommand\kone{1}
\newcommand\kzero{0}

\begin{defi}\label{def:KAT}
\begin{sloppypar}
A \dt{Kleene algebra with tests}~\cite{Kozen97} (\dt{KAT}) is a structure $(\kK,\kB,\kplus,\kdot,\kstar,\kneg,\kone,\kzero)$
such that
(a) $\kK$ is a set and $\kB\subseteq \kK$ (elements of $\kB$ are called \dt{tests}).
(b) $\kB$ contains $\kone$ and $\kzero$, and is closed under the operations $\kplus,\kdot,\kneg$,
       and these satisfy the laws of Boolean algebra, with $\kone$ as true
and $\kdot$ as conjunction.
(c) $\kK$ is an idempotent semiring and the 
following hold for all $x,y,z$ in $\kK$.
\[\begin{array}{lcl@{\hspace{5em}}lcl}
\kone \kplus x\kdot x^\kstar &=& x^\kstar 
& y \kplus x\kdot z \leq z &\imp& x^\kstar\kdot y \leq z \\
\kone \kplus x^\kstar\kdot x &=& x^\kstar 
& y \kplus z\kdot x \leq z &\imp& y\kdot x^\kstar \leq z 
\end{array}
\]
The ordering $\leq$ is defined by \graybox{$x\leq y$} iff $x+y=y$.
The operator $;$ binds tighter than $+$.
\end{sloppypar}
\end{defi}

In a \dt{relational model}~\cite{Kozen1996}, $\kK$ is some set of relations on some set $\Sigma$,
with $\kzero$ the empty relation, 
$\kone$ the identity relation on $\Sigma$,
$+$ union of relations,
$\kdot$ relational composition, and $*$ reflexive-transitive closure. 
Moreover  $\kB$ is a set of \dt{coreflexives}, i.e., subsets of the identity relation $\kone$,
and $\kneg$ is complement with respect to $\kone$.
In a relational model, $\leq$ is set inclusion.
The \dt{relational model for GCL}, denoted \graybox{$\relKAT$}, is the relational model comprising all relations on $\Sigma$ where $\Sigma$ is the set of variable stores.\footnote{KAT is complete for relational models
 and there are other completeness results for KAT~\cite{Kozen1996}.  We make no use of those results.  All we need about KAT is that it is sufficient for our \autoref{thm:normEquiv}.  That result is like 
\cite{Kozen97} which uses KAT to prove that every command is simulated by one with a single loop. 
}

\paragraph{Representing programs in KAT}

In this paper we work with $\relKAT$ and also with equational hypotheses formalized in terms of KAT expressions which are usually defined with respect to given finite sets of primitive tests and actions~\cite[Sect.\ 2.3]{Kozen1996}.
We only need the special case where the actions are primitive commands and the tests are primitive boolean expressions.\footnote{Unlike in \cite{Kozen1996}, we have infinitely many primitives, 
because the GCL grammar generates infinitely many boolean and arithemetic expressions, but this causes no problems.}
Define the \dt{KAT expressions} as follows.\footnote{The primitives $\underline{bprim}$, $\underline{\lassg{}{x}{e}}$, and $\underline{\havc{x}}$  are written this way to avoid confusion with their counterparts in GCL syntax.}
Recall that $bprim$ stands for primitive boolean expressions in GCL. 
\[ 
\begin{array}{lcl}
\KB &::=& \underline{bprim} \mid \KB + \KB \mid \KB\mathbin{;}\KB \mid \neg \KB \mid 1 \mid 0 \\
\KE &::=& \underline{\lassg{}{x}{e}} \mid \underline{\havc{x}} \mid
\KE + \KE \mid \KE\mathbin{;}\KE \mid \KE^* \mid \KB
\end{array}
\]
We refer to terms given by $\KB$ as KAT boolean expressions and by $\KE$ as KAT expressions.


Given any model, and mappings of the primitive expressions to elements in the model, one obtains an interpretation of all KAT expressions in the model~\cite[Sect.\ 2.3]{Kozen1996}.
To be precise, let $\anyKAT$ be a model $(\kK^\anyKAT,\kB^\anyKAT,\kplus^\anyKAT,\kdot^\anyKAT,\kstar^\anyKAT,\kneg^\anyKAT,\kone^\anyKAT,\kzero^\anyKAT)$.
Given interpretations $\underline{bprim}^\anyKAT$, $\underline{x:=e}^\anyKAT$,
and $\underline{\havc{x}}^\anyKAT$ 
in $\anyKAT$ for each primitive KAT expression,
we get an interpretation $\KE^\anyKAT$ for every $\KE$, 
defined homomorphically as usual. 
For example: 
$(\KE_0+\KE_1)^\anyKAT \eqdef \KE_0^\anyKAT \,+^\anyKAT\,\KE_1^\anyKAT$.
Henceforth we never write $+^\anyKAT$ because the reader can infer whether the symbol
$+$ is meant as syntax in a KAT expression or as an operation of a particular KAT.

The \dt{GCL-to-KAT translation} \graybox{$\mkt{-}$} 
maps GCL commands (resp.\ boolean expressions) to KAT expressions ($\KE$)
(resp.\ KAT boolean expressions ($\KB$)).
For boolean expressions it is defined by structural recursion as follows:
 \[
 \mkt{bprim} \eqdef \underline{bprim} 
 \qquad \mkt{\neg e} \eqdef  \kneg\mkt{e} 
 \qquad \mkt{e\land e'} \eqdef  \mkt{e}\kdot\mkt{e'} 
 \qquad\mkt{e\lor e'} \eqdef  \mkt{e}\kplus\mkt{e'} 
 \]
For commands the translation is defined by mutual recursion with the definition of $\mkt{gcs}$:
 \vspace*{-1ex}
\begin{small}
\begin{equation}\label{eq:def:mkt}
\begin{array}{l@{\;}c@{\;}l@{\hspace*{3em}}l@{\;}c@{\;}l}
\mkt{\lassg{}{x}{e}} & \eqdef  & \underline{x:=e} 
  & \mkt{c;d} & \eqdef  & \mkt{c}\kdot\mkt{d} \\
\mkt{\havc{x}} &\eqdef& \underline{\havc{x}} 
  & \mkt{\lifgc{}{gcs}} & \eqdef  & \mkt{gcs} \\
\mkt{\skipc} & \eqdef  & 1 
  & \mkt{\ldogc{}{gcs}} & \eqdef  & \mkt{gcs}^\kstar \kdot \kneg\mkt{\enab(gcs)}  \\[1ex]
\mkt{gcs} & \eqdef  & \multicolumn{4}{l}{\quant{\kplus}{e,c}{(e\gcto c) \in gcs}{\mkt{e}\kdot\mkt{c}} }
\end{array} 
\end{equation}
\end{small}
This is an adaptation of the well known translation of imperative programs into KAT~\cite{Kozen97}.

We define interpretations in $\relKAT$ for the primitive expressions, as follows:
\[\begin{array}{lclllcl}
\underline{\lassg{}{x}{e}}^\relKAT & \eqdef & \means{\lassg{}{x}{e}} 
& \qquad &  
\underline{bprim}^\relKAT & \eqdef & \{(s,s) \mid \means{bprim}(s) = true \} \\
\underline{\havc{x}}^\relKAT  &\eqdef & \means{\havc{x}}
\end{array}\]
As with an interpretation in any model, this induces an interpretation
$\KE^\relKAT$ for any KAT expression $\KE$.  

\begin{lem}
\label{lem:correctInterp} 
\upshape
For all commands $c$ we have $\mkt{c}^\relKAT = \means{c}$.
For all boolean expressions $e$ we have 
$\mkt{e}^\relKAT = \{(s,s) \mid \means{e}(s) = true \}$.
\end{lem}
\begin{proof}
The proof is by induction on $e$ and then on $c$.
The base cases are by definition.
A secondary induction is used for the loop case.  
\end{proof}



\paragraph{Command equivalence and soundness of HL+ and RHL+.}

For specifications where the pre- and post-condition are expressible as tests in the KAT,
one can express correctness judgments~\cite{Kozen00}.
In our setting, for boolean expressions $e_0$ and $e_1$ we have that
$ \models c:\spec{e_0}{e_1}$ is equivalent to the equation 
\( \mkt{e_0}\kdot\mkt{c}\kdot\kneg\mkt{e_1} = 0 \) being true in $\relKAT$.
In this paper we do not use the KAT formulation for correctness judgments in general, but we do use it in a limited way.

Let $H$ be a set of equations between KAT expressions.
We write \graybox{$H \proves \KE_0 = \KE_1$} to say $\KE_0=\KE_1$ 
is provable by equational reasoning from hypotheses $H$ plus 
the axioms of KAT (\autoref{def:KAT}).

The idea for the equivalence condition $c\kateq d$ in rules \rn{Rewrite} (\autoref{fig:HLplus}) and \rn{rRewrite} (\autoref{fig:RHL})
is that it should mean $H\proves \mkt{c}=\mkt{d}$ for a suitable set of hypotheses
that axiomatize the semantics of some primitive boolean expressions and commands.
To prove our main results we only need a few axioms (as detailed in \autoref{sec:KATequiv}).
For the sake of a straightforward presentation we 
formulate equivalence in terms of a larger set of axioms.

\begin{defi}\label{def:Hyp}
Define \graybox{$\Hyp$} to be the set of equations 
given by:
(a) the equation $\mkt{e}=0$ for every boolean expression $e$
such that $e \imp\mathit{false}$ is valid;
(b) the equation $\mkt{e_0};\mkt{x:=e};\kneg\mkt{e_1}=\kzero$
for all assignments $x:=e$
and boolean expressions $e_0$, $e_1$ 
such that $e_0\imp \subst{e_1}{x}{e}$ is valid;
(c) the equation 
$\mkt{e_0};\mkt{\havc{x}};\kneg\mkt{e_1}=0$ 
for $x,e_0,e_1$ such that $e_0\imp \allSet{x}{e_1}$ is valid.
\end{defi}

\begin{defi}[command equivalence]\label{def:kateq}
Define $\kateq$ by 
\( \graybox{$c \kateq d$} \eqdef \Hyp \proves \mkt{c} = \mkt{d} \). 
\end{defi}

An example is the equivalence
\( \ldogc{}{ e_0 \gcto c } \;\kateq\;
   \ldogc{}{ e_0 \gcto c;\ldogc{}{e_0\land e_1\gcto c}} \)
which holds for any $c,e_0,e_1$.  It is used in the loop tiling example,
see (\ref{eq:c12}).

\begin{rem}\label{rem:KAT}
It is straightforward to present a deductive system for $\kateq$,
based on $\Hyp$, the axioms of KAT, and the rules of equational logic.  
For practical purposes an alternative is to leverage the fact that the hypotheses are all equations of the form $\KE=0$.  For any finite set $H$ of such equations, entailments $H\proves \KE_0 = \KE_1$ are decidable in PSPACE~\cite{KozenKATcomplex}.
Many practical cases of $\kateq$ require no hypotheses at all.  For the equivalences used in our normal form theorem, the requisite hypotheses are a finite set 
of KAT-consequences of $\Hyp$, 
syntactically determined by the relevant command $c$ as detailed in 
\autoref{sec:KATequiv}. 
\qed
\end{rem}

Having defined $\kateq$, 
the key ingredient of the rules \rn{Rewrite} and \rn{rRewrite},
we have completed the definition of HL+ and RHL+.

\begin{lem}\label{lem:Hyp}
\upshape
Every equation in $\Hyp$ holds in $\relKAT$.
\end{lem}
This is an easy consequence of the definition of $\Hyp$.

\begin{thm}\label{prop:HLsound}
\upshape
All the rules of HL+ are sound.
\end{thm}
\begin{proof}
The proofs are straightforward using the definitions, and induction for the loop rule.
The only rule which is not standard is \rn{Rewrite} which we prove as follows.  
Suppose $c \kateq d$ holds.
That means $\Hyp\proves\mkt{c}=\mkt{d}$ so
$\mkt{c}^\anyKAT = \mkt{d}^\anyKAT$ in any model $\anyKAT$ that satisfies
the equations $\Hyp$.
So by \autoref{lem:Hyp} we have $\mkt{c}^\relKAT=\mkt{d}^\relKAT$.
Hence $\means{c}=\means{d}$ by \autoref{lem:correctInterp}.
Suppose the premise of \rn{Rewrite} holds,
i.e., $\models c: \spec{P}{Q}$.
This is a condition on $\means{c}$ 
---see (\ref{eq:valid})--- 
so we have $\models d: \spec{P}{Q}$.
\end{proof}

\begin{thm}
\upshape
All the rules of RHL+ (\autoref{fig:RHL}) are sound.
\end{thm}
\begin{proof} 
The proofs are straightforward using the definitions, and induction for the loop rule.
(Similar loop rules are proved sound in~\cite{Beringer11,BNNN19} and in the long version of~\cite{NagasamudramN21}.)
The proof of \rn{rRewrite} is similar to the proof of \rn{Rewrite} for \autoref{prop:HLsound}.
\end{proof}

\subsection{Automaton normal form} 


\begin{figure}[t]
\begin{footnotesize}
\(\begin{array}{lcl}
\addPC(\lskipc{n}) &\eqdef& \spc  n \,; \lskipc{n} \\
\addPC(\lassg{n}{x}{e}) &\eqdef& \spc  n \,; \lassg{n}{x}{e} \\
\addPC(\lhavc{n}{x}) &\eqdef& \spc  n \,; \lhavc{n}{x} \\
\addPC(c;d) &\eqdef& \addPC(c) \,; \addPC(d) \\
\addPC(\lifgc{n}{gcs}) &\eqdef& \spc n \,; \lifgc{n}{(\addPC_0(gcs))} \\
\addPC(\ldogc{n}{gcs}) &\eqdef& \spc n \,; \ldogc{n}{(\addPC_1(n,gcs))} \\[1ex]
\addPC_0(e\gcto c) &\eqdef& e\gcto \addPC(c) \\[.5ex]
\addPC_0(e\gcto c \gcsep gcs) &\eqdef& e\gcto \addPC(c) \gcsep \addPC_0(gcs)\\[1ex]
\addPC_1(n,e\gcto c) &\eqdef& e\gcto \addPC(c);\spc n \\[.5ex]
\addPC_1(n,e\gcto c \gcsep gcs) &\eqdef& e\gcto \addPC(c);\spc n \gcsep \addPC_1(n,gcs)
\end{array}\)
\end{footnotesize}
\caption{Definition of $\addPC$.}\label{fig:addPC}
\end{figure}

Choose a variable name $pc$.
\autoref{fig:addPC} defines $\addPC$, a function from commands to commands 
that adds $pc$ as in the example $c0^+$ on page~\pageref{page:c0}.
The definition of $\addPC$ is by structural recursion, with mutually recursive 
helpers $\addPC_0$ and $\addPC_1$, and is written using the
abbreviations in (\ref{eq:pcabbrev}).
For \keyw{if} commands, $\addPC_0$ maps $\addPC$ over the guarded commands,
and for \keyw{do}, $\addPC_1$ additionally adds a trailing assignment to set $pc$ to 
the loop label.

\begin{lem}\label{lem:addPC}
\upshape
If $pc$ does not occur in $c$ then $ \erase(pc,\addPC(c)) \kateq c$
and $\ghost(pc,\addPC(c))$ holds.
\end{lem}
\begin{proof}
To show $\erase(pc,\addPC(c)) \kateq c$, 
we must show $\Hyp\proves\mkt{\erase(pc,\addPC(c))} = \mkt{c}$. 
In fact we can show $\proves\mkt{\erase(pc,\addPC(c))} = \mkt{c}$. 
To do so, go by induction on $c$, using that $\mkt{\skipc} = \kone$
and $\kone$ is the unit of sequence.
The proof of $\ghost(pc,\addPC(c))$ is also a straightforward induction on $c$.
\end{proof}


\begin{figure}[t]
\begin{footnotesize}
\begin{mathpar}
\inferrule{}{
\norm{\lskipc{n}}{f}{
(\tpc n\gcto \spc f)}
}

\inferrule{}{
\norm{\lassg{n}{x}{e}}{f}{ 
(\tpc  n \gcto \lassg{}{x}{e}; \spc  f)}
}

\inferrule{}{
\norm{\lhavc{n}{x}}{f}{(\tpc n \gcto \havc{x};\spc f)}
}

\inferrule{
\norm{c_0}{f}{gcs_0} \\ \norm{c_1}{f}{gcs_1} 
}{ 
\norm{\lifgc{n}{e_0\gcto c_0 \gcsep e_1\gcto c_1}}{f}{
  \tpc  n \land e_0 \gcto \spc \lab(c_0)  \gcsep
  \tpc  n \land e_1 \gcto \spc \lab(c_1)  \gcsep
  gcs_0 \gcsep gcs_1
}}

\inferrule{
\norm{c}{\lab(d)}{gcs_0} \\
\norm{d}{f}{gcs_1}
}{
\norm{c;d}{f}{gcs_0\gcsep gcs_1}
}

\inferrule{
\norm{c}{n}{gcs}
}{
\norm{\ldogc{n}{e\gcto c}}{f}{
  \tpc  n\land e \gcto \spc \lab(c)  \gcsep
  \tpc  n\land\neg e \gcto \spc f \gcsep gcs 
}}

\inferrule{
\norm{c_0}{n}{gcs_0} \\ \norm{c_1}{n}{gcs_1} 
}{
\norm{\ldogc{n}{e_0\gcto c_0 \gcsep e_1\gcto c_1}}{f}{
  \tpc  n\land e_0 \gcto \spc \lab(c_0)  \gcsep
  \tpc  n\land e_1 \gcto \spc \lab(c_1)  \gcsep
  \tpc  n\land\neg (e_0 \lor e_1) \gcto \spc f \gcsep
  gcs_0 \gcsep gcs_1
}}

\end{mathpar}
\end{footnotesize}
\vspace*{-2ex}
\caption{Normal form bodies.} 
\label{fig:norm}
\end{figure}

\autoref{fig:norm} defines the ternary relation $\norm{c}{m}{gcs}$
to relate a command and a label to the gcs that will be the body of its normal form.
To be precise, the relation depends on a choice of $pc$ variable but we leave this implicit, as it is in the notations $\tpc \_$ and $\spc \_$ of (\ref{eq:pcabbrev}).
We can write ``$\norm{c}{m}{gcs}$ (for $pc$)'' to make the choice explicit.
For readability, \autoref{fig:norm} gives special cases for if and do. 
The general form is relegated to \autoref{sec:app:nfbodies}.

An example is $\norm{\lassg{4}{x}{x-1}}{2}{(\tpc 4\gcto x:=x-1;\spc 2)}$.
For the running example, we have $\norm{c0}{6}{gcs}$ where $gcs$ is the body of example $d0$ on page~\pageref{page:d0page}.



There are six kinds of transition in the small step semantics.
Each kind has a corresponding form of guarded command in the normal form.
We spell them out for later reference. 

\begin{lem}[guarded commands of a normal form]\label{lem:nfCases}
\upshape
Suppose $\okf(c,f)$ and $\norm{c}{f}{gcs}$.
Every guarded command in $gcs$ has one of these six forms:
\begin{itemize}
\item $\tpc k\gcto \spc m$, 
for some $k,m$ such that $\sub(k,c)$ is $\lskipc{k}$ and $m=\fsuc(k,c,f)$.
\item $\tpc k\gcto \lassg{}{x}{e};\spc m$,
for some $k,m,x,e$ such that $\sub(k,c)$ is $\lassg{k}{x}{e}$ 
and $m = \fsuc(k,c,f)$.
\item $\tpc k\gcto \havc{x};\spc m$,
for some $k,m,x$ such that $\sub(k,c)$ is $\lhavc{k}{x}$ 
and $m = \fsuc(k,c,f)$.
\item $\tpc k\land e\gcto \spc m$,
for some $k,m,e,d,gcs_0$ such that $\sub(k,c)$ is $\lifgc{k}{gcs_0}$ 
and $m=\lab(d)$ where $e\gcto d$ is in $gcs_0$.

\item $\tpc k\land e\gcto \spc m$, 
for some $k,m,e,d,gcs_0$ such that $\sub(k,c)$ is $\ldogc{k}{gcs_0}$
and $m=\lab(d)$ where $e\gcto d$ is in $gcs_0$.

\item $\tpc k\land \neg\enab(gcs_0)\gcto \spc m$, 
for some $k,m,gcs_0$ such that $\sub(k,c)$ is 
$\ldogc{k}{gcs_0}$ and $m=\fsuc(k,c,f)$.
\end{itemize}
\end{lem}

\begin{lem}\label{lem:normExists} 
\upshape
For all $c$ and $f$, there is some $gcs$ with $\normSmall{c}{f}{gcs}$.
\end{lem}
\begin{proof}
Straightforward structural induction on $c$.
\end{proof}
In fact $gcs$ is uniquely determined by $c$, $m$, and the chosen variable $pc$.  But none of our results depend on uniqueness.

\begin{defi}\label{def:autnf}
An \dt{automaton normal form} of a command $c$ with $\lab(c)=n$ and label $f\notin\labs(c)$, for chosen variable $pc$, is 
the command 
\(\: \spc  n ; \ldogc{}{ gcs } \: \)
where $\normSmall{c}{f}{gcs}$.
\end{defi}
This does not require $c$ to be $\ok$, but the normal form is only useful if $\okf(c,f)$.

As noted in \autoref{sec:intro}, we are not using the term ``normal form'' 
in the sense of term rewriting systems~\cite{BaaderNipkow99}.
It is not the case that semantically equal programs have 
identical automaton normal forms.
The important property is the theorem to follow.

\paragraph{Normal form equivalence theorem.}

Finally we are ready for the main result of \autoref{sec:KATnf},
which loosely speaking says every command is equivalent to one in automaton normal form.

\begin{thm}\label{thm:normEquiv} 
\upshape
If $\okf(c,f)$, $pc\notin vars(c)$, and 
$\normSmall{c}{f}{gcs}$ (for $pc$) 
then
\begin{equation}\label{eq:normEquiv}
\spc n; \ldogc{}{gcs} \kateq \addPC(c); \spc f   
\quad\mbox{where $n=\lab(c)$.}
\end{equation}
\end{thm}
\begin{proof}
By definition of $\kateq$ we must prove
$\Hyp\proves \mkt{\spc n; \ldogc{}{gcs}} = \mkt{\addPC(c); \spc f}$
using laws of KAT.
In fact we do not need all of $\Hyp$; the proof shows that
$\nfax(pc,c,f)\proves \mkt{\spc n; \ldogc{}{gcs}} = \mkt{\addPC(c); \spc f}$
where $\nfax(pc,c,f)$ is the relevant finite set of axioms for $c$
(detailed in~\autoref{sec:KATequiv}).\footnote{As remarked following 
   Def.~\ref{def:kateq}, owing to the form of our hypotheses any instance 
   is decidable (in PSPACE). But this does not help prove the theorem,
   where we have infinitely many instances to prove.
}
These include the following:
\\
\begin{tabular}[t]{ll}
(setTest)&
$\mkt{ \spc  i }\kdot\mkt{ \tpc  i } = \mkt{ \spc  i }$ for $i$ in $\labs(c)\union\{f\}$. \\
(diffTest)& 
$\mkt{ \tpc i } \kdot \mkt{ \tpc j } = \kzero$  for 
$i$ and $j$ in $\labs(c)\union\{f\}$ such that $i\neq j$. 
\end{tabular}
\\
By equational reasoning these yield consequences such as 
\\
\begin{tabular}[t]{ll}
(diffTestNeg)& $\mkt{\tpc i} = \mkt{\tpc i};\neg\mkt{\tpc j}$ 
for $i\neq j$ with $i,j$ in $\labs(c)\union\{f\}$. \\
(nf-enab-labs)& $\mkt{\enab(gcs)} = \mkt{ \quant{\lor}{i}{i\in\labs(c)}{ \tpc  i }}$.
\end{tabular}

\smallskip

The theorem's proof goes by rule induction on $\normSmall{c}{f}{gcs}$.
There is one normal form rule per command form, so we proceed by cases on those forms.
In each case, aside from unfolding definitions of $\addPC$, $\mkt{-}$, etc.,
we use only KAT reasoning and the $\nfax$ hypotheses,
together with induction hypotheses for subprograms.
The lengthy details can be found in \cite[Appendix~D]{BNN23v5}; here we show just one case.

\textbf{Case $c$ is $\lassg{n}{x}{e}$}.  
We have $\normSmall{\lassg{n}{x}{e}}{f}{(\tpc n\gcto \lassg{n}{x}{e};\spc f)}$.
Operationally, the loop $\ldogc{}{\tpc n\gcto \lassg{n}{x}{e};\spc f}$
iterates exactly once.  This is reflected in our proof of (\ref{eq:normEquiv}) for this case, in which we unroll the loop once;
see the calculation in \autoref{fig:assignCalc}.
Note: in hints we do not mention associativity, unit law for 1, etc.
\end{proof}

\begin{figure}[t]
\[\begin{array}{lll}
  & \mkt{\spc n ; \ldogc{}{ \tpc n \gcto \lassg{n}{x}{e}; \spc f} } \\
= & \hint{def $\mkt{-}$, see (\ref{eq:def:mkt}), and def $\enab$ } \\ 
  & \mkt{\spc n}; (\mkt{\tpc n} ; \mkt{\lassg{n}{x}{e}};\mkt{\spc f})^*;  \neg\mkt{\tpc n } \\
= & \hint{star unfold, distrib} \\
  & \mkt{\spc n}; \neg\mkt{\tpc n }  + \mkt{\spc n };  \mkt{\tpc n};  \mkt{\lassg{n}{x}{e}};\mkt{\spc f};  (\mkt{\tpc n};  \mkt{\lassg{n}{x}{e}};\mkt{\spc f})^*;  \neg\mkt{\tpc n } \\
= &\hint{left term is 0, using axiom (setTest) and lemma (diffTestNeg)} \\
  & \mkt{\spc n}; \mkt{\tpc n}; \mkt{\lassg{n}{x}{e}};\mkt{\spc f}; (\mkt{\tpc n};\mkt{\lassg{n}{x}{e}};\mkt{\spc f})^* ;\neg\mkt{\tpc n} \\
= &\hint{(setTest) for $n$}\\
  & \mkt{\spc n};  \mkt{\lassg{n}{x}{e}};\mkt{\spc f}; (\mkt{\tpc n}; \mkt{\lassg{n}{x}{e}};\mkt{\spc f})^*; \neg\mkt{\tpc n} \\
= &\hint{(setTest) for $f$} \\
  & \mkt{\spc n};  \mkt{\lassg{n}{x}{e}};\mkt{\spc f}; \mkt{\tpc f };  (\mkt{\tpc n};  \mkt{\lassg{n}{x}{e}};\mkt{\spc f})^*;  \neg\mkt{\tpc n } \\
= &\hint{(diffTest) with $f\neq n$ from $\okf(c,f)$;} \\    
  & \hint{KAT fact $p;(q;a)^* = p$ if $p$; $q=0$ (any $p,q,a$)}\\ 
  & \mkt{\spc n};  \mkt{\lassg{n}{x}{e}};\mkt{\spc f};  \mkt{\tpc f }; \neg\mkt{\tpc n } \\
= &\hint{(diffTestNeg), $f\neq n$ by $\okf(c,f)$} \\
  & \mkt{\spc n};  \mkt{\lassg{n}{x}{e}};\mkt{\spc f};  \mkt{\tpc f } \\
= &\hint{(setTest)} \\
  & \mkt{\spc n};  \mkt{\lassg{n}{x}{e}};\mkt{\spc f} \\
= & \hint{def $\mkt{-}$} \\
  & \mkt{\spc n ; \lassg{n}{x}{e}; \spc f }  \\
= & \hint{def $\addPC$} \\
  & \mkt{\addPC(\lassg{n}{x}{e}); \spc f } 
  \end{array}
\]
\caption{Proof of assignment case for \autoref{thm:normEquiv}.}
\label{fig:assignCalc}
\end{figure}

\begin{restatable}{lem}{lemnfEnabLabs}\label{lem:nf-enab-labs-X} 
\upshape
If $\okf(c,f)$ and $\norm{c}{f}{gcs}$ then
$\means{\enab(gcs)} = \means{\quant{\lor}{i}{i\in\labs(c)}{ \tpc  i }}$.
\end{restatable} 
\begin{proof}   
This follows from the provable equation (nf-enab-labs) mentioned in the proof of \autoref{thm:normEquiv},
together with \autoref{lem:correctInterp}.
\end{proof}

\section{Floyd completeness}\label{sec:autUnary}

In this section we put the normal form equivalence theorem to work showing that
any IAM proof of a unary correctness judgment can be translated to one in HL+.
This sets a pattern that guides the proofs of alignment completeness.  It also
gives a way to prove completeness of HL+ in the sense of Cook.

\begin{thm}\label{thm:FloydComplete}
\upshape
Suppose $\okf(c,f)$.
Suppose $an$ is a valid annotation of $\aut(c,f)$ for $\spec{P}{Q}$ 
and $P,Q,an$ are finitely supported.
Then $c: \spec{P}{Q}$ can be proved in HL+, using only assertions derived from $an$.
\end{thm}
The phrase ``derived from $an$'' is deliberately vague,
as is the similar result of Nagasamudram and Naumann~\cite{NagasamudramN21}.
As sketched by example in \autoref{sec:overview},
our HL+ proof uses only a single instance of the \rn{Do} rule and no instance of the \rn{If} rule.
More importantly, the judgments use only assertions derived from those of $an$ in simple ways.
In particular, we use boolean combinations of the following:
assertions $an(i)$, boolean expressions that occur in $c$,
and equality tests $pc=n$ of the program counter variable and numeric literals;
and we use substitution instances $\subst{an(i)}{x}{e}$ for assignments
$x:=e$ that occur in $c$.
The assumption about finite support is a technicality.
It holds for assertions expressed by formulas in any usual assertion language.

\begin{proof} 
Choose variable $pc$ that is fresh with respect to $c$, $P$, $Q$, and $an$ (i.e., for all $i$, $an(i)$ is independent from $pc$).
Existence of such a variable is ensured by the assumption of finite support.
By \autoref{lem:normExists} we have some $gcs$
with $\normSmall{c}{f}{gcs}$.
By \autoref{thm:normEquiv} we have
\begin{equation}\label{eq:nfEqu}
\spc n; \ldogc{}{gcs} \: \kateq \: \addPC(c); \spc f 
\quad\mbox{where $n=\lab(c)$.}
\end{equation}
For a loop invariant to reason about the normal form, with an eye on
the example we might try this formula:
\( 1 \leq pc \leq f \land \quant{\land}{i}{0\leq i\leq f}{pc=i \imp an(i)} \).
But this only makes sense if the labels form a contiguous sequence, which
we do not require.  There is no need to reason arithmetically about labels.
We define the invariant $I$ as follows:
\[ I: \quad
\quant{\lor}{i}{i\in\labs(c)\union\{f\}}{\tpc i} \land 
\quant{\land}{i}{i\in\labs(c)\union\{f\}}{\tpc i \imp an(i)} 
\]
The disjunction says the current value of $pc$ is in $\labs(c)\union\{f\}$.

The next step is to obtain proofs of 
\begin{equation}\label{eq:DOprem}
b:\spec{I\land e}{I} \qquad \mbox{for each $e\gcto b$ in $gcs$ }
\end{equation}
To do so, first note that we have for any $m$ that 
\begin{equation}\label{eq:set}
\spc m : \spec{an(m)}{an(m)\land \tpc m} 
\end{equation}
using rules \rn{Asgn} and \rn{Conseq}, because by freshness $pc$ is not in $an(m)$.
(We are not writing explicit $\proves$ for provability of correctness judgments.)
Second, note that by definition of $I$ we have valid implications
\begin{equation}\label{eq:Ian}
I\land \tpc n \imp an(n) \qquad\mbox{and}\qquad an(n)\land \tpc n \imp I 
\qquad\mbox{for any $n$ in $\labs(c)\union\{f\}$}
\end{equation}
Now go by the possible cases of $b$ in (\ref{eq:DOprem}),
which are given by \autoref{lem:nfCases}.
\begin{itemize}
\item 
$b$ has the form 
$\tpc n\gcto \lassg{}{x}{e};\spc m$, where $\sub(n,c)$ is $\lassg{n}{x}{e}$ 
and $m = \fsuc(n,c,f)$.

To show: $\lassg{}{x}{e};\spc m : \spec{I\land \tpc n}{I}$.
By rule \rn{Asgn} we have 
$\lassg{}{x}{e} : \spec{\subst{an(m)}{x}{e}}{an(m)}$.
By the VC in \autoref{fig:VC} we have
$an(n)\imp \subst{an(m)}{x}{e}$,
so using \rn{Conseq} we get
$\lassg{}{x}{e} : \spec{an(n)}{an(m)}$.
By fact (\ref{eq:set}) we have
$\spc m : \spec{an(m)}{an(m)\land \tpc m}$, so
by rule \rn{Seq} we have $\lassg{}{x}{e};\spc m : \spec{an(n)}{an(m)\land \tpc m}$.
So by \rn{Conseq} using both implications in fact (\ref{eq:Ian}) we get 
$\lassg{}{x}{e};\spc m : \spec{I\land \tpc n}{I}$.
\end{itemize}
The other cases are similar (see~\cite[Appendix~D.4]{BNN23v5}).

Having established the premises of rule \rn{Do}, we get its conclusion:
\[  \ldogc{}{gcs} : \spec{I}{I\land \neg\enab(gcs)} \]
By \autoref{lem:nf-enab-labs-X} and definition of $I$,
$I\land \neg\enab(gcs)$ is equivalent to $I\land \tpc f$, 
so by consequence we get 
\( \ldogc{}{gcs} : \spec{I \land \tpc n}{I\land \tpc f} \). 
Now $I\land \tpc n$ is equivalent to $an(n)\land \tpc n$.
We have $P\imp an(n)$ and $an(f)\imp Q$ 
because $an$ is an annotation for the spec $\spec{P}{Q}$.
So by consequence we get 
\( \ldogc{}{gcs} : \spec{P \land \tpc n}{Q} \).
By the assignment rule and consequence using that $pc$ is fresh for $P$ we get
$\spc n:\spec{P}{P\land\tpc n}$,
so using the sequence rule we get
\[ \spc n; \ldogc{}{gcs} : \spec{P}{Q} \]
Then \rn{Rewrite} using (\ref{eq:nfEqu}) yields 
\( \addPC(c); \spc f : \spec{P}{Q} \).
By \autoref{lem:addPC} and freshness of $pc$ we have that
$pc$ is ghost in $\addPC(c); \spc f$. Also, $P$ and $Q$ are independent from $pc$.
So by rule \rn{Ghost} we get 
\( \erase(pc,\addPC(c);\spc f) : \spec{P}{Q} \).
Now using \autoref{lem:addPC} together with the general law $c;\skipc\kateq c$
and transitivity of $\kateq$,\footnote{Both of which are easily
derived using the definition of $\kateq$.} 
we have that $\erase(pc,\addPC(c); \spc f) \kateq c$,
so by \rn{Rewrite}
we get $c: \spec{P}{Q}$.
\end{proof}

\begin{cor}[HL+ is Cook complete]
\upshape
If $\models c:\spec{P}{Q}$ and $P,Q$ are finitely supported then 
there is a proof of $c:\spec{P}{Q}$ in HL+.  
\end{cor}
\begin{proof}
Suppose $\models c:\spec{P}{Q}$.  
Without loss of generality assume $c$ has ok labels and $\okf(c,f)$ for some $f$.
By completeness of IAM (\autoref{prop:IAM}) there is a valid and finitely supported annotation of $\aut(c,f)$ for $P,Q$.
So by \autoref{thm:FloydComplete} there is a proof in HL+ of $c:\spec{P}{Q}$.  
\end{proof}

\section{RHL+ is alignment complete and Cook complete}\label{sec:acomplete}

In \autoref{sec:acompleteX} we prove alignment completeness.
In \autoref{sec:unaryRevisited} we prove Cook completeness and consider connections with unary logic, in particular the well known use of sequential alignment to obtain Cook completeness from a complete unary logic.  

\subsection{Alignment completeness of RHL+}\label{sec:acompleteX}

Our main result for $\forall\forall$ properties says that 
given any IAM-style proof for a program alignment automaton, one can construct an RHL+ proof.
Conditions (b) and (c) in the theorem say there is an IAM-style proof.

\begin{thm}\label{thm:acomplete}
\upshape
Suppose we have the following.
\\
(a) $\okf(c,f)$ and $\okf(c',f')$.
\\
(b) $an$ is a valid annotation of $\aprod(\aut(c,f),\aut(c',f'),L,R,J)$
for $\spec{\S}{\T}$.
\\
(c) $\breve{an}(i,j)\imp L \lor R \lor J \lor [\fin|\fin']$ for all control points $(i,j)$ of 
$\aprod(\aut(c,f),\aut(c',f'),L,R,J)$.
\\
(d) $an$, $\S$, $\T$, $L$, $R$, and $J$ all have finite support.
\\
Then the judgment $c\sep c': \rspec{\S}{\T}$ has a proof in RHL+.
\end{thm}
The proof of the Theorem yields a deductive proof that uses only relational assertions derived in simple ways from the relations $an(i,j)$ of the annotation together with $L$, $R$, and $J$.
Specifically, the proof uses boolean combinations 
of the annotation's assertions, conjunctions with conditional tests in the code (and with $L$ and $R$), 
and substitutions for expressions in assignment commands.

Restriction (a) in \autoref{thm:acomplete} is just a technicality.
The $\okf$ condition says labels of $c$ are unique and do not include $f$. 
Labels have no effect on program semantics so they can always be chosen to satisfy the condition.  Restriction (d) certainly holds when specs are given by formulas in some assertion language;
it is a technicality to ensure that a fresh $pc$ variable can be chosen for application of \autoref{thm:normEquiv}.

\begin{proof}
Suppose $\S,\T,c,c',f,f',L,R,J$ and $an$ satisfy the hypotheses (a)--(d) of the theorem.
Choose variable $pc$ that is fresh with respect to $\S,\T,c,c',an,L,R,J$.
To be precise: $\indep(pc|pc,\S)$, $\indep(pc|pc,\T)$, $pc$ does not occur in $c$ or $c'$,
and $L$, $R$, $J$ are independent from $pc$ on both sides,
as is $an(i,j)$ for all $i,j$.  
Existence of such a variable is ensured by hypothesis (d) of finite support.

By Lemma~\ref{lem:normExists} there are $gcs$ and $gcs'$ such that 
$\normSmall{c}{f}{gcs}$ and $\normSmall{c'}{f'}{gcs'}$.
Let $n=\lab(c)$ and $n'=\lab(c')$.
By Theorem~\ref{thm:normEquiv} we have
\begin{equation}\label{eq:nfEquLR}
\begin{array}{l}
\spc n; \ldogc{}{gcs} \: \kateq \: \addPC(c); \spc f 
\\
\spc n'; \ldogc{}{gcs'} \: \kateq \: \addPC(c'); \spc f' 
\end{array}
\end{equation}
Define store relation $\Q$ to be $\Q_{an}\land\Q_{pc}$ where 
\[ \begin{array}{l}
\Q_{an}: \qquad
   \quant{\land}{i,j}{i\in \labs(c)\union\{f\} \land j\in \labs(c')\union\{f'\}
   }{\bothF{\tpc i\sep \tpc j} \imp an(i,j)} 
\\
\Q_{pc}: \qquad 
   \quant{\lor}{i,j}{i\in \labs(c)\union\{f\} \land j\in \labs(c')\union\{f'\}
   }{\bothF{\tpc i\sep \tpc j}}
   \end{array}
\]
We will derive 
\begin{equation}\label{eq:C}
 \ldogc{}{gcs} \sep \ldogc{}{gcs'} : \rspec{\Q}{\Q\land\neg\leftF{\enab(gcs)}\land\neg\rightF{\enab(gcs')}}  
 \end{equation}
using rule \rn{rDo} instantiated with $\Q:=\Q$, $\Lrel:=\encode{L}$, and $\R:=\encode{R}$.
The side condition of \rn{rDo} is
\begin{equation}\label{eq:side} \Q \imp 
(\leftex{\enab(gcs)} = \rightex{\enab(gcs')})
          \lor (\encode{L} \land \leftF{\enab(gcs)})
          \lor (\encode{R} \land \rightF{\enab(gcs')})
\end{equation}
To prove (\ref{eq:side}), first rewrite $\Q$ using distributivity and renaming dummies, to the equivalent form
\[
   \quant{\lor}{i,i'}{}{ 
     \bothF{\tpc i\sep \tpc i'} \land 
     \quant{\land}{k,k'}{}{ 
         (\bothF{\tpc k\sep\tpc k'} \imp an(k,k')) }}
\]
where we omit that $i,k$ range over $\labs(c)\union\{f\}$
and $i',k'$ range over $\labs(c')\union\{f'\}$.
This implies 
\[
   \quant{\lor}{i,i'}{}{  \bothF{\tpc i\sep\tpc i'} \land an(i,i'       ) }
\] 
Thus by hypothesis (c), any $\Q$-state satisfies
$\encode{L} \lor \encode{R} \lor \encode{J} \lor \bothF{\tpc f\sep\tpc f'}$.
We show each of these disjuncts implies the right side of (\ref{eq:side}).
\begin{itemize}
\item $\encode{L}$ implies $\encode{L}\land\leftF{\enab(gcs)}$ 
because (i) $L$ is a set of states of the alignment automaton,
with control on the left ranging over $\labs(c)\union\{f\}$,
(ii) by liveness (Def.~\ref{def:alignProd}), $L$ allows transitions 
by $\aut(c,f)$, and so excludes control being at $f$,
and (iii) $\leftF{\enab(gcs)}$ means control on the left is in $\labs(c)$,
             by Lemma~\ref{lem:nf-enab-labs-X}.
\item $\encode{R}$ implies $\encode{R}\land\rightF{\enab(gcs')}$
for reasons symmetric to the $L$ case
\item $\encode{J}$ implies $\leftF{\enab(gcs)}$ and $\rightF{\enab(gcs')}$ are both true
(using liveness and Lemma~\ref{lem:nf-enab-labs-X} again),
so $\leftF{\enab(gcs)} = \rightF{\enab(gcs')}$
\item $\bothF{\tpc f\sep\tpc f'}$ implies 
both $\leftF{\enab(gcs)}$ and $\rightF{\enab(gcs')}$ are false (again using  Lemma~\ref{lem:nf-enab-labs-X}) 
so $\leftF{\enab(gcs)} = \rightF{\enab(gcs')}$
\end{itemize}
So the side condition (\ref{eq:side}) of \rn{rDo} is proved.
Before proceeding to prove the premises for \rn{rDo}, note that we can prove
\begin{equation}\label{eq:setR}
\spc m \sep \spc m' : \rspec{\P}{\P\land \leftF{\tpc m}\land\rightF{\tpc m'}} 
\end{equation}
for any $\P$ in which $pc$ does not occur, and any $m,m'$,
using rules \rn{rAsgn} and \rn{rConseq}.
Also, by definition of $\Q$ we have valid implications 
\begin{equation}\label{eq:IanRel}
\begin{array}[t]{l} 
\Q\land \bothF{\tpc m\sep\tpc m'} \imp an(m,m') 
\qquad\mbox{and}\qquad 
an(m,m') \land \bothF{\tpc m\sep\tpc m'} \imp \Q \\
\mbox{for any $m$ in $\labs(c)\union\{f\}$ and 
$m'$ in $\labs(c')\union\{f'\}$} 
\end{array}
\end{equation}
From condition (c) of the theorem we get  
\( 
an(i,j)\land\bothF{\tpc i\sep\tpc j}\imp\encode{L}\lor\encode{R}\lor\encode{J}\lor\bothF{\tpc \fin\sep\tpc\fin'}
\)
for all $i,j$ (by definitions), and hence
\begin{equation}\label{eq:dliftJ}
an(i,j)\land\bothF{\tpc i\sep\tpc j}\land\neg\encode{L}\land\neg\encode{R}\imp \encode{J}
\quad\mbox{for all $i,j$ with $i\neq\fin$ or $j\neq\fin'$}
\end{equation}

There are three sets of premises of \rn{rDo} for the loops in (\ref{eq:C}), with these forms:
\\
\begin{tabular}{ll}
(left-only) &
$b\sep \skipc : \rspec{\Q\land \leftF{e}\land\encode{L} }{\Q}$
for each $e\gcto b$ in $gcs$ 
\\
(right-only) &
$\skipc\sep b' : \rspec{\Q\land \rightF{e'}\land\encode{R} }{\Q}$
for each $e'\gcto b'$ in $gcs'$ 
\\
(joint) & 
$b\sep b' : \rspec{\Q\land \bothF{e\sep e'} \land \neg\encode{L} \land \neg\encode{R}}{\Q}$
for each $e\gcto b$ in $gcs$ and $e'\gcto b'$ in $gcs'$
\end{tabular}
\\
By Lemma~\ref{lem:nfCases}, the guarded commands in $gcs$ and $gcs'$ have six possible forms,
so there are six left-only cases to consider, six right-only, and 36 joint ones.
We start with the latter.

\paragraph{Joint cases}

For each of the six possibilities for $e\gcto b$ in $gcs$ for $c$,
we must consider it with each of the six possibilities for $e'\gcto b'$ in $gcs'$ for $c'$.
We give the argument for one case, with skip on both sides.
\begin{itemize}
\item $\spc m\sep \spc m' : \;
\rspec{\Q\land \bothF{\tpc k\sep \tpc k'} \land \neg\encode{L} \land \neg\encode{R}}{\Q}$ , where \\
$\sub(k,c) = \lskipc{k}$, $m=\fsuc(k,c,f)$,
$\sub(k',c') = \lskipc{k'}$, $m=\fsuc(k',c',f')$.

By (\ref{eq:setR}) (and freshness of $pc$) we have
$\spc m \sep \spc m' : \rspec{an(m,m')}{an(m,m')\land \bothF{\tpc m\sep\tpc m'}}$.

So by \rn{rConseq} using the second implication in (\ref{eq:IanRel}) we have
\begin{equation}\label{eq:mm}
\spc m \sep \spc m' : \rspec{an(m,m')}{\Q}
\end{equation}
By the first implication in (\ref{eq:IanRel}) we have
$\Q\land\bothF{\tpc k\sep\tpc k'}\imp an(k,k')$.
So using (\ref{eq:dliftJ}) we get 
$\Q\land\bothF{\tpc k\sep\tpc k'}\land\neg\encode{L}\land\neg\encode{R}\imp \encode{J}$.
By validity of the annotation, we have the VC in the first row of \autoref{fig:RVCjo},
i.e., $J\land \breve{an}(k,k')\imp \hat{an}(m,m')$.
Then by Lemma~\ref{lem:liftRVC} we get the $pc$-encoded form 
$\encode{J}\land\bothF{\tpc k\sep\tpc k'}\land an(k,k')\imp an(m,m')$.
So this is valid:
\[ \Q\land \bothF{\tpc k\sep\tpc k'} \land \neg\encode{L} \land \neg\encode{R}
\imp an(m,m') \]
Using this with \rn{rConseq} and (\ref{eq:mm}) yields
$\spc m\sep \spc m' : \;
\rspec{\Q\land \bothF{\tpc k\sep \tpc k'} \land \neg\encode{L} \land \neg\encode{R}}{\Q}$.
\end{itemize}
We refrain from spelling out details of the remaining joint cases (more can be found in \cite{BNN23v5}).  
The arguments are all similar: every case uses a VC and rule \rn{rConseq},
together with (\ref{eq:setR}).
Cases that involve an assignment in the original program $c$ or $c'$ also
use rules \rn{rAsgnAsgn}, \rn{rSkipAsgn}, or \rn{rAsgnSkip}.
Cases that involve havoc use the corresponding rules.

\paragraph{Left-only cases.}

These cases are proved using the same rules as the joint cases,
plus one additional rule: \rn{rDisj}, in the form \rn{rDisjN} derived from it (see \autoref{fig:derivedRHL}).  
This is needed due to the following complication.
The joint cases determine a starting pair and ending pair of control points,
which determines which VC to appeal to.
The left-only cases do not determine a control point on the right side; instead
we have VCs for each possible point on the right (\autoref{fig:RVClo-encoded}).
So we go by cases on the possible control points on the right, for which purpose
we make the following observation.
In virtue of the conjunct $\Q_{pc}$ of $\Q$, we have that $\Q$ is equivalent to this 
disjunction over control points:
\[ 
\quant{\lor}{i,j}{i\in \labs(c)\union\{f\} \land 
                    j\in \labs(c')\union\{f'\} }{ \Q^{i,j} }
\]
where $\Q^{i,j}$ is defined to say control is at those points:
\( \graybox{$\Q^{i,j}$} \eqdef \Q\land\bothF{\tpc i\sep\tpc j} \).
Now we have the equivalence
\[ \Q\land\leftF{\tpc i}
 \iff 
\quant{\lor}{j}{j\in \labs(c')\union\{f'\} }{ \Q^{i,j} }
\quad\mbox{for any $i$}
 \]
With this we can proceed to prove the left-only premises.  
We give only the assignment case.
\begin{itemize}
\item $\lassg{}{x}{e};\spc m \sep \skipc : \rspec{\Q\land \leftF{\tpc k}\land\encode{L} }{\Q}$, where 
$\sub(k,c) = \lassg{k}{x}{e}$ and $m = \fsuc(k,c,f)$.

In accord with the discussion above, we have that
$\Q\land \leftF{\tpc k}$ is equivalent to 
$\quant{\lor}{j}{}{ \Q^{k,j} }$ (omitting the range $j\in \labs(c')\union\{f'\}$), 
so the goal can be obtained by \rn{rConseq} 
from 
\[ \lassg{}{x}{e};\spc m\sep \skipc : \rspec{\quant{\lor}{j}{}{ \Q^{k,j} }\land\encode{L} }{\Q}
\]
In turn, this can be obtained by derived rule \rn{rDisjN} from judgments 
\begin{equation}\label{eq:Y}
\lassg{}{x}{e};\spc m \sep \skipc : \rspec{\Q^{k,j} \land \encode{L} }{\Q} 
\quad\mbox{for every $j$.}
\end{equation}
for all $j$ (in range $j\in \labs(c')\union\{f'\}$ that we continue to omit). 
It remains to prove (\ref{eq:Y}) for arbitrary $j$.  

By \rn{rAsgnSkip} and \rn{rConseq} (using that $\rightF{\tpc j}$ is independent from $pc$ on the left) we get
\[ \spc m\sep\skipc : \rspec{ an(m,j)\land\rightF{\tpc j} }{ an(m,j) \land \bothF{\tpc m\sep\tpc j}} \]
from which using (\ref{eq:IanRel}) we get  
\[ \spc m\sep\skipc : \rspec{ an(m,j)\land\rightF{\tpc j} }{ \Q } \]
By \rn{rAsgnSkip}, using that $\rightF{\tpc j}$ is independent from $x$ because $pc$ is fresh, we can prove
\[ x:=e\sep\skipc : \rspec{ \subst{an(m,j)}{x|}{e|} \land \rightF{\tpc j}}{an(m,j)\land\rightF{\tpc j}} 
\]
By derived rule \rn{rSeqSkip} (\autoref{fig:derivedRHL}),
from the above we get 
\[ x:=e;\spc m\sep\skipc : \rspec{ \subst{an(m,j)}{x|}{e|} \land \rightF{\tpc j}}{\Q} 
\]
The disjunction over $j$ was introduced so that we can appeal to a VC,
specifically the lifted VC for $((k,j),(m,j))$.  
It is an instance of the second line in \autoref{fig:RVClo-encoded} and it says this is valid:
\( \encode{L}\land\bothF{\tpc k\sep\tpc j}\land an(k,j) \imp \subst{an(m,j)}{x\sep}{e\sep}
\),
so by \rn{rConseq} we get 
\[ x:=e;\spc m\sep\skipc : \rspec{ \encode{L}\land\bothF{\tpc k\sep\tpc j}\land an(k,j) }{\Q} 
\]
By definitions we have $\Q^{k,j} \land \encode{L} \imp \encode{L}\land\bothF{\tpc k\sep\tpc j}\land an(k,j) $.
Using this with \rn{rConseq} yields (\ref{eq:Y}) and we are done with this case.
\end{itemize}
The other left-only cases are similar.  
The right-only cases are symmetric with the left-only cases.
We omit them all and proceed.

\paragraph{Finishing the proof.}

Having proved the premises and the side condition (\ref{eq:side}), rule \rn{rDo} yields 
(\ref{eq:C}).   The remaining steps are similar to corresponding steps in the proof of
the Floyd completeness Theorem~\ref{thm:FloydComplete} and we spell them out.

Using \autoref{lem:nf-enab-labs-X} twice, and the definition of $\Q$,
we have 
\[ \Q\land\neg\leftF{\enab(gcs)}\land\neg\rightF{\enab(gcs')} 
\imp \Q\land\bothF{\tpc f\sep\tpc f'} \]
So, using the first implication in (\ref{eq:IanRel}) 
and assumption (b) of the theorem (which says $an(f,f')\imp\T$ since $an$
is an annotation for $\rspec{\S}{\T}$),
we can use 
\rn{rConseq} with (\ref{eq:C}) to get
\[
 \ldogc{}{gcs} \sep \ldogc{}{gcs'} : \rspec{\Q}{\T}
\]
Using the second implication in (\ref{eq:IanRel})
and assumption (b) (which says $\S\imp an(n,n')$ 
since $n,n'$ are the initial control points),
we have $\bothF{\tpc n\sep \tpc n'} \land \S \imp \Q$,
so by \rn{rConseq} we get 
\[
 \ldogc{}{gcs} \sep \ldogc{}{gcs'} : \rspec{\bothF{\tpc n\sep \tpc n'} \land \S}{\T}
\]
By freshness assumption for $pc$, by (\ref{eq:setR})
we have a proof of $\spc n\sep \spc n': \rspec{\S}{\bothF{\tpc n\sep\tpc n'} \land \S}$.
So by \rn{rSeq} we get
\[
 \spc n;\ldogc{}{gcs} \sep \spc n';\ldogc{}{gcs'} : \rspec{\S}{\T}
\]
Now using rule \rn{rRewrite} with the 
equivalences (\ref{eq:nfEquLR}) we get 
\[
\addPC(c); \spc f \sep \addPC(c'); \spc f' : \rspec{\S}{\T}
\]
By freshness of $pc$ and Lemma~\ref{lem:addPC}, 
it has the ghost property for both 
$\addPC(c); \spc f$ and $\addPC(c'); \spc f'$,
and does not occur in $\S$ or $\T$, so by rule \rn{rGhost} we get 
\begin{equation}\label{eq:H}
\erase(pc,\addPC(c); \spc f) \sep \erase(\addPC(c'); \spc f') : \rspec{\S}{\T}
\end{equation}
By definition of $\erase$, we have
$\erase(pc,\addPC(c); \spc f) = \erase(pc,\addPC(c)); \skipc$ 
and also 
\\
$\erase(\addPC(c'); \spc f') = \erase(pc,\addPC(c')); \skipc$. 
So using Lemma~\ref{lem:addPC} together with the general law $c;\skipc\kateq c$
and transitivity of $\kateq$, we have 
\[ \erase(pc, \addPC(c); \spc f) \kateq c \qquad\mbox{and}\qquad 
   \erase(pc, \addPC(c'); \spc f') \kateq c' \]
Using these equivalences with \rn{rRewrite}, from (\ref{eq:H}) we obtain 
$c  \sep c'  : \rspec{\S}{\T}$. 
\end{proof}

\subsection{Cook completeness revisited}\label{sec:unaryRevisited}

Cook completeness can be proved as a consequence of alignment completeness.

\begin{thm}[Cook completeness of RHL+]\label{thm:CookCompleteRHL}
\upshape
Suppose $\models c\sep c': \rspec{\S}{\T}$ and $\S,\T$ are finitely supported.
Then there is a proof of $c\sep c': \rspec{\S}{\T}$ in RHL+.
\end{thm}
\begin{proof}
Suppose $\models c\sep c': \rspec{\S}{\T}$, and assume wlog
that $\okf(c,f)$ and $\okf(c',f')$. 
Using Lemma~\ref{lem:autConsistent} 
we have $\aut(c,f), \aut(c',f') \models \rspec{\S}{\T}$.
By \autoref{cor:relIAMcomplete} there are $L,R,J,an$ such that 
$an$ is a valid annotation of $\aprod(\aut(c,f),\aut(c',f'),L,R,J)$ for $\spec{\S}{\T}$ and for any $i,j$ we have that
$\breve{an}(i,j)\imp L \lor R \lor J \lor [\fin|\fin']$ 
which is condition (c) of Theorem~\ref{thm:acomplete}.
Moreover these are finitely supported as required by condition (d).
We have conditions (a) and (b) as well, so by the theorem we 
get a proof of 
$c\sep c': \rspec{\S}{\T}$ in RHL+.
\end{proof}

Prior Cook completeness results for RHLs were based on a left-first sequential alignment rule like our \rn{rLRseq} in \autoref{fig:derivedRHL},
together with unary HL and a way to represent or interpret the one-sided judgments
$c\sep\skipc:\rspec{\P}{\Q}$ and 
$\skipc\sep c':\rspec{\Q}{\R}$ as unary judgments in HL.
It is instructive to consider that approach in our setting.
But first we indulge in little detour.

Our formulation of RHL+ does not include unary judgments but it subsumes HL+ in the following sense.
A unary judgment $c:\spec{P}{Q}$ can be represented by
the relational judgment $c\sep\skipc:\rspec{\leftF{P}}{\leftF{Q}}$,
as well as by $\skipc\sep c:\rspec{\rightF{P}}{\rightF{Q}}$.
That is, the judgments express the same semantic property of $c$.  
So if the judgment $c:\spec{P}{Q}$ is valid then we have
$\models c\sep\skipc:\rspec{\leftF{P}}{\leftF{Q}}$,
hence by \autoref{thm:CookCompleteRHL}
one can prove $c\sep\skipc:\rspec{\leftF{P}}{\leftF{Q}}$ in RHL+.
Furthermore, any proof in HL+ gives rise to a proof in RHL+ with the same structure and intermediate assertions.  This is because every rule in HL+, when translated to the representation $c\sep\skipc:\rspec{\leftF{P}}{\leftF{Q}}$, is a derivable rule in RHL+. 
Indeed, they are instances of the more general one-sided rules \rn{rAsgnSkip},
\rn{rIfSkip}, etc.\ in \autoref{fig:RHL} and \autoref{fig:derivedRHL}.

Having completed the detour we return to the reduction of relational reasoning to unary.
To this end we need to interpret judgments of the forms
$c\sep\skipc:\rspec{\P}{\Q}$ and 
$\skipc\sep c':\rspec{\Q}{\R}$ ---which are one-sided in terms of the code, but which involve general relational formulas--- as unary judgments.  
Such an interpretation is easier to formulate in terms of syntactic formulas.
By renaming the variables of $c$ and $c'$ to be disjoint, and renaming the variables 
of the relational formulas accordingly, one can consider that $\Q$ simply \emph{is} a unary formula.\footnote{Renaming can be minimized by assuming at the outset that the considered programs $c,c'$ are acting on separable parts of the store~\cite{BartheDArgenioRezk,BartheDR11}, but some complications are inevitable for programs acting on the heap~\cite{Naumann06esorics,Beringer11}.}

In the present context, using shallow embedding of relations and assertions, let us postulate the following:
Any store relation $\Q$ has an encoding $\codeleft{\Q}$ as a predicate on variable stores 
where the variables on which the considered program acts are ``on the left'',
and an encoding $\coderight{\Q}$ where those variables are ``on the right''.\footnote{Such encodings are slightly tricky to formalize in our setting with stores as total maps on all variables. 
We sketch the idea in terms of a simpler setting where stores are finite maps,
written like $\{x\scol 1,y\scol 7,z\scol 12\}$.  Suppose that for the relevant variables for the considered programs and specs are $x,y,\dots$.  Assume some bijection to a disjoint set of variables $\hat{x},\hat{y},\dots$.
So a pair of stores, say $(\{x\scol 1, y\scol 2, z\scol 3\},\{x\scol 4, y\scol 5, z\scol 6\})$
has encodings  
$\codeleft{(\{x\scol 1, y\scol 2, z\scol 3\},\{x\scol 4, y\scol 5, z\scol 6\})} 
\eqdef \{x\scol 1, y\scol 2, z\scol 3, \hat{x}\scol 4, \hat{y}\scol 5, \hat{z}\scol 6\}$
and 
$\coderight{(\{x\scol 1, y\scol 2, z\scol 3\},\{x\scol 4, y\scol 5, z\scol 6\})} 
\eqdef \{ \hat{x}\scol 1, \hat{y}\scol 2, \hat{z}\scol 3, x\scol 4, y\scol 5, z\scol 6\}$.
Then $\codeleft{\P} \eqdef \{ \codeleft{(s,t)} \mid (s,t)\in \P \}$.
To do something similar for total map stores,
one should restrict to finitely supported relations $\P$ so 
the definitions of $\codeleft{\P}$ and $\coderight{\P}$ can exploit unused variables.
}
That is, the encodings satisfy the following, for all $c,\P,\Q$.
\begin{equation}\label{eq:encodeUnary}
\begin{array}[t]{l}
\models c \sep \skipc : \rspec{\P}{\Q}  \quad\mbox{iff}\quad
\models c : \spec{\codeleft{\P}}{\codeleft{\Q}}
\\
\models \skipc \sep c : \rspec{\P}{\Q}  \quad\mbox{iff}\quad
\models c : \spec{\coderight{\P}}{\coderight{\Q}} 
\end{array}
\end{equation}
To put these tedious details to work, we need the following which can be proved straightforwardly using semantic weakest preconditions or strongest postconditions. For later reference we introduce notation for weakest preconditions of program pairs:
$\graybox{$\WP(c\sep c')(\R)$} \eqdef \{ (s,s') \mid \all{t,t'}{\means{c}\,s\,t \land \means{c'}\,s'\,t' \imp (t,t') \in \R}\}$.  
Note that 
$\models c \sep c' : \rspec{\P}{\R}$ iff 
$\P\imp\WP(c\sep c')(\R)$.
\begin{lem} \label{lem:aa-decomp}
  \upshape
  For any $c,c',\P$ and $\R$, the following are equivalent:
  \begin{enumerate}
  \item $\models c \sep c' : \rspec{\P}{\R}$
  \item There is a $\Q$ such that $\models c \sep \skipc : \rspec{\P}{\Q}$ and $\models \skipc \sep c' : \rspec{\Q}{\R}$.
  \item There is a $\Q$ such that $\models \skipc \sep c' : \rspec{\P}{\Q}$ and $\models c \sep \skipc : \rspec{\Q}{\R}$.
  \end{enumerate}
\end{lem}
A direct consequence is the soundness of the following rules, which are analogous to 
\rn{rLRseq} and \rn{rRLseq}.
\begin{mathpar}
\inferrule[uLRseq]{
c : \spec{\codeleft{\P}}{\codeleft{\Q}} \\
c': \spec{\coderight{\Q}}{\coderight{\R}} 
}{
c\sep c' : \rspec{\P}{\R}
}

\inferrule[uRLseq]{
c' : \spec{\coderight{\P}}{\coderight{\Q}}  \\
c  : \spec{\codeleft{\Q}}{\codeleft{\R}}
}{
c\sep c' : \rspec{\P}{\R}
}
\end{mathpar}
Now we obtain a completeness result akin to those in the literature:
rule \rn{uLRseq}, together with HL, comprises a Cook complete logic for relational judgments.
The proof is as follows.
Suppose 
$\models c\sep c' : \rspec{\P}{\R}$ is valid.  
By Lemma~\ref{lem:aa-decomp}, there is a $\Q$ such that 
$\models c \sep \skipc : \rspec{\P}{\Q}$ and 
$\models \skipc \sep c' : \rspec{\Q}{\R}$.
By (\ref{eq:encodeUnary}) we have 
$\models c : \spec{\codeleft{\P}}{\codeleft{\Q}}$ and 
$\models c' : \spec{\coderight{\Q}}{\coderight{\R}}$.
By completeness of HL these judgments are provable.
Application of rule \rn{rLRseq} proves 
$c\sep c' : \rspec{\P}{\R}$.
A symmetric proof establishes Cook completeness of \rn{uRLseq} plus HL.

In accord with the detour about embedding HL in RHL+, the preceding considerations lead to an alternate proof of \autoref{thm:CookCompleteRHL} along the following lines.
First show that RHL+ is complete for one-sided judgments (since HL is). 
Second, the rules \rn{rSeq} and \rn{rRewrite} suffice to derive
\rn{rRLseq} (in fact using only command equivalences of the form 
$c;\skipc\kateq c$ and $\skipc;c\kateq c$).
Finally, proceed by an argument similar to the preceding paragraph.

\section{The \texorpdfstring{$\forall\exists$}{∀∃} logic ERHL+}\label{sec:ERHL}


In this section, we consider a standalone deductive system for the
$\aespecSym$ judgment, called ERHL+, which involves only the $\aespecSym$
judgment together with assertion validity and command equivalence $\kateq$
just like RHL+.  Unary correctness is subsumed because
$c:\spec{P}{Q}$ is valid iff $c\sep\skipc: \aespec{\leftF{P}}{\leftF{Q}}$ is
valid.  
On the other side, 
$\skipc\sep c: \aespec{\rightF{P}}{\rightF{Q}}$ is valid iff
the \dt{forward underapproximation} judgment
$c: \espec{P}{Q}$ is valid.  
This is defined by 
\begin{equation}\label{eq:espec}
\mbox{\graybox{$\models c : \espec{P}{Q}$} iff for any $s \in P$ there exists $t \in Q$ such that $\means{c}\,s\,t$. }
\end{equation}
This has been called \emph{possible correctness}~\cite{Hoare:wp}
and more recently the \emph{existential} Hoare triple of~\cite{DickersonYZD22} and the \emph{sufficient incorrectness} triple of~\cite{Ascari24}.\footnote{O'Hearn's incorrectness judgment~\cite{OHearn2019} ---backwards underapproximation--- 
is similarly obtained by a backwards version of (\ref{eq:aespec}),
as noted in~\cite{AntonopoulosEtal2022popl}.}


\begingroup 
\newlength{\specHeight}
\settoheight{\specHeight}{\(\aespec{P}{P}\)}
\newcommand{\myStrut}{\rule{0pt}{\specHeight}}

\begin{figure}[t]
\begin{footnotesize}
\begin{mathpar}
\inferrule[eSkipHav]{}{ 
  \skipc\sep\havc{x}:\aespec{(\someSet{\smSep x}{\P})}{\P}
}

\inferrule[eHavSkip]{}{ 
  \havc{x}\sep\skipc : \aespec{(\allSet{x\smSep}{\P})}{\P}
}

\inferrule[eDo]{ 
 {\begin{array}{l}
   c\sep \skipc : \aespec{\Q\land \leftF{e}\land\Lrel }{\Q}
       \quad\mbox{for all $e\gcto c$ in $gcs$} 
   \\
   \skipc\sep c' : \aespec{\Q\land \rightF{e'}\land\R\land V=v }{\Q\land V\prec v}
   \quad\mbox{for all $e'\gcto c'$ in $gcs'$ and all $v\in D$} 
   \\
  c\sep c' : \aespec{\Q\land \leftF{e}\land\rightF{e'} \land \neg\Lrel \land \neg\R}{\Q}
  \quad\mbox{for all $e\gcto c$ in $gcs$ and $e'\gcto c'$ in $gcs'$} 
  \\
  \myStrut 
  \Q\imp (\leftex{\enab(gcs)} = \rightex{\enab(gcs')}  
           \lor (\Lrel \land \leftF{\enab(gcs)})
           \lor (\R \land \rightF{\enab(gcs')}))
 \\
   \end{array}}
}{
   \dogc{gcs} \Sep \dogc{gcs'} : \aespec{\Q}{\Q\land \neg\leftF{\enab(gcs)}\land\neg\rightF{\enab(gcs')}}
}
\end{mathpar}
\end{footnotesize}
\vspace*{-1ex}
\caption{Rules of ERHL+.
In \rn{eDo}, $(D,\prec)$ is well-ordered and
$V$ is a total function $(\Var\to\Z)\times(\Var\to\Z)\to D$.
Rules \rn{eRewrite}, \rn{eGhost}, \rn{eSkip}, \rn{eSkipAss}, \rn{eAssSkip}, \rn{eSeq}, \rn{eIf}, \rn{eConseq}, \rn{eDisj}, and \rn{eFalse} are the same as those with corresponding names in \autoref{fig:RHL} but for $\aespecSym$.
}
\label{fig:ERHL}
\end{figure}

\endgroup


The rules of ERHL+ appear in \autoref{fig:ERHL}.  Most have the same form as corresponding rules in \autoref{fig:RHL},
although the soundness proofs are different in detail.  
If $d$ is deterministic then $\models c\sep d : \aespec{\P}{\Q}$ implies
$\models c\sep d : \rspec{\P}{\Q}$.  If $d$ denotes a domain-total\footnote{Domain-totality means the command may terminate from any state, not that it must.
For example,  $\ifgc{true\gcto diverge \gcsep true \gcto \skipc}$ is domain-total.}
relation then
$\models c\sep d : \rspec{\P}{\Q}$ implies $\models c\sep d : \aespec{\P}{\Q}$.  This hints at
why some proof rules for $\aespecSym$ are the same as those in
\autoref{fig:RHL} for the $\rspecSym$ judgment. 

The rule \rn{eSkipHav} reflects the $\forall\exists$ nature of the $\aespecSym$ judgment, 
existentially quantifying $x$ on the right state
whereas \rn{eHavSkip} universally quantifies it on the left.
Note that \rn{eIf} and \rn{eDo} do \emph{not} existentially quantify over guarded commands on the right, as one might guess at first.  This would be unsound, because property (\ref{eq:aespec}) universally quantifies over all pairs of initial states; we return to this later.  
A distinguishing feature of the $\aespecSym$ judgment is that it does not validate the rule of conjunction which is sound for the $\forall\forall$ judgment.  For example, 
both $\skipc\sep\havc{x} : \aespec{true}{\rightF{x<0}}$ and 
$\skipc\sep\havc{x} : \aespec{true}{\rightF{x>0}}$ are valid 
but not $\skipc\sep\havc{x}:\aespec{true}{\rightF{x<0 \land x>0}}$.

Besides \rn{eSkipHav}, another noticeable difference from RHL+ is \rn{eDo}.
Like the rule \rn{rDo} in RHL+, \rn{eDo} captures a conditional alignment of iterations directed by the
relations $\Lrel$ and $\R$.  The side condition connects the invariant $\Q$ with $\Lrel$ and $\R$ in a way that ensures adequacy in the sense of covering all iterations.  The rule differs
from \rn{rDo} in its treatment of right-only iterations.  The $\forall\exists$
judgment requires termination on the right, and the rule relies on the standard
approach of showing a variant decreases.\footnote{As an alternative to
universally quantifying over $v\in D$ in the metalanguage, it is possible to
formulate rule \rn{eDo} using a fresh program variable to snapshot the initial value of $V$, with side condition $\Q\imp V\geq 0$.
This is sufficient because for completeness it suffices to take $D$ to
be $\nat$; this is shown in the proof of \autoref{thm:FiltSoundCompl}.} 
The variant $V$ maps pairs of stores to some well-ordered set $D$.
In accord with the shallow embedding of
relational assertions, for any value $v$ in $D$ we write $V=v$ to denote the
relation $\{(s,s') \mid V(s,s') = v \}$.
Likewise for $V\prec v$.  

Formally the premise for $\skipc\sep c'$ in \rn{eDo} is a $D$-indexed set of premises,
but this is just an artifact of shallow embedding.   
Think of $v$ as a logical variable in a single premise,
universally quantified over the judgment.


\begin{figure}[t]
\begin{footnotesize}
\begin{mathpar}
\inferrule[eHavHav]{}{
\havc{x}\sep \havc{x'} : \aespec{(\allSet{x|}{(\someSet{|x'}{\P})})}{\P}
}

\inferrule[eAsgnAsgn]{}{
x:=e \sep x':=e' : \aespec{\subst{\R}{x|x'}{e|e'}}{\R}
}

\inferrule[eSkipDo]{ 
   \skipc\sep c : \aespec{\Q\land \rightF{e}\land V=v }{\Q\land V\prec v}
   \quad\mbox{for all $e\gcto c$ in $gcs$ and all $v\in D$} 
}{
   \skipc \Sep \dogc{gcs} : \aespec{\Q}{\Q\land\neg\rightF{\enab(gcs)}}
}
\end{mathpar}
\end{footnotesize}
\vspace*{-1ex}
\caption{Some derived rules of ERHL+.
Rules \rn{eLRseq}, \rn{eRLseq}, \rn{eAlgnIf}, \rn{eSeqSkip}, \rn{eIfSkip}, \rn{eDoSkip}, and
\rn{eDisjN} are the same as those with corresponding names in \autoref{fig:derivedRHL} but for $\aespecSym$.
}\label{fig:derivedERHL}
\end{figure}

Some derived rules are in \autoref{fig:derivedERHL}.
Rule \rn{eHavHav} is proposed in~\cite{AntonopoulosEtal2022popl}.
It expresses that $\P$ in the initial state must be total as a relation from $x$ on the left to $x'$ on the right.
It can be derived as follows.
Instantiate \rn{eHavSkip} as 
$\havc{x}\sep\skipc : \aespec{(\allSet{x|}{(\someSet{|x'}{\P})})}{\someSet{|x'}{\P}}$.
Then use \rn{eSeq} with 
$\skipc\sep\havc{x}:\aespec{\someSet{|x}{P}}{P}$ (from \rn{eSkipHav})
to get 
\[ \havc{x};\skipc \sep \skipc;\havc{x'} : \aespec{(\allSet{x|}{(\someSet{|x'}{\P})})}{\P} \]
Now obtain \rn{eHavHav} by \rn{eRewrite} using skip unit laws.

Aside from \rn{eSkipDo}, the other derived rules in \autoref{fig:derivedERHL}
are like those in \autoref{fig:derivedRHL} and can be
derived using \rn{eRewrite}.  For instance, 
\rn{eAsgnAsgn} can be derived using 
\rn{eSkipAsgn}, \rn{eAsgnSkip}, \rn{eSeq} and \rn{eRewrite}.
The $\forall\exists$ versions of \rn{rLRseq} and \rn{rRLseq} 
are both derivable, but \rn{eRLseq} is less useful
than \rn{eLRseq} because the premises of \rn{eRLseq} amount to a strong $\exists\forall$ property.
In \autoref{sec:unaryRevisitedE}
we show that \rn{eLRseq} is the basis of a Cook complete logic whereas \rn{eRLseq} is not.

Rule \rn{eSkipDo} is derived as follows.  
Observe that $\skipc\kateq\dogc{\mathit{false}\gcto\skipc}$.
So we get the conclusion of \rn{eSkipDo} by \rn{eRewrite} from 
$\dogc{\mathit{false}\gcto\skipc}
\Sep \dogc{gcs} : \aespec{\Q}{\Q\land\neg\rightF{\enab(gcs)}}$.
This we prove using \rn{eDo} with $\Lrel,\R:=\mathit{false},\mathit{true}$.
The right-only premises have the form 
$\skipc\sep c : \aespec{\Q\land \rightF{e}\land\mathit{true}\land (V=d) }{\Q\land (V\prec d)}$.
They follow from the premises of \rn{eSkipDo} using \rn{eConseq} to 
add the conjunct $\mathit{true}$.  
The left-only and joint premises are proved by \rn{eFalse}.
The side condition of this instantiation of \rn{eDo} is 
$\Q\imp(\leftex{\mathit{false}}=\rightex{\enab(gcs)})\lor\rightF{\enab(gcs)}$ and the consequent simplifies to true.

\begin{exa}\label{eg:c3ded}


Recall the example adapted from Unno et  al.~\cite{UnnoTerauchiKoskinen21}, described on page~\pageref{pg:c3} in \autoref{sec:overview}.  The judgment $c3\sep c3 : \aespec{\Agr\mathit{low}}{\Agr x}$ specifies possibilistic noninterference.  We construct a deductive proof in ERHL+ following heuristics mentioned in \autoref{sec:overview}.

The proof is presented in a goal directed style.  
It starts by an application of \rn{eIf}, yielding four obligations, corresponding to the four combinations of the guards $\mathit{high}\neq 0$ and $\mathit{high}=0$.
  \begin{enumerate}
  \item $\lhavc{2}{x}; \lifgc{3}{\ldots} \sep \lhavc{2}{x};
    \lifgc{3}{\ldots} :
    \aespec{\Agr\mathit{low} \land \bothF{\mathit{high}\neq 0\sep
        \mathit{high}\neq 0}}{\Agr x}$.

    \noindent
    We prove this using a lockstep alignment. Rule \rn{eSeq} is
    instantiated with $\Agr\mathit{low}\land\Agr x$ at the intermediate point.  The judgment
    $\lhavc{2}{x}\sep\lhavc{2}{x} :
    \aespec{\Agr\mathit{low}\land\bothF{\mathit{high}\neq
        0\sep\mathit{high}\neq 0}}{\Agr\mathit{low}\land\Agr x}$ is obtained 
using \rn{eHavHav} and then \rn{eConseq} with the valid implication 
$\Agr\mathit{low}\imp(\allSet{x\smSep}{(\someSet{\smSep x}{\Agr\mathit{low}\land\Agr x})})$.
The postcondition $\Agr x$ is chosen to serve as  filtering condition.
The judgment 
    $\lifgc{3}{\ldots}\sep\lifgc{3}{\ldots} : \aespec{\Agr\mathit{low}\land\Agr x}{\Agr x}$
is proved using the lockstep rules $\rn{eAlgnIf}$
and $\rn{eAlgnDo}$ with loop invariant $\Agr x$.
    
  \item $\begin{array}{@{}l}
   \lassg{7}{x}{\mathit{low}};\lhavc{8}{b};\ldogc{9}{b\neq0\gcto\lassg{11}{x}{x+1}\ldots}\ \sep \\
   \lassg{7}{x}{\mathit{low}};\lhavc{8}{b};\ldogc{9}{b\neq0\gcto\lassg{11}{x}{x+1}\ldots}
  \end{array}
  : \aespec{\Agr\mathit{low} \land \bothF{\mathit{high}=0\sep\mathit{high}=0}}{\Agr x}$

  \noindent
  This judgment, too, is proved using a lockstep alignment.  Key to
  the proof is having $\Agr x \land \Agr b$ as the relational
  invariant for the two loops at control point 9.  In this derivation
  \rn{eHavHav} is applied twice with a post-relation that includes $\Agr b$ (again, for filtering).
  
  \item $\lhavc{2}{x}; \lifgc{3}{\ldots}\sep \lassg{7}{x}{\mathit{low}};\ldots :
    \aespec{\Agr\mathit{low} \land \bothF{\mathit{high}\neq 0\sep\mathit{high}=0}}{\Agr x}$.

    \noindent
    We prove this using the left-first sequential alignment rule
    \rn{eLRseq} with intermediate assertion
    $\Agr\mathit{low}\land\leftF{x\geq\mathit{low}}$, for which the
    premises are:
    \begin{enumerate}
    \item
      $\lhavc{2}{x}; \lifgc{3}{\ldots} \sep \skipc :
      \aespec{\Agr\mathit{low} \land
        \bothF{\mathit{high}\neq 0 \sep \mathit{high}=0}}
      {\Agr\mathit{low}\land\leftF{x\geq\mathit{low}}}$

      \noindent This judgment is derived  using the left-side rules \rn{eSeqSkip}, \rn{eHavSkip},
      \rn{eIfSkip}, and \rn{eDoSkip} 
with invariant $\Agr \mathit{low}\land\leftF{x<\mathit{low}}$.

    \item
      $\skipc \sep \lassg{7}{x}{\mathit{low}}; \lhavc{8}{b}; \ldogc{10}{b\neq 0\gcto
        \ldots} :
      \aespec{\Agr\mathit{low}\land\leftF{x\geq\mathit{low}}}{\Agr x}$

      \noindent This is proved using \rn{eSkipSeq} twice, to compose
      the following three judgments.
      First in the sequence is $\skipc\sep\lassg{7}{x}{\mathit{low}} :     \aespec{\Agr\mathit{low}\land\leftF{x\geq\mathit{low}}}{\Agr\mathit{low}\land\bothF{x\geq\mathit{low}\sep x=\mathit{low}}}$, proved using \rn{eSkipAssg} and \rn{eConseq}.  Second is $\skipc\sep\lhavc{8}{b}:
\aespec{\Agr\mathit{low}\land\bothF{x\geq\mathit{low}\sep x=\mathit{low}}}{\rightF{b\geq0}\land(\rightex{b}=\leftex{x}-\rightex{x})}$, proved by \rn{eSkipHav} and \rn{eConseq}, noting that the precondition implies $(\someSet{\smSep b}{\rightF{b\geq0}\land(\rightex{b}=\leftex{x}-\rightex{x})})$.  Third is $\skipc\sep\ldogc{9}{b\neq 0 \gcto \lassg{10}{x}{x+1}; \lhavc{11}{b}} : \aespec{\rightF{b\geq0}\land (\rightF{b}=\leftF{x}-\rightF{x})}{\Agr x}$, proved using \rn{eSkipDo} with invariant $\Q := \rightF{b\geq0}\land(\rightF{b}=\leftF{x}-\rightF{x})$ and variant $V := abs(\rightF{b})$.  The loop body is proved using \rn{eSkipSeq}, \rn{eSkipAssg}, and \rn{eSkipHav}; \rn{eSkipHav} is instantiated with post-relation  $\rightF{b}=\leftF{x}-\rightF{x}$.
  \end{enumerate}
    
  \item $\lassg{7}{x}{\mathit{low}}; \lhavc{8}{b}; \ldogc{9}{\ldots}
     \sep \lhavc{2}{x};\lifgc{3}{\ldots} :
     \aespec{\Agr\mathit{low} \land \bothF{\mathit{high}=0\sep\mathit{high}\neq 0}}{\Agr x}$.

     \noindent We prove this by left-first sequential alignment again, instantiating \rn{eLRseq} with intermediate assertion $\Agr\mathit{low} \land \leftF{x\geq\mathit{low}}$.  
  \end{enumerate}
\end{exa}

\begin{thm} 
\upshape
All the rules of ERHL+ (\autoref{fig:ERHL}) are sound.
\end{thm}
\begin{proof}
Soundness proofs of most ERHL+ rules are straightforward.  Soundness of
\rn{eRewrite} holds for the same reason \rn{Rewrite} and \rn{rRewrite} are
sound: the property (\ref{eq:aespec}) is about program semantics which is preserved by $\kateq$.
The soundness proof for \rn{eGhost} relies on the fact that the $\ghost$
condition ensures erasure of a ghost variable does not influence
termination.

For \rn{eIf}, assume we have the premises:
$ c\sep c' : \rspec{\R\land \leftF{e}\land\rightF{e'}}{\S} $
for all $e\gcto c$ in $gcs$ and $e'\gcto c'$ in $gcs'$.
To show the conclusion $\ifgc{gcs} \sep \ifgc{gcs'} : \rspec{\R}{\S}$,
consider states $s,s',t$ such that
$(s,s') \in \R$ and $\means{\ifgc{gcs}}\, s\, t$.  We must show there
is $t'$ such that $\means{\ifgc{gcs'}}\, s'\, t'$ and
$(t,t') \in \S$.
By semantics there is $e\gcto c$ in $gcs$ such that
$s\in\means{e}$ and $\means{c}\,s\,t$.
By the $\totalIf$ condition for $\ifgc{gcs'}$,
there is some $e'\gcto c'$ in $gcs'$ such that $s'\in\means{e'}$.
By the premise 
$ c\sep c' : \rspec{\R\land \leftF{e}\land\rightF{e'}}{\S} $
there is some $t'$ with $\means{c'}\,s'\,t'$ and $(t,t')\in\S$.

For \rn{eDo} we sketch the argument for loops that have a single guarded command;
a detailed proof is given in \cite{BNN23v5}.
The goal is to show 
$\models \dogc{e\gcto c} \Sep \dogc{e'\gcto c'} : \aespec{\Q}{\Q\land \neg\leftF{e}\land\neg\rightF{e'}}$
given the side condition
$\Q\imp (\leftex{e} = \rightex{e'}  
           \lor (\Lrel \land \leftF{e})
           \lor (\R \land \rightF{e'}))$
and premises
$\models c\sep \skipc : \aespec{\Q\land \leftF{e}\land\Lrel }{\Q}$,
$\models\skipc\sep c' : \aespec{\Q\land \rightF{e'}\land\R\land (V=v) }{\Q\land (V\prec v)}$
(for all $v$), and 
$\models c\sep c' : \aespec{\Q\land \leftF{e}\land\rightF{e'} \land \neg\Lrel \land \neg\R}{\Q}$.
Consider states $s,s',t$ such that
$(s,s') \in \Q$ and $\means{\dogc{e\gcto c}}\, s\, t$.  We must show there
exists a $t'$ such that $\means{\dogc{e'\gcto c'}}\, s'\, t'$ and
$(t,t') \in \Q \land\neg\leftF{e}\land\neg\rightF{e'}$.  We proceed by rule induction
on $\means{\dogc{e\gcto c}}\, s\, t$, keeping $s'$ arbitrary.
In the base case, the run is already terminated: $s \in \neg\leftF{e}$ and $s = t$.  
Existence of the required $t'$ is proved by well-founded induction on $V(s,s')$, as follows. 
If $s' \in \neg\rightF{e'}$, we are done by letting $t' := s'$. 
Otherwise, using $(s,s')\in\Q$, $s\in\neg\leftF{e}$, and the side condition we get the precondition of the right-side premise for $c'$.  Applying that premise yields some $t''$ with $(s,t'')\in\Q$ 
and $V(s,t'')\prec V(s,s')$
and then the inner induction hypothesis yields the required $t'$. 

In the inductive case, we have $s \in \leftF{e}$, $\means{c}\, s\, u$, and
$\means{\dogc{e\gcto c}}\, u\, t$ for some $u$.  The inductive hypothesis says
that for any $u'$, if $(u,u') \in \Q$, then there is a $t'$ such that 
$\means{\dogc{e'\gcto c'}}u't'$ and 
$(t,t')\in\Q\land\neg\leftF{e}\land\neg\rightF{e'}$.
If $(s,s')$ satisfy the precondition for the left-only or joint premise then applying the premise
yields $t'$ (reached from $s'$ by zero or one iterations of $c'$) such that $(u,t')\in\Q$ whence the inductive hypothesis yields our goal.  Otherwise, by $(s,s')\in\Q$ and the side condition, $(s,s')$ satisfies the precondition for the right-only premise.  But that does not immediately enable use of the inductive hypothesis.  So we show by well founded induction on $V(s,s')$ that there exists some $t''$,
reached by some number of iterations of $c'$, with $(s,t'')\in\Q$ 
and 
$(s,t'')$ satisfies either the left-only or joint premise.
Either of those premises, together with the main induction hypothesis and reachability of $t''$,
yields the goal.
\end{proof}

\subsection{Digression on control determinacy}\label{sec:controldet}

Our GCL has two forms of nondeterminacy.  The language has havoc, which makes an unboundedly nondeterministic choice of a value.  This can serve to model randomization as well as input data.  
It also has nondeterminacy in terms of control. 
This does not necessarily lead to nondeterministic outcomes.
For example, the command 
\( 
\lifgc{1}{true\gcto \lassg{2}{x}{y} \gcsep 
           true\gcto \lassg{3}{x}{y}}
\) 
satisfies 
$\rspec{\leftex{y}=\rightex{y}}{\leftex{x}=\rightex{x}}$
but its control flows nondeterministically to either the point labelled 2 or the point labelled 3.
On the other hand, $d3'$ below has nondeterministic outcomes.
\[ 
    d3\eqdef \ifgc{true\gcto x:=0} \quad\qquad
    d3'\eqdef \ifgc{true\gcto x:=0 \gcsep true\gcto x:=1}
\]

A command $c$ is called \dt{control deterministic} provided that 
the guards are mutually exclusive,
for every if- and do-command in $c$. 
A control deterministic $\ifgc{e_0\gcto c_0\gcsep e_1\gcto c_1\gcsep\ldots}$
is equivalent to the nested if-else $\ifc{e_0}{c_0}{\ifc{e_1}{c_1}{\ldots}}$.
A control deterministic do-command can be represented similarly.

For a command $\ifgc{e_0\gcto c_0\gcsep e_1\gcto c_1\ldots}$ that is not control deterministic, one can make it so in the form
$\ifgc{e_0\gcto c_0\gcsep e_1\land\neg e_0\gcto c_1\ldots}$
but this may eliminate some behaviors, as is the case for $d3'$ above.
To retain all behaviors one can use havoc with an extra variable
that serves to prophesize the choice, as in 
\[ 
 d3'' \eqdef 
   \havc{z};\ifgc{true\land z=0\gcto x:=0 \gcsep true\land z\neq 0\gcto x:=1} 
\]
where $z$ is fresh.
Conventional if/else and while commands are control deterministic, 
and the normal form construction of \autoref{def:autnf} preserves control determinacy.
So there is little reason to dwell on programs that are not control deterministic.  Nonetheless we briefly consider the following limitation of ERHL+ for such programs. 

This $\forall\exists$ judgment is valid:
$d3\sep d3':\aespec{true}{\Agr x}$.   
But rule \rn{eIf} is not directly applicable,
because  one of the premises would be $x:=0\sep x:=1 : \aespec{true\land true\land true}{\Agr x}$
which is false.
One might guess to change the rule simply by existentially quantifying the right side guarded commands but this is unsound.
For example, consider
\[ 
    d4\eqdef \ifgc{true\gcto x:=0} \quad\qquad
    d4'\eqdef \ifgc{x\geq 0\gcto x:=0 \gcsep x\leq 0\gcto x:=1} .
\]
For every guarded command on the left, there is one on the right that relates according to the 
spec $\aespec{true\land\rightF{e}}{\Agr x}$ where $e$ is the guard.
In particular:
  $x:=0\sep x:=0: \aespec{true\land\rightF{x\geq 0}}{\Agr x}$.
But it is not the case that $d4\sep d4'$ satisfies 
$\aespec{true}{\Agr x}$.
What is needed is to existentially quantify a subset of guarded commands on the right that covers all cases. 
\[
\inferrule*[right=eIfX]{
\parbox{.85\textwidth}{
  For all $e\gcto c$ in $gcs$ there is some $g\subseteq gcs'$ such that
    $\R\land\leftF{e}\imp\enab(g)$ and \\
    \hspace*{4em} $c\sep c' : \aespec{\R\land \leftF{e}\land\rightF{e'}}{\S}$ 
     for all $e'\gcto c'$ in $g$
}
}{ \ifgc{gcs} \Sep \ifgc{gcs'} : \aespec{\R}{\S}  }
\]
The side condition $\R\land\leftF{e}\imp\enab(g)$ ensures the set $g$ covers all cases.
Note that there is no such set that can be used 
to show the invalid judgment $d4\sep d4': \aespec{true}{\Agr x}$.
Rule \rn{eIfX} does yield $d3\sep d3':\aespec{true}{\Agr x}$.   
A similar issue arises for the joint premises in rule \rn{eDo},
and one can address it in the same way as \rn{eIfX}.
The rule \rn{eDoX} replaces the joint premise of \rn{eDo} with the following:
\[ \begin{array}[t]{l}
     \mbox{For all $e\gcto c$ in $gcs$ there is some $g\subseteq gcs'$ such that
                   $\Q\land\leftF{e}\land\neg\Lrel\land\neg\R\imp\enab(g)$}\\ 
     \mbox{\quad and for all $e'\gcto c'$ in $g$ we have
        $c\sep c' : \aespec{\Q\land \leftF{e}\land\rightF{e'}\land\neg\Lrel\land\neg\R}{\Q}$}
\end{array}
\]

We choose the simpler rules \rn{rIf} and \rn{rDo} to streamline the presentation, at the cost of restricting to  control deterministic programs when necessary,
specifically \autoref{thm:aeAlignComp}(a).
We conjecture the restriction can be dropped, using rules \rn{eIfX} and \rn{eDoX}. 
There is little practical motivation because conventional control structures (if/else and while) are control deterministic.

\section{Filtered alignment automata and alignment completeness of ERHL+}\label{sec:eacomplete}

This section introduces a form of alignment automaton suited to $\forall\exists$ properties.  The logic ERHL+ is shown, in 
\autoref{sec:aecompleteSubsect}, to be alignment complete with respect to these automata.

\subsection{Filtered alignment automata}\label{sec:filt}

For $\forall\exists$ reasoning, the form of alignment given by \autoref{def:alignProd} is unsatisfactory: if the right program is nondeterministic, there needs to be a way to keep some but not all its transitions for right-only and joint steps of the product.
An example is $c3$ in \autoref{sec:overview}, which we explore further in the sequel.
First we adapt \autoref{def:alignProd} of alignment
product to include an additional state relation which serves to filter product executions.

\begin{defi}\label{def:aeAlignProd}
  Suppose $\aprod(A,A',L,R,J)$ is an alignment automata as in
  \autoref{def:alignProd}.  Let \dt{keep set} $K$ be a set of states, 
  i.e., $K\subseteq (Ctrl \times Ctrl') \times (Sto \times Sto')$.  The \dt{filtered alignment
    automaton} \graybox{$\aprod(A,A',L,R,J,K)$} is 
$((Ctrl\times Ctrl'), (Sto\times Sto'), (\init,\init'), (\fin,\fin'),
  \biTrans_K) $
where \graybox{$\biTrans_K$} is defined by:
  $\sigma\biTrans_K \tau$ iff $\tau \in K$ and
  $\sigma \biTrans \tau$ (with $\sigma,\tau$ ranging over states).
\end{defi}
Here the un-subscripted $\biTrans$ refers to the relation in \autoref{def:alignProd}.
Note that $K$ is used in $\biTrans_K$ to filter states that the product steps
to.  As a consequence, $[\init\mid\init']\lor K$ is always a true-invariant of $\biTrans_K$.
Apart from this use of $K$, the transition relation is the same as 
the relation $\biTrans$ for unfiltered alignment automata (\autoref{def:alignProd}).

\begin{exa} 
Fig.~\ref{fig:filteredProductEx} shows a filtered product for $c3$, the running example on $\forall\exists$ properties adapted from~\cite{UnnoTerauchiKoskinen21}---
see  page~\pageref{pg:c3} in \autoref{sec:overview}, and \autoref{fig:c3aut}.  
Fig.~\ref{fig:filteredProductEx} depicts possible transitions of the product, given $L,R,J$ conditions that capture the following alignment: lockstep if the two executions agree on the test $\mathit{high}\neq0$ (abbreviated as $h$ in Fig.~\ref{fig:filteredProductEx}); left-first sequential otherwise. This is similar to the deductive proof in \autoref{eg:c3ded}.
Automata edges are labeled with tests or commands.  We write $\bothF{c}$, $\leftF{c}$, and $\rightF{c}$ to mean that $c$ takes place on both sides, on the left, and on the right, respectively.  The label $f$ abbreviates the final control point 12.  Filter $K$ is $\mathit{true}$ everywhere except at control points marked in the figure with green dashed boxes.  For example, the step from $(2,2)$ to $(3,3)$ which havocs $x$ on both sides is filtered by $\Agr x$.  The step to control point $(f,9)$ is filtered so that the value of $\rightF{b}$ is the same as the difference between the values of $x$ on both sides.

The figure does not depict transitions that cannot take place.  In particular, vertices for control points $(f,5)$ and $(f,6)$ are missing.  These vertices correspond to the diverging loop in the example taking place on the right.  The filter $\Agr x$ on $(f,3)$, in conjunction with invariant
$[f|3]\imp\leftF{x\geq\mathit{low}}$ 
ensures that the transition from $(f,4)$ to $(f,5)$ which would be guarded by $\rightF{x<\mathit{low}}$ cannot occur.
\qed
\end{exa}

\newlength\branchdist%
\setlength{\branchdist}{24em}
\begingroup
\newcommand{\vhigh}{\ensuremath{\mathit{high}}}
\newcommand{\vlow}{\ensuremath{\mathit{low}}}

\begin{figure}[t]
\begin{tikzpicture}
  [-{Latex[length=2mm]},every state/.style={fill=gray!10},initial text=$ $,auto,
  node distance=2.5cm,
  line width=0.1mm,
  scale=0.7,
  transform shape]

  \node[state] (n1) {$1,1$};
  \node[state, right of=n1, xshift=6em, yshift=.7\branchdist] (n2) {$2,2$};
  \node[state, right of=n1, xshift=6em, yshift=-.7\branchdist] (n7) {$7,7$};

  \node[state, right of=n2] (n3) {$3,3$};
  \node[state, right of=n3, xshift=2em, yshift=2.5em] (n4) {$4,4$};
  \node[state, right of=n3, xshift=2em, yshift=-2.5em] (n5) {$5,5$};
  \node[state, below of=n5, yshift=2em] (n6) {$6,6$};

  \node[state, right of=n7, xshift=2em] (n8) {$8,8$};
  \node[state, right of=n8, xshift=2em] (n9) {$9,9$};
  \node[state, below of=n9, yshift=1em, xshift=-4em] (n10) {\footnotesize$10,10$};
  \node[state, below of=n9, yshift=1em, xshift=4em] (n11) {\footnotesize$11,11$};

  \node[state, right of=n1,xshift=40em] (fin) {\footnotesize$f,f$};

  \draw
  (n1) edge[bend left] node[above, xshift=-1em] {$\bothF{h}$} (n2)
  (n1) edge[bend right=35] node[below, xshift=-1em] {$\bothF{\neg h}$} (n7)
  (n2) edge node[above] {$\bothF{\havc{x}}$} (n3)

  (n3) edge[bend left] node[above,xshift=-.5em] {$\bothF{x\geq\vlow}$} (n4)
  (n3) edge[bend right] node[below,xshift=-.5em] {$\bothF{x<\vlow}$} (n5)

  (n4) edge[bend left=45] node[above,xshift=-2.5em,yshift=1em] {$\bothF{\skipc}$} (fin)

  (n5) edge[bend right] node[left] {$\bothF{\mathit{true}}$} (n6)
  (n6) edge[bend right] node[right] {$\bothF{\skipc}$} (n5)
  (n5) edge[bend left] node[above,yshift=2em,xshift=-2em] {$\bothF{\neg\mathit{true}}$} (fin)

  (n7) edge node[above] {$\bothF{x:=\vlow}$} (n8)
  (n8) edge node[above] {$\bothF{\havc{b}}$} (n9)

  (n9) edge[bend right] node[left,yshift=-.5em] {$\bothF{b\neq0}$} (n10)
  (n10) edge[bend right] node[above,yshift=.5em] {$\bothF{x:=x+1}$} (n11)
  (n11) edge[bend right] node[left] {$\bothF{\havc{b}}$} (n9)

  (n9) edge[bend right=40] node[below,yshift=-2.25em,xshift=-1em] {$\bothF{b=0}$} (fin)
  ;  

  \node[state, right of=n1, yshift=.25\branchdist] (n2_7) {$2,7$};
  \node[state, right of=n2_7,xshift=-1em] (n3_7) {$3,7$};
  \node[state, right of=n3_7,xshift=-1em,yshift=2.5em] (n4_7) {$4,7$};
  \node[state, right of=n3_7,xshift=-1em,yshift=-2.5em] (n5_7) {$5,7$};
  \node[state, below of=n5_7,yshift=2em] (n6_7) {$6,7$};
  
  \node[state, right of=n5_7,xshift=-1em,yshift=2.5em] (nfin_7) {$f,7$};

  \node[state, right of=nfin_7] (nfin_8) {$f,8$};
  \node[state, right of=nfin_8] (nfin_9) {$f,9$};
  \node[state, below of=nfin_9,xshift=-4em,yshift=1em] (nfin_10) {\footnotesize$f,10$};
  \node[state, below of=nfin_9,xshift=4em,yshift=1em] (nfin_11) {\footnotesize$f,11$};

  \node[state, right of=n1, yshift=-.25\branchdist] (n7_2) {$7,2$};
  \node[state, right of=n7_2] (n8_2) {$8,2$};
  \node[state, right of=n8_2] (n9_2) {$9,2$};
  \node[state, below of=n9_2, xshift=-4em,yshift=1em] (n10_2) {\footnotesize$10,2$};
  \node[state, below of=n9_2, xshift=4em,yshift=1em] (n11_2) {\footnotesize$11,2$};
  \node[state, right of=n9_2] (nfin_2) {$f,2$};
  \node[state, right of=nfin_2] (nfin_3) {$f,3$};
  \node[state, right of=nfin_3] (nfin_4) {$f,4$};

  \draw
  (n1) edge[bend left] node[below,yshift=-.2em,xshift=1.2em] {$\bothF{h\sep \neg h}$} (n2_7)
  (n2_7) edge node[above,yshift=.5em] {$\leftF{\havc{x}}$} (n3_7)
  (n3_7) edge[bend left] node[above,xshift=-.5em,yshift=.5em] {$\leftF{x\geq\vlow}$} (n4_7)
  (n3_7) edge[bend right] node[below,xshift=-1.25em] {$\leftF{x<\vlow}$} (n5_7)
  (n5_7) edge[bend right] node[left] {$\leftF{\mathit{true}}$} (n6_7)
  (n6_7) edge[bend right] node[right] {$\leftF{\skipc}$} (n5_7)
  (n5_7) edge[bend right] node[below] {$\leftF{\neg\mathit{true}}$} (nfin_7)
  (n4_7) edge[bend left] node[above] {$\leftF{\skipc}$} (nfin_7)
  (nfin_7) edge node[above,yshift=.5em] {$\rightF{x:=\vlow}$} (nfin_8)
  (nfin_8) edge node[above,yshift=.5em] {$\rightF{\havc{b}}$} (nfin_9)
  (nfin_9) edge[bend right] node[left,yshift=-.25em] {$\rightF{b\neq0}$} (nfin_10)
  (nfin_10) edge[bend right] node[above,yshift=.5em] {$\rightF{x:=x+1}$} (nfin_11)
  (nfin_11) edge[bend right] node[left] {$\rightF{\havc{b}}$} (nfin_9)
  (nfin_9) edge[bend left] node[left,yshift=1.5em] {$\rightF{b=0}$} (fin)
  ;

  \draw
  (n1) edge[bend right] node[above,yshift=.75em,xshift=1em] {$\bothF{\neg h\sep h}$} (n7_2)
  (n7_2) edge node[above,yshift=.5em] {$\leftF{x:=\vlow}$} (n8_2)
  (n8_2) edge node[above,yshift=.5em] {$\leftF{\havc{b}}$} (n9_2)
  (n9_2) edge[bend right] node[left,yshift=-.25em] {$\leftF{b\neq0}$} (n10_2)
  (n10_2) edge[bend right] node[above,yshift=.5em] {$\leftF{x:=x+1}$} (n11_2)
  (n11_2) edge[bend right] node[left] {$\leftF{\havc{b}}$} (n9_2)
  (n9_2) edge node[above,yshift=.5em] {$\leftF{b=0}$} (nfin_2)
  (nfin_2) edge node[above,yshift=.5em] {$\rightF{\havc{x}}$} (nfin_3)
  (nfin_3) edge node[above,yshift=.5em] {$\rightF{x\geq\vlow}$} (nfin_4)
  (nfin_4) edge[bend right] node[left] {$\rightF{\skipc}$} (fin)
  ;

  \node[draw,rectangle,thick,dashed,fill=green!10,below of=n3,yshift=3em,xshift=-2em] (k1) {\Large$\Agr x$};
  \draw[black!60!green,thick,decorate,decoration={snake,amplitude=.5mm,segment length=4mm,post length=2mm}] (k1) -- ([yshift=-2mm]n3.center);

  \node[draw,rectangle,thick,dashed,fill=green!10,above of=n9,yshift=-3em,xshift=2em] (k2) {\Large$\Agr b$};
  \draw[black!60!green,thick,decorate,decoration={snake,amplitude=.5mm,segment length=4mm,post length=2mm}] (k2) -- ([yshift=2mm]n9.center);

  \node[draw,rectangle,thick,dashed,fill=green!10,above of=nfin_9,yshift=-2em,xshift=4em] (k3) {\Large$\rightF{b}=\leftF{x}-\rightF{x}$};
  \draw[black!60!green,thick,decorate,decoration={snake,amplitude=.5mm,segment length=4mm,post length=2mm}] (k3.south) -- ([yshift=2mm,xshift=2mm]nfin_9.center);

  \node[draw,rectangle,thick,dashed,fill=green!10,below of=nfin_3,yshift=3em,xshift=3em] (k4) {\Large$\Agr x$};
  \draw[black!60!green,thick,decorate,decoration={snake,amplitude=.5mm,segment length=4mm,post length=2mm}] (k4) -- ([yshift=-2mm,xshift=2mm]nfin_3.center);  

\end{tikzpicture}
\caption{Filtered product for example $c3$.
Non-trivial filter conditions are shown in green dashed boxes; the filter is $\mathit{true}$ at all other control points.
Notation \graybox{$\bothF{e}$} abbreviates $\bothF{e\sep e}$,
and we use similar notation for commands.
}
\label{fig:filteredProductEx}
\end{figure}
\endgroup

\paragraph{Adequacy of filtered alignment automata}

\begin{defi}  
  Let $\aprod(A,A',L,R,J,K)$ be a filtered alignment automata and let
  $\Q \subseteq (Sto \times Sto')$.  The filtered automata is \dt{$\Q$-adequate in
  the $\forall\exists$ sense} provided for all $(s,s') \in \Q$ and $t$ with
  $(\init,s) \trans^* (\fin,t)$, there exists a $t'$ such that
  $((\init,\init'),(s,s')) \biTrans^*_K ((\fin,\fin'),(t,t')).$
\end{defi}


\begin{lem}\label{lem:aeAdequate}
\upshape
Given automata $A$,$A'$ and store relations $\Q$, $\S$,
if $\aprod(A,A',L,R,J,K)$ is $\Q$-adequate in the $\forall\exists$ sense and
$\aprod(A,A',L,R,J,K)\models\spec{\Q}{\S}$, then 
$A,A'\models\aespec{\Q}{\S}$.
\end{lem}
\begin{proof}
To prove $A,A'\models\aespec{\Q}{\S}$, consider any $s,s',t$ such that 
$(\init,s)\trans^*(\fin,t)$ and $(s,s') \in \Q$.
By $\Q$-adequacy there is a state $t'$ such that
$((\init,\init'),(s,s'))\biTrans_K^*((\fin,\fin'),(t,t'))$.  By taking the right
projection of this trace, and destuttering, we obtain a trace
$(\init',s')\trans'^*(\fin',t')$ of $A'$.  
(Stuttering steps can arise from left-only steps of the product.)
Now from $\aprod(A,A',L,R,J,K)\models\spec{\Q}{\S}$ we
have $(t,t')\in\S$. 
\end{proof}

The lemma suggests an approach for verifying $\forall\exists$
properties: to show $A,A'\models\aespec{\Q}{\S}$,
construct a filtered product, prove it is adequate and
satisfies the partial correctness spec
$\spec{\Q}{\S}$.
The latter amounts to proving a $\forall\forall$ property of the set of trace pairs represented by the product.
As in the $\forall\forall$ setting, a proof method needs to connect adequacy with annotations and alignment conditions,
in a way that can be checked modularly.
To this end, we state the key definition and then explain its elements.

\begin{defi}\label{def:adequateFiltering}
  Given $\aprod(A,A',L,R,J,K)$ with annotation $an$, 
  we say the automaton has \dt{adequate filtering for $an$} iff there exists a function
  $V : (Ctrl\times Ctrl') \times (Sto\times Sto') \to D$ where $(D,\prec)$ is
  a well-ordered set, such that the following four conditions hold.
  (Where $\trans,\trans'$ are the transition relations of $A,A'$.)
\begin{samepage}
\begin{description}
  \item[\emph{Left-permissive}] 
    For any $n,n',s,s',m,t$,
    if $(s,s')\in an(n,n')$ and $((n,n'),(s,s'))\in L$ and $(n,s)\trans(m,t)$ then $((m,n'),(t,s'))\in K$.
  \item[\emph{Joint-productive}] 
For any $n,n',s,s',m,t$, if $(s,s')\in an(n,n')$ and 
      $((n,n'),(s,s')) \in J$ and $(n,s)\!\trans\!(m,t)$ then there are $m',t'$ such that
      $(n',s')\!\trans'\!(m',t')$ and $((m,m'),(t,t')) \in K$.
  \item[\emph{Right-productive}] 
\begin{sloppypar}
For any $n,n',s,s'$, if
   $(s,s')\in an(n,n')$ and $((n,n'),(s,s'))\in R$ then there are $m',t'$ such that $(n',s')\!\trans'\!(m',t')$
      and $((n,m'),(s,t')) \in K$ and $V((n,m'),(s,t')) \prec V((n,n'),(s,s'))$.
\end{sloppypar}
\item[\emph{Enabled}] For any $n,n'$, the implication $\breve{an}(n,n')\imp L \lor R \lor J \lor [\fin|\fin']$ is valid.
\end{description}
  \end{samepage}
\end{defi}
We sometimes say ``$K$ is an \dt{adequate filtering for $an$}'' to emphasize the key role of $K$.
We briefly explain why these conditions ensure adequacy in the $\forall\exists$ sense, if the annotation $an$ is valid.
First, the \emph{enabled} condition ensures the underlying $L,R,J$-product 
can always perform left-only, right-only, or joint steps; this is the same as
in \autoref{sec:acomplete} for $\forall\forall$.
But transitions by $\biTrans_K$ are filtered by the keep set $K$,
so we need to ensure that it keeps enough.
Transitions of $A$ can only be covered by the alignment product using LO or JO steps.  
The \emph{left-permissive} condition applies to states $((n,n'),(s,s'))$ 
where the product is poised to perform an LO step, and requires that it be allowed by $K$.  
The \emph{joint-productive} condition applies if the product is ready to take a JO step.  It requires that for any $A$ transition, there is some $A'$ transition such that the joint step is allowed by $K$.  
Finally, the \emph{right-productive} condition ensures that
for RO steps, $K$ is not just keeping divergent traces of $A'$. 
When the product is poised to take a RO step, $K$ must allow some transition
that decreases the value of the variant $V$.

The following results parallel \autoref{lem:manifest}, \autoref{cor:relIAM}, and
\autoref{cor:relIAMcomplete}.

\begin{lem}\label{lem:FiltAdequate}
\upshape 
Let $an$ be a valid annotation of 
$\aprod(A,A',L,R,J,K)$ for $\spec{\P}{\Q}$
and suppose $K$ is an adequate filtering for $an$.
Then $\aprod(A,A',L,R,J,K)$ is $\P$-adequate in the $\forall\exists$ sense.
\end{lem}
\begin{proof}
First we make a simple observation that pertains to any automaton $A$
and annotation $an$ for some spec $\spec{P}{Q}$.
If $an$ is valid, $s$ is in $P$, and $(\init,s)\trans^*(m,t)$,
then $t\in an(m)$.  
Thus for the product $\aprod(A,A',L,R,J,K)$, the condition $an(n,n')$ holds whenever the product is at control point $(n,n')$. 
Hence $L\lor R\lor J\lor[\fin|\fin']$ is an invariant of the product,
owing to the enabled condition in \autoref{def:adequateFiltering} of adequate filtering.

To show that $\aprod(A,A',L,R,J,K)$ is $\P$-adequate in the $\forall\exists$ sense,
suppose $(s,s') \in \P$ and suppose $(\init,s) \trans^* (\fin,t)$, i.e., there is a terminated trace
$\tau$ of $A$ from $s$ to $t$.
We must show that the product can simulate $\tau$, without getting stuck, until it reaches $(\fin,\fin')$.
Because $L\lor R\lor J\lor[\fin|\fin']$ is invariant,
and using the liveness of $L$, $R$, and $J$ as required by \autoref{def:alignProd}
(via \autoref{def:aeAlignProd}),
the product can keep taking steps until it reaches $(\fin,\fin')$.
The LO and JO steps move forward simulating $\tau$, thus eventually matching all of $\tau$,
unless the product diverges taking only RO steps.  
Right-productivity ensures that only finitely many RO steps can happen before $L\lor J\lor[\fin|\fin']$ holds.

Making this precise requires a slightly intricate induction hypothesis.  A detailed proof can be found in \cite{BNN23v5}.
\end{proof}

\begin{restatable}[semantic soundness and completeness of filtered automata]{thm}{thmFiltSoundCompl} \label{thm:FiltSoundCompl}
\upshape
We have 
$A,A'\models\aespec{\P}{\Q}$ iff there are $L,R,J,K$ and a 
valid annotation $an$ of $\aprod(A,A',L,R,J,K)$ for $\spec{\P}{\Q}$ 
such that $K$ is an adequate filtering for $an$.
Moreover, if $A,A'$ act on variable stores,
and $A,A',\P,\Q$ are finitely supported, 
then there are finitely supported such $L,R,J,K,an$ (and witness $V$).
\end{restatable}
\begin{proof}[Proof (Sketch)]
The proof is by mutual implication.  
For right-implies-left the argument is easy.
If $an$ is a valid annotation of $\aprod(A,A',L,R,J,K)$ for $\aespec{\P}{\Q}$
and $K$ is an adequate filtering for $an$,
then by Lemma~\ref{lem:FiltAdequate} the product is $\P$-adequate.
Since $an$ is valid for $\spec{\P}{\Q}$ we have
$\aprod(A,A',L,R,J,K)\models\spec{\P}{\Q}$ by Lemma~\ref{prop:IAM}. 
So by Lemma~\ref{lem:aeAdequate} we have $A,A'\models\aespec{\P}{\Q}$.

For left-implies-right,
suppose $A,A'\models\aespec{\P}{\Q}$. 
The idea is to use a left-first sequential alignment automaton such that,
once the left-only execution has finished, with final store $t$,
the filter condition only keeps states that are in some terminating run of 
$A'$ that ends in a store $t'$ such that $(t,t')\in\Q$.
The variant is defined in terms of shortest terminating runs that end in $\Q$. 
The details are somewhat complicated and relegated to an appendix
for the interested reader (\autoref{sec:app:proofs}).
\end{proof}

\paragraph{Verification conditions for filtered alignment automata.}

Let $an$ be an annotation for the filtered alignment automaton
$\aprod(A,A',L,R,J,K)$ for $\spec{\S}{\T}$.  For each pair of control points
$(n,n')$, $(m,m')$ of the product, the corresponding VC is given by
definition (\ref{eq:VC}) instantiated by $\biTrans_K$.
Restricting attention to filtered alignment automata of programs, the VCs can be
expressed in a form similar to those for (unfiltered) alignment automata as
given in \autoref{fig:RVClo} and \autoref{fig:RVCjo}.  The key difference is
in how the keep set $K$ is handled.
To facilitate the definitions, we define for any control points $(n,n')$
the state set $\graybox{$K(n,n')$} \eqdef \{ ((i,i'),(s,s')) \mid ((n,n'),(s,s')) \in K \}$.  

\begin{figure}[t]
\begin{footnotesize}
\begin{tabular}{lll}
if $\sub(n',c')$ is\ldots \!\!\!     &
and $m'$ is\ldots &
then the VC for $((n,n'),(n,m'))$ is equivalent to $\ldots$
\\\hline
$\lskipc{n'}$ & $\fsuc(n',c',f')$ & $R \land \breve{an}(n,n') \land K(n,m') \imp \hat{an}(n,m')$
\\
$\lassg{n'}{x'}{e'}$ & $\fsuc(n',c',f')$
& $R \land \breve{an}(n,n') \land \subst{K(n,m')}{\smSep x'}{\smSep e'}
  \imp \subst{\hat{an}(n,m')}{\smSep x'}{\smSep e'}$
\\
$\lhavc{n'}{x'}$ & $\fsuc(n',c',f')$
& $\all{v'\in\Z}{R \land \breve{an}(n,n') \land \subst{K(n,m')}{|x'}{|v'} \imp \subst{\hat{an}(n,m')}{|x'}{|v'}}$
\\
$\lifgc{n'}{gcs'}$ & $\lab(d')$ where $e' \gcto d'$ in $gcs'$
& $R \land \breve{an}(n,n') \land \rightF{e'} \land K(n,m') \imp \hat{an}(n,m')$
\\
$\ldogc{n'}{gcs'}$ & $\lab(d')$ where $e' \gcto d'$ in $gcs'$
& $R \land \breve{an}(n,n') \land \rightF{e'} \land K(n,m') \imp \hat{an}(n,m')$
\\
$\ldogc{n'}{gcs'}$ & $\fsuc(n',c',f')$
& $R \land \breve{an}(n,n') \land \neg\rightF{\enab(gcs')} \land K(n,m') \imp \hat{an}(n,m')$
\\[1ex]
\multicolumn{3}{l}{
In all other cases, there are no transitions from $(n,n')$ to $(n,m')$ so the VC is $true$ by definition.}
\end{tabular}
\end{footnotesize}
\vspace*{-1ex}
\caption{The right-only VCs for annotation $an$ of
  $\aprod(\aut(c,f),\aut(c',f'),L,R,J,K)$.}
\label{fig:EVCro}
\end{figure}

\autoref{fig:EVCro} lists right-only VCs for the filtered product
$\aprod(\aut(c,f),\aut(c',f'),L,R,J,K)$.  
As an example, consider the VC for right-only
$\lhavc{n'}{x'}$.  It says that for any value of $x'$, if $R$ and
$\breve{an}(n,n')$ hold, and that value of $x'$ will be ``kept'' by $K$ when
control reaches $(n,m')$, then the annotation at $(n,m')$ holds for that value
of $x'$.  Left-only VCs are similar.
\autoref{fig:EVCjo} lists a selected set of joint VCs.

\setlength{\dashlinedash}{.5ex}
\setlength{\dashlinegap}{1ex}

\begin{figure}[t]
\begin{footnotesize}
\begin{tabular}{lll}
if\hspace*{-.5ex} $\begin{array}[t]{l} 
    \sub(n,c) \\
    \sub(n',c') 
    \end{array}$
    \hspace*{-2ex}
    \!\!\!are\ldots\hspace*{-2ex}     & 
and $\begin{array}[t]{l} m \\ m'\end{array}$ are\ldots &
then the VC for $((n,n'),(m,m'))$ is equivalent to$\ldots$ \\\hline

$\lassg{n}{x}{e}$ & $\fsuc(n,c,f)$ 
& $J \land \breve{an}(n,n')\land \rightF{e'} \land \subst{K(m,m')}{x|}{e|} \imp \subst{\hat{an}(m,m')}{x|}{e|}$ 
\\
$\lifgc{n'}{gcs'}$ & $\lab(d')$ for $e'\gcto d'$ in $gcs'$
\\\hdashline

$\lhavc{n}{x}$ & $\fsuc(n,c,f)$ & $\all{v\in\Z}{J \land \breve{an}(n,n') \land \subst{K(m,m')}{x|x'}{v|e'} \imp \subst{\hat{an}(m,m')}{x|x'}{v|e'}}$
\\
$\lassg{n'}{x'}{e'}$ & $\fsuc(n',c',f')$
\\\hdashline
$\lhavc{n}{x}$ & $\fsuc(n,c,f)$ & $\all{v\in\Z}{J \land \breve{an}(n,n') \land \rightF{e'} \land \subst{K(m,m')}{x|}{v|} \imp \subst{\hat{an}(m,m')}{x|}{v|}}$
\\
$\ldogc{n'}{gcs'}$ & $\lab(d')$ for $e'\gcto d'$ in $gcs'$
\\\hdashline
$\lhavc{n}{x}$ & $\fsuc(n,c,f)$ & $\all{v,v'\in\Z}{J \land \breve{an}(n,n') \land \subst{K(m,m')}{x|x'}{v|v'} \imp \subst{\hat{an}(m,m')}{x|x'}{v|v'}}$
\\
$\lhavc{n'}{x'}$ & $\fsuc(n',c',f')$
\\[.5ex]
\multicolumn{3}{l}{Omitted: the other 32 cases with nontrivial VCs.}
\end{tabular}
\end{footnotesize}
\vspace*{-1ex}
\caption{Selected joint VCs for annotation $an$ of $\aprod(\aut(c,f),\aut(c',f'),L,R,J,K)$.}
\label{fig:EVCjo}
\end{figure}

\begin{figure}[t]
\begin{footnotesize}
\begin{tabular}{lll}
if $\sub(n',c')$ is\ldots\!\!     & 
and $m'$ is\ldots &
then the encoded VC for $((n,n'),(n,m'))$ is $\ldots$ 
\\\hline
$\lskipc{n'}$ & $\fsuc(n',c',f')$ & $\encode{R}\land\bothF{\tpc n\sep\tpc n'} \land an(n,n') \land \proj{K}(n,m') \imp an(n,m')$ 
\\
$\lassg{n'}{x'}{e'}$ & $\fsuc(n',c',f')$ & $\encode{R}\land\bothF{\tpc n\sep\tpc n'}\land an(n,n')\land\subst{\proj{K}(n,m')}{\smSep x'}{\smSep e'}\imp \subst{an(n,m')}{\smSep x'}{\smSep e'}$ 
\\
$\lhavc{n'}{x'}$ & $\fsuc(n',c',f')$ & $\all{v'\in\Z}{\encode{R}\land\bothF{\tpc n\sep\tpc n'}\land an(n,n')
                                       \land \subst{\proj{K}(n,m')}{\smSep x'}{\smSep v'}
                                       \imp \subst{an(n,m')}{\smSep x'}{\smSep v'}}$ 
\hspace*{-1ex} 
\\
$\lifgc{n'}{gcs'}$ & $\lab(d')$ 
& $\encode{R}\land\bothF{\tpc n\sep\tpc n'}\land an(n,n')\land \rightF{e} \land \proj{K}(n,m') \imp an(n,m')$ 
\\ 
& for $e'\gcto d'$ in $gcs'$ \\
$\ldogc{n'}{gcs'}$ & $\lab(d')$ 
& $\encode{R}\land\bothF{\tpc n\sep\tpc n'}\land an(n,n')\land \rightF{e} \land \proj{K}(n,m') \imp an(n,m')$ 
\\ 
& for $e'\gcto d'$ in $gcs'$ \\
$\ldogc{n'}{     gcs'}$ & $\fsuc(n',c',f')$ 
& $\encode{R}\land\bothF{\tpc n\sep\tpc n'}\land an(n,n')\land \neg\rightF{\enab(gcs')} \land \proj{K}(n,m') \imp an(n,m')$ 
\hspace*{-2ex} 
\end{tabular}
\end{footnotesize}
\vspace*{-1ex}
\caption{The right-only $pc$-encoded VCs for $an$ and  $\aprod(\aut(c,f),\aut(c',f'),L,R,J,K)$.}
\label{fig:EVCro-encoded}
\end{figure}

\begin{figure}[t]
\begin{footnotesize}
\begin{tabular}{lll}
if $\begin{array}[t]{l} 
    \sub(n,c) \\
    \sub(n',c') 
    \end{array}$
    \hspace*{-2.5ex}are\ldots\!\!     & 
and\!$\begin{array}[t]{l} m \\ m' \end{array}$\!are\ldots\!\! &
then the encoded VC for $((n,n'),(m,m'))$ is $\ldots$ \\\hline

$\lskipc{n}$ &
$\fsuc(n,c,f)$ & $\encode{J}\land\bothF{\tpc n\sep\tpc n'} \land an(n,n') \land \proj{K}(m,m') \imp an(m,m')$ 
\\
$\lskipc{n'}$ &
$\fsuc(n',c',f')$ & 
\\\hdashline

$\lassg{n}{x}{e}$ &
$\fsuc(n,c,f)$ & $\encode{J}\land\bothF{\tpc n\sep\tpc n'} \land an(n,n') \land \subst{\proj{K}(m,m')}{x|x'}{e|e'} \imp \subst{an(m,m')}{x|x'}{e|e'}$ 
\\
$\lassg{n'}{x'}{e'}$ &
$\fsuc(n',c',f')$ & 
\\\hdashline

$\lhavc{n}{x}$ & $\fsuc(n,c,f)$ &
$\all{v\in\Z}{\encode{J}\land\bothF{\tpc n\sep\tpc n'}\land an(n,n')\land \subst{\proj{K}(m,m')}{x|x'}{v|e'} \imp \subst{an(m,m')}{x|x'}{v|e'}}$
  \\
$\lassg{n'}{x'}{e'}$ & $\fsuc(n',c',f')$ &
\end{tabular}
\end{footnotesize}
\vspace*{-1ex}
\caption{Selected $pc$-encoded joint VCs for $an$ and $\aprod(\aut(c,f),\aut(c',f'),L,R,J,K)$.}
\label{fig:EVCjo-encoded}
\end{figure}

For the alignment completeness result for ERHL+, we rely on encoded VCs that use
$pc$-encoded versions of $L, R, J$, and a convenient encoding of $K$.
For the keep set encoding, define
\[\graybox{\(\proj{K}(m,m')\)} = \{ (s,s') \mid ((m,m'),(s,s')) \in K \}\]
Corresponding to Lemma~\ref{lem:liftRVC}, we have the following for filtered
alignment automata.

\begin{lem}[$pc$-encoded VCs for filtered alignment automata]\label{lem:liftEVC} \upshape
  Let $an$ be an annotation of $\aprod(\aut(c,f),\aut(c',f'),L,R,J,K)$ for commands $c,c'$.
  Suppose $pc$ is a fresh variable in the sense that it does not occur in $c$ or $c'$,
  and all of $L,R,J$ and any $an(i,j)$ are independent from $pc$ on both sides.
  Then each of the right-only relational VC of \autoref{fig:EVCro} implies the corresponding
  condition on store relations in \autoref{fig:EVCro-encoded}.  Each joint VC in
  \autoref{fig:EVCjo} 
  implies the corresponding condition in \autoref{fig:EVCjo-encoded}.
  Similarly for left-only VCs.
\end{lem}

The conditions in \autoref{fig:EVCro-encoded} are obtained using the $pc$-encoded
$\encode{R} \land \bothF{\tpc n\sep \tpc n'}$ in place of $R \land [n|n']$, $an$ in place of $\hat{an}$,
and $\proj{K}$ in place of $K$.  The conditions in \autoref{fig:EVCjo-encoded} are obtained similarly.
Note that \autoref{fig:EVCjo-encoded} lists only a few $pc$-encoded joint VCs but all 36 VC have encodings.
The proof of \autoref{lem:liftEVC} is similar to that of 
\autoref{lem:liftRVC}.

\subsection{Alignment completeness of ERHL+}\label{sec:aecompleteSubsect}

Almost all the ground work has been laid to state and prove alignment completeness of ERHL+ with respect to filtered alignment automata. 
It remains to address the issue of control determinacy of programs that occur on the right in $\forall\exists$ judgments, 
as discussed in \autoref{sec:controldet}.
A straightforward argument connects the property to automata as follows.\footnote{If $A$ and $A'$ are control deterministic then so is
any $\aprod(A,A',\dots)$, but we do not need this fact.} 

\begin{lem}\label{lem:controlDet}
\upshape
If $c$ is control deterministic then 
$\aut(c,f)$ is control deterministic in the sense that for any states
$(n,s)$, $(m_0,t_0)$, and $(m_1,t_1)$,
if $(n,s)\trans (m_0,t_0)$ and 
   $(n,s)\trans (m_1,t_1)$ then $m_0=m_1$.
\end{lem}

The alignment completeness theorem restricts the right-side program to be control deterministic.  To this end, define \graybox{$\okfd(c,f)$} to mean $c$ is control deterministic and $\okf(c,f)$.
It is used in assumption (a) of the theorem below.
Assumptions (b) and (c) constitute an IAM-style proof.
Assumption (d) is a technicality to ensure that a fresh $pc$ variable can be chosen for application of \autoref{thm:normEquiv}.

\begin{restatable}{thm}{thmaeAlignComp}\label{thm:aeAlignComp}
\upshape
Suppose we have the following.\\
(a) $\okf(c,f)$ and $\okfd(c',f')$. \\
(b) $an$ is a valid annotation of $\aprod(\aut(c,f),\aut(c',f'),L,R,J,K)$
  for $\spec{\S}{\T}$. \\
(c) $K$ is an adequate filtering for $an$ and $\aprod(\aut(c,f),\aut(c',f'),L,R,J,K)$. \\
(d) $an$, $\S$, $\T$, $L$, $R$, $J$,
and the witness $V$ for adequate filtering, all have finite support.
\\
Then the judgment $c\sep c' : \aespec{\S}{\T}$ has a proof in ERHL+.
\end{restatable}
As in the case of \autoref{thm:acomplete}, the only relational assertions used in the proof are those derived from 
$an(i,j)$ $L$, $R$, $J$, and $K$.
\begin{proof}[Proof (Sketch)]
The lengthy proof is similar to that of \autoref{thm:acomplete} for RHL+.  We start by
choosing a fresh variable $pc$, transform $c,c'$ to their automata normal
forms, and apply rule \rn{eDo}.  The loop invariant $\Q$ is identical to the
one defined in the proof of \autoref{thm:acomplete}.  The main difference
between the two proofs is in how each of the premises of the loop rule are
established.  As in the proof for RHL+, we derive premises of \rn{eDo} by the
assignment and sequence rules, and \rn{eConseq}, using VCs provided by $an$.
However, VCs of filtered automata have antecedents involving $K$, see
Fig~\ref{fig:EVCro} and \autoref{fig:EVCjo}.  So to exploit the VCs, in
applications of \rn{eConseq}, we rely on the fact that $K$ is an adequate
filtering.  Proofs of left-only and right-only premises use \rn{eDisj} 
in the same way \rn{rDisj} is used in the proof of \autoref{thm:acomplete}.  For right-only premises, we
additionally need to reason about the variant, for which we rely on the
right-productivity condition of \autoref{def:adequateFiltering}.
Owing to the assumption that $c'$ is control deterministic,
the automaton $\aut(c',f')$ is control deterministic (\autoref{lem:controlDet}).
This is important 
because when appealing to right-productivity or joint-productivity there is a
unique control point witnessing the existential in those properties.

Finally, having used \rn{eDo} to show
that the automata normal forms of $c$ and $c'$ satisfy $\aespec{\S}{\T}$, we
finish the proof using \rn{eRewrite} 
(with \autoref{thm:normEquiv}) 
and \rn{eGhost} to conclude the same judgment for $c,c'$,
as in the proof of \autoref{thm:acomplete}.
\end{proof}

\subsection{Cook completeness revisited for ERHL+}\label{sec:unaryRevisitedE}

\begin{thm}\label{thm:CookERHL}
\upshape
The logic ERHL+ is Cook complete, for control deterministic programs.
\end{thm}
\begin{proof}
The proof is like Cook completeness for RHL+.
If $\models c\sep c': \aespec{\S}{\T}$
and $\S,\T$ are finitely supported
then there is a valid annotation and adequate filtered alignment automaton 
for the automata of $c$ and $c'$, by \autoref{thm:FiltSoundCompl}.
Suppose $c'$ is control deterministic.
Then by \autoref{thm:aeAlignComp} there is a proof in ERHL+.
\end{proof}

By inspection of the proof of \autoref{thm:FiltSoundCompl},
that theorem still holds if the well-ordered set $D$ in \autoref{def:adequateFiltering}
is restricted to be $\nat$ with the usual order.
So Cook completeness holds even if rule \rn{eDo} of ERHL+ is restricted to use $\nat$. Cognoscenti may note that the presence of unbounded nondeterminacy (as with our havoc command) necessitates use of ordinals beyond $\omega$ for proving always-termination~\cite{AptPlotkin86}; 
but the $\forall\exists$ judgment is about existence of terminating runs.

In parallel to the discussion in \autoref{sec:unaryRevisited}, the reader can check that,
using the representation $c\sep\skipc:\aespec{\leftF{P}}{\leftF{Q}}$ for $c:\spec{P}{Q}$,
the rules of HL+ can be derived in ERHL+.
Furthermore, rules of forward underapproximation logic (e.g.,~\cite{DickersonYZD22,Ascari24}) can be derived using the representation
$\skipc\sep c:\aespec{\rightF{P}}{\rightF{Q}}$ for $c:\espec{P}{Q}$
(see for example \rn{eSkipDo} in \autoref{fig:derivedERHL}).
The difference between $\forall\forall$ and $\forall\exists$ judgments is 
evident when we consider proving Cook completeness using sequential alignment.
First, compared with \autoref{lem:aa-decomp} 
the following result is only for left-before-right.

\begin{lem} \label{lem:ae-decomp}
  \upshape
  For any $c,c',\P$ and $\R$, we have $\models c \sep c' : \aespec{\P}{\R}$ iff there is a $\Q$ such that $\models c \sep \skipc : \aespec{\P}{\Q}$ and $\models \skipc \sep c' : \aespec{\Q}{\R}$.
\end{lem}
\begin{proof}
  For the if direction, consider $s,s',t$ with $(s,s') \in \P$ and $\means{c}\,s\,t$.  By the assumption $\models c \sep \skipc : \aespec{\P}{\Q}$ 
we have $(t,s') \in \Q$.  Thus, by the assumption about $\skipc\sep c'$, there is a $t'$ such that $(t,t') \in \R$ and we are done.

  For the only-if direction, let $\Q := \{ (t,s') \mid \some{s}{(s,s') \in \P \land \means{c}\,s\,t} \}$.  This $\Q$ is the strongest relation that holds after executing $c$ on the left from $\P$-related states.  
It is straightforward to show $\models c \sep \skipc : \aespec{\P}{\Q}$.  To show $\models \skipc \sep c' : \aespec{\Q}{\R}$, consider $s,s',t$ with $(s,s') \in \Q$ and $\means{\skipc}\,s\,t$ which implies $s=t$ by semantics.  By definition of $\Q$ there is $u$ such that $\means{c}\,u\,t$ and $(u,s') \in \P$.  By the assumption $\models c \sep c' : \aespec{\P}{\R}$, there is a $t'$ such that $\means{c'}\,s'\,t'$ and $(t,t') \in \R$.  Thus, $\models \skipc \sep c' : \aespec{\Q}{\R}$.
\end{proof}
Analogous to (\ref{eq:encodeUnary}) we can connect with the unary
over- and under-approximate judgments using relations encoded as predicates 
on variable stores:
\begin{equation}\label{eq:encodeUnaryE}
\begin{array}[t]{l}
\models c \sep \skipc : \aespec{\P}{\Q}  \quad\mbox{iff}\quad
\models c : \spec{\codeleft{\P}}{\codeleft{\Q}}
\\
\models \skipc \sep c : \aespec{\P}{\Q}  \quad\mbox{iff}\quad
\models c : \espec{\coderight{\P}}{\coderight{\Q}} 
\end{array}
\end{equation}
Consider these rules that correspond to the derivable rules \rn{eLRseq} and \rn{eRLseq}, which reduce a relational property to unary properties.
\begin{mathpar}
\inferrule[ueLRseq]{
c : \spec{\codeleft{\P}}{\codeleft{\Q}} \\
c': \espec{\coderight{\Q}}{\coderight{\R}} 
}{
c\sep c' : \aespec{\P}{\R}
}

\inferrule[ueRLseq]{
c' : \spec{\coderight{\P}}{\coderight{\Q}}  \\
c  : \espec{\codeleft{\Q}}{\codeleft{\R}}
}{
c\sep c' : \aespec{\P}{\R}
}
\end{mathpar}
Soundness of \rn{ueLRseq} follows from \autoref{lem:ae-decomp}.
In fact \rn{ueRLseq} is also sound, but by contrast with 
\autoref{lem:ae-decomp} there is not an equivalence but only an implication.

\begin{exa}\label{exa:counter}
In general, $\models c\sep c': \aespec{\P}{\R}$ does not imply that there
exists  $\Q$ such that both $\models \skipc \sep c' : \aespec{\P}{\Q}$ and $\models c \sep \skipc : \aespec{\Q}{\R}$.
Here is a counter-example.  
We have 
\( \models \havc{x}\sep\havc{x} : \aespec{\mathit{true}}{\Agr x} \).
Note that, because $\skipc$ is domain total and deterministic (per comments at start of \autoref{sec:eacomplete}), 
we have for any $c,\P,\R$ that $c\sep \skipc:\aespec{\P}{\R}$
iff $\P\imp \WP(c\sep\skipc)(\R)$.
Now suppose there is a $\Q$ such that (i) $\models \skipc\sep\havc{x} : \aespec{\mathit{true}}{\Q}$ and (ii) $\models \havc{x}\sep\skipc : \aespec{\Q}{\Agr x}$.  By (ii), $\Q \imp \WP(\havc{x}\sep\skipc)(\Agr x)$.
But $\WP(\havc{x}\sep\skipc)(\Agr x) 
  \iff (\all{v}{\subst{\Agr x}{x|}{v|}}) 
  \iff (\all{v}{v=\rightex{x}})
  \iff \mathit{false}$.  
Thus $\Q = \mathit{false}$.
Note that by definition of the $\forall\exists$ judgment there are no
$c,d,\P$ such that $\models c\sep d:\aespec{\P}{\mathit{false}}$
(unless $c$ has no executions from $\P$, which is not the case in the counter-example).
So $\Q = \mathit{false}$ contradicts (i).
\qed
\end{exa}


Now we can show Cook completeness, for $\forall\exists$ properties, with the single \rn{ueLRseq} rule. 
Assume we have a complete logic for $\especSym$ judgments.\footnote{Such logics do exist.  For example, sufficient incorrectness logic~\cite{Ascari24} is shown to be complete.}  Suppose $\models c \sep c' : \aespec{\P}{\R}$ holds.  By Lemma~\ref{lem:ae-decomp} there is a $\Q$ such that 
$\models c \sep \skipc : \aespec{\P}{\Q}$ and 
$\models \skipc \sep c' : \aespec{\Q}{\R}$.  
So we have 
$\models c : \spec{\codeleft{\P}}{\codeleft{\Q}}$ and 
$\models c' : \espec{\coderight{\Q}}{\coderight{\R}}$.  
By completeness of the unary logics, these specs are provable.
One application of \rn{ueLRseq} yields
$c \sep c' : \aespec{\P}{\R}$.

By contrast with the situation for $\forall\forall$,
completeness does not hold for \rn{ueRLseq} plus unary 
($\specSym,\especSym$)
rules.
Consider the counterexample $\havc{x}\sep\havc{x} :\aespec{\mathit{true}}{\Agr x}$ in \autoref{exa:counter}.
It cannot be proved using \rn{ueRLseq}
because there is no $\Q$ with which to instantiate the premises. 

For the record, Cook completeness of ERHL+ (\autoref{thm:CookERHL}) can be proved for $\forall\exists$ judgments without restriction to judgments where the right side program is control deterministic. The details are in~\cite{BNN23v5}.


\section{Related work}\label{sec:related}


Trace logic~\cite{BartheEGGKM2019} reasons about $\forall\forall$ properties by way of constraints between arbitrary different points in the traces rather than restricting to points that progress in the manner of a schedule--what we call alignment.
In principle, alignments may be defined by arbitrary strategy functions and the like~\cite{KovacsSF13,BaumeisterCBFS21,ClochardMP20}.
In the following we confine attention to assertion-oriented work rather than works using global conditions on traces.

Cook's completeness result~\cite{Cook78} is in terms of a formal language of assertions.  The result is relative to provability of assertion entailments, and depends on expressivity of the assertion language
(good explanations can be found in~\cite{AptOld3} and~\cite{WinskelBook}).  In recent years, and especially in the context of machine-checked theories, it has become common to sidestep these issues by way of shallow embedding, as we have done.  What is left is exactly the standard notion of completeness for a logic; we use the term ``Cook completeness'' for contrast with alignment completeness.

Cook completeness has been proved for several $\forall\forall$ relational Hoare logics (e.g.,~\cite{Beringer11,SousaD2016,BartheGHS17,WangDilligLahiriCook})
using in each case the semantic completeness of sequential alignment
and relying on completeness of HL along the lines we sketch in \autoref{sec:unaryRevisited}.
For alignment completeness, we are not aware of prior results besides those
of Nagasamudram and Naumann~\cite{NagasamudramN21}.
They give several $\forall\forall$ alignment completeness results for very
specialized automata forms that account for alignments given by particular proof rules. 
Their proofs of alignment completeness are quite different from ours: owing to the specialized structure of the automata considered, they are able to construct deductive 
proofs that follow the structure of the source programs without any rewriting.
The alignment completeness result for RHL+ was presented in our unpublished preprint~\cite{BNN22},
which is superceded by the present article (and~\cite{BNN23v5}). 

Nagasamudram and Naumann~\cite{NagasamudramN21} 
introduce the term Floyd completeness and use it for a 
result like our \autoref{thm:FloydComplete}.
The practical importance of both their and our Floyd completeness results is that the deductive proof does not require more expressive assertions than used in the IAM proof.  

The conditionally aligned loop rule (our \rn{rDo} in \autoref{fig:RHL})
appears first in Beringer's work~\cite{Beringer11}. 
Variations appear in Barthe et al~\cite{BartheGHS17},
in the full version of Banerjee et al~\cite{BNN16}, and in~\cite{BNNN19}.
Our adaptation for $\forall\exists$ (\rn{eDo} in \autoref{fig:ERHL}) is new.
Several recently published logics only support lockstep alignment of loops,
or lockstep until one terminates (even~\cite{DickersonYZD22} published subsequent to~\cite{BartheGHS17}).  
Beutner's~\cite{Beutner24} loop rule for $\forall\exists$, named \emph{loop-counting}, generalizes the lockstep pattern, catering for alignments that relate $n$ unrollings of a loop with $m$ unrollings of the other, for fixed $n$ and $m$.  The rule requires loops being related to terminate simultaneously, disallowing alignments that reason in terms of lockstep iterations up to the point where one loop terminates and then reason about the remaining iterations of the other.  Due to the side-condition on termination, the rule doesn't require a variant.  Although fixing $n$ and $m$ independent of data can handle some examples, it is too restrictive for others as discussed 
in \autoref{exa:majorize}.   A primary motivation for Beutner's work is automated verification, for which the loop-counting rule is shown to be effective.
  
We are not aware of unary HLs that feature a rewriting rule,
but verification tools often use correctness-preserving rewriting.
The RHLs of~\cite{BartheGHS17} and~\cite{BNN16} each feature a rewriting rule and a custom set of rules for command equivalence (with a relational precondition, in~\cite{BartheGHS17}).
(The judgments of~\cite{BartheGHS17} are probabilistic but alignment is still central.) 
Rewriting by command equivalence is combined with relational reasoning 
in an extension of KAT called BiKAT~\cite{AntonopoulosEtal2022popl}.
Rewriting (unfoldings)
is used by Strichman and Veitsman~\cite{StrichmanV2016}
to improve alignment in regression verification of recursive functions.
 Verifiable C~\cite{cao2018vst} includes a proof rule that reassociates sequences, as does~\cite{Beutner24}.

The $\forall\forall$ property is also known as 2-safety~\cite{TerauchiA2005}.
Cartesian Hoare logic~\cite{SousaD2016,PickFG18} reasons about $k$-safety, for $k$ that is fixed throughout a proof.
D'Osualdo et al~\cite{DosualdoFD2022} develop a logic for $k$-safety that features rules for combining judgments with varying $k$;
they have a Cook completeness result based essentially on sequential alignment
as explained in~\cite[appendix C]{DosualdoFD2022}.
The system of \opcit\ includes a rewriting rule, but based on semantic refinement 
(i.e., refinement is not formalized in a deductive system but rather must be proved
in the metalogic).  
Our results should generalize to $k>2$, but the case of 2 admits simpler notations and suffices to illuminate the issues we address. 

Except as noted, none of the preceding works  consider $\forall\exists$ properties.  
Hawblitzel et al~\cite{hawblitzelklr13} use the term relative termination for $\forall\exists$ judgments.  The logic of Benton~\cite{Benton:popl04} is for cotermination of deterministic programs,
which can be expressed by a pair of $\forall\exists$ judgments of the form
$c\sep d : \aespec{\P}{\Q}$ and $d\sep c : \aespec{\P^\circ}{\Q^\circ}$
(where $\P^\circ$ is the converse of $\P$).
Rinard gives a logical formulation of verification conditions for a $\forall\exists$ relation, to prove correctness of compiler transformations acting on control flow graphs~
\cite{RinardMarinov99,RinardCredible99}.

Antonopoulos et al~\cite{AntonopoulosEtal2022popl} derive some deductive rules for $\forall\exists$
(and also the variation known as backward simulation), in a KAT-based algebraic framework.
Their ``witness'' technique for $\forall\exists$ reasoning is akin to our filtering $K$ condition,
obtaining existence by filtering behaviors of a $\forall\forall$ automaton, 
by contrast with works that literally construct witness executions~\cite{LamportS21,UnnoTerauchiKoskinen21}.
An early formulation of relational verification using automata is~\cite{BartheCK13} which introduces a notion of asymmetric product for verifying $\forall\exists$ properties. 

To our knowledge the first published deductive system for general pre-post $\forall\exists$ properties is RHLE~\cite{DickersonYZD22}.
It is based on HL together with unary rules for forward underapproximation
---the judgment we write as $c:\espec{P}{Q}$, see (\ref{eq:espec}).
The core rules of RHLE reduce relational reasoning to sequential unary reasoning.
Soundness is proved.
The system includes rules for modular reasoning using unary procedure specs,
and uses over- and under-approximate semantics of procedure calls.
Loops are handled using the mostly lockstep rules like in Sousa and Dillig's work~\cite{SousaD2016}.
The use of unary procedure specs gives rise to nondeterminacy, motivating the use of $\forall\exists$ judgements.  By contrast, Eilers et al~\cite{EilersMH18} and Banerjee et al~\cite{BNNN19} use relational procedure specs for $\forall\forall$ properties.
Subsequent to developing our results we became aware of Beutner's Forall-Exists Hoare logic (FEHL) for $\forall\exists$ properties~\cite{Beutner24}.  FEHL features rules similar to RHLE, and includes the more general loop rule (loop-counting) mentioned earlier in this section.  It also includes a couple of rules for rewriting with sequence associativity and skip unit law.  Like in RHLE, the rules in FEHL rely on HL and a unary logic for forward underapproximation ($\especSym$).  Cook completeness is obtained for FEHL via Cook completeness of HL and a logic for forward underapproximation, using a variation of our rule \rn{ueLRseq}.  
Owing to the restrictive treatment of loops it is unlikely that FEHL or RHLE is alignment complete.
The purpose of both FEHL and RHLE is for use in automated search and both works provide a search algorithm and experimental results.
  
ReLoC~\cite{FruminKB18} is a logic for contextual refinement of higher order concurrent programs, a specific $\forall\exists$ property.
It is not a freestanding deductive system but rather it is shallow embedded in Iris~\cite{JungKJBBD18} (which in turn is implemented in the Coq proof assistant).  So one can, e.g., negate the refinement judgment and express some forms of conditional refinement.

Turning to IAM-style verification, 
the work of~\cite{ChurchillP0A19,ShemerGSV19,UnnoTerauchiKoskinen21} can be seen as various techniques for finding
adequate alignment conditions ($L,R,J$) and annotations expressible in SMT-supported assertion languages.  
Churchill et al also use testing to evaluate whether a candidate $(L,R,J)$ is manifestly adequate.  
Our example $c3$ (\autoref{sec:overview}) is from~\cite{UnnoTerauchiKoskinen21}; they formulate adequacy conditions for $\forall\exists$
properties, whereas the others cited only address $\forall\forall$.
Unno et al address the right-only progress condition for $\forall\exists$ by finding a well-founded relation between transition states; this is not directly representable in a logic of pre/post relations (though it may be in a logic like RHTT~\cite{NanevskiBG13} where postconditions constrain two initial + two final states),
so we need variant function $V$ and a fresh snapshot variable in rule \rn{eDo}.
Instead of filtering, Unno et al require that choices on right be given by a function of the left state together with prophecy variables about the final state on the left.
Our formalization enables use of prophecy for both final and intermediate values.
More recently, Itzhaky et al.~\cite{ItzhakySV24} develop a method for verifying $\forall\exists$ properties via reduction to constrained Horn clauses (CHCs).  Their technique solves simultaneously for alignments, relational invariants, and witness functions for right side executions.  The natural encoding of the $L,R,J$ adequacy condition as a first-order formula is not Horn, so Itzhaky et al.\ use multiple steps to transform it to a set of equi-satisfiable CHCs.  This enables using a single CHC solver query when searching for solutions, as opposed to prior approaches~\cite{UnnoTerauchiKoskinen21,ShemerGSV19} that require multiple queries or the use of specialized solvers.

Results like our adequacy Proposition~\ref{prop:adequate-sound}
have been proved for several notions of alignment product that are similar to ours.
Our $L,R,J$ corresponds to the ``alignment predicate'' in Churchill et al~\cite{ChurchillP0A19},
the ``composition function'' in Shemer et al~\cite{ShemerGSV19},
and ``scheduler'' in Unno et al~\cite{UnnoTerauchiKoskinen21}.
These works focus on automated search for good alignments and annotations 
using solvers for restricted assertion languages.

Sequential alignment has been used to prove completeness results for product automata~\cite{Francez83,UnnoTerauchiKoskinen21}
as in \autoref{thm:FiltSoundCompl}.

Beutner and Finkbeiner~\cite{BeutnerF22CAV} formulate temporal $\forall\exists$ properties in terms of games, so a proof involves a strategy whereby the $\exists$ player produces a witnessing execution.  They represent programs by transition systems, and combine the search for a strategy with search for an alignment (called reduction, cf.~\cite{FarzanV19}) and filtering conditions (called restrictions), also represented as a game. 
Itzhaky et al's approach~\cite{ItzhakySV24} to $\forall\exists$ verification using CHCs is shown to be sound, and complete with respect to this game semantics for transition systems with bounded nondeterminism.  However, this game semantics itself is incomplete.
Beutner and Finkbeiner show~\cite{BeutnerF22CSF} that with the inclusion of prophecies, a game based method is complete for finite state systems and specifications in  synchronous HyperLTL. 

Our normal form is like that of Hoare et al~\cite{HoareHS93} which uses an
explicit program counter variable.
They prove every program can be reduced to normal form using a set of refinement laws, demonstrating a kind of completeness of the laws.  
Kozen~\cite{Kozen97} uses KAT to prove that every command is simulated by one with a single loop. That result is refined in~\cite{GrathwohlKM14} using KAT augmented with finite mutable state; in principle this would provide an alternate way to define $\kateq$, cf.~\autoref{rem:KAT}.



\section{Future work}\label{sec:future}

Because our development makes no commitment or unusual demands on the assertion language, the results should be applicable to richer data types including dynamically allocated pointer structures, and deductive systems including separation logic.  Our program equivalence $\kateq$ makes a minimal demand on reasoning about data: the ability to specify that primitive expressions and commands do not interfere with a fresh ghost variable. 

For languages with procedures, alignment of calls is important to facilitate
use of relational specs~\cite{GodlinS08,EilersMH20,BNNN19}.  A theory of alignment completeness of such languages would involve the complications of automata representation for programs and some notion of modular IAM.

A key question is what are good criteria for relational logics.
Cook completeness is clearly important.
Alignment completeness is an additional criterion that accounts for widely used rules of RHL by connection with the fundamental IAM.
For unary HL, the corresponding notion is Floyd completeness.  These notions are not the only sensible criteria. 
 
One obvious criterion is essentially aesthetic: the core set of rules should be general and orthogonal or minimal in some sense.  The model is HL for simple imperative programs. There is one rule for each program construct.  
In addition there are so-called \emph{structural} rules.
To capture IAM reasoning it suffices to have a single structural rule, \rn{Conseq}; this is formalized in the Floyd completeness result of~\cite{NagasamudramN21}.
Principles beyond IAM, such as framing, conjunctive splitting and auxiliary variables, which facilitate modular reasoning, are embodied by additional rules often included in HL.  For example, the rule of conjunction enables to prove correctness of quicksort by first proving the permutation property and then proving sortedness (see~\cite[chapter 5]{AptOld3}).  History and prophecy variables extend the expressive power of assertions: Although an assertion pertains to the store at a particular control point (or aligned pair thereof), auxiliary variables are used to express relationships with computation steps at other control points.  

For our main results we focus on sets of rules (RHL+ and ERHL+) with only the structural rules needed for alignment completeness ---rewriting, ghost elimination, and disjunction.  For our main results one could also severely restrict the other rules: Inspecting the proofs of alignment completeness one may see we could drop the \rn{rIf/eIf} rules and others, and rely on a specialized form of \rn{rDo/eDo} that only applies to the patterns that appear in the normal form (\autoref{lem:nfCases}).
Of course the aesthetic criterion guides us to include a single, general rule for each program construct. 

Adopting highly specialized rules tailored to alignment completeness is a bad idea because it would force all proofs to go by maximal rewriting to normal form, and all proofs to be IAM style, largely abandoning the syntax-oriented benefits of Hoare logic. This suggests another criterion: a good logic should support proofs that are modular and natural.
One way to make ``natural'' more precise is to connect it with alignment completeness.
We pose this open problem: 
\emph{Prove alignment completeness by some technique that minimizes the use of rewriting and instead preserves the original program structure as much as possible.}
Although definitions \ref{def:alignProd} and \ref{def:aeAlignProd} admit strange and gratuitously complicated alignment conditions, as may arise using automated inference techniques, examples suggest that for deductive proofs it should usually suffice to rewrite just enough to make the control structures similar.
For that matter, if an automaton has one of the forms in~\cite{NagasamudramN21} then little rewriting should be needed
as shown in that article.
To derive rules for common patterns, such as those in \autoref{fig:derivedRHL} and \autoref{fig:derivedERHL}, 
it suffices to do rewrites by very minimal laws such as the 
unit law for skip.
Finding a notion of minimality for rewriting, and proving alignment-completeness with minimal rewrites, might have practical value to combine automata-based and deductive methods.  

A key criterion that can be made precise is known as \dt{adaptation completeness}~\cite{Kleymann99,AptO19}.
For unary correctness it says that if the spec $\spec{P}{Q}$ is semantically entailed by the spec $\spec{R}{S}$, in the sense that 
$\models c:\spec{R}{S}$ implies $\models c:\spec{P}{Q}$ for any $c$,
then the proof rules are sufficient to derive 
$c: \spec{P}{Q}$ from $c: \spec{R}{S}$.
The \rn{Conseq} rule is useful for this purpose but not adaptation complete, nor is the Adaptation rule of~\cite{Hoare71} but complete rules are known~\cite{AptO19}. 

A generalization of adaptation completeness emerges in the context of relational reasoning.
As an example, D'Osualdo et al~\cite{DosualdoFD2022}
consider the following hypotheses about unknown commands 
(or command variables) $c$ and $d$.
First, $c\sep c : \rspec{\Agr x}{\Agr x}$ and 
$d\sep d : \rspec{\Agr x}{\Agr x}$, i.e., they are deterministic with respect to variable $x$. 
Second, they commute in the sense that 
$(c; d)\sep(d;c) : \rspec{\Agr x}{\Agr x}$.
Third, they may terminate from any state, i.e.,
$c:\espec{\mathit{true}}{\mathit{true}}$ and 
$d:\espec{\mathit{true}}{\mathit{true}}$.
Equivalently, $\skipc \sep c : \aespec{\mathit{true}}{\mathit{true}}$
and $\skipc \sep d : \aespec{\mathit{true}}{\mathit{true}}$.
It follows semantically that 
\( (c;d;d) \sep (d;d;c) : \rspec{\Agr x}{\Agr x} \).
One expects to show this by transitively composing the judgments
$(c;d;d) \sep (d;c;d) : \rspec{\Agr x}{\Agr x}$ 
and 
$(d;c;d) \sep (d;d;c) : \rspec{\Agr x}{\Agr x}$.
D'Osualdo et al posit that this is beyond the reach of RHLs and introduce a somewhat different deductive system that handles the example.
As noted earlier, $\forall\forall$ judgments do not transitively compose in general,
owing to the possibility that the middle program diverges.  So D'Osualdo et al introduce a special judgment for termination.  Another approach is to leverage $\forall\exists$ judgments as in this sound rule.
\[
  \inferrule*[right=rTrans]
  {c\sep d : \rspec{\P}{\Q} \\
  \skipc \sep d : \aespec{\P}{\mathit{true}} \\
  d\sep c' : \rspec{\R}{\S}}
  {c\sep c' : \rspec{\P;\R}{\Q;\S}}  
\]
Here we write $\P;\R$ for composition of relations 
and note that $\Agr x;\Agr x$ is $\Agr x$.

In~\cite{BNN23v5} we extend RHL+ and ERHL+ with this rule and a few others adapted from D'Osualdo et al, which suffice to prove the examples.
This includes reasoning about idempotence where one execution of a command is related to a sequence of its executions, which is beyond the conventional notion of alignment.
Although D'Osualdo et al show their system works nicely on a range of examples, they neither state nor establish the general property 
which we dub \dt{entailment completeness}.  Meaning: If a judgment follows semantically from a set $H$ of judgments, then it can be derived from $H$ in the logic.  
To our knowledge the only entailment completeness result for a program logic is that of Kozen and Tiuryn~\cite{KozenT01} for propositional Hoare logic.\footnote{In more powerful 
systems like dynamic logic or embeddings in higher order logic, entailment can expressed by a formula so entailment completeness reduces to ordinary completeness. This can be done in KAT~\cite{Kozen00} and in propositional dynamic logic~\cite[Thm.~7.7]{HarelKozenTiuryn00}.}
For relational correctness, we suspect that this open problem can be solved for some system that combines $\forall\forall$ and $\forall\exists$ judgments.

To see how entailment completeness motivates some well known structural rules, here is an example which in particular shows the need for prophecy variables to prove $\forall\exists$ judgments.
Assuming that $c\sep c: \aespec{\Agr w}{\Agr w}$, and $c$ neither reads nor writes $y$, 
this is valid:
\begin{equation}\label{eq:prophSpec}
c; \havc{x} \sep \havc{y}; c : \aespec{\Agr w}{\Agr w \land \leftex{x}=\rightex{y}} 
\end{equation}
Typically, as in \autoref{eg:c3ded}, 
we align right-side havocs together with, or following, left-side havocs, so the postcondition on the existential side can refer to the left side.  
Here, it is not convenient to align $\havc{y}$ later than $\havc{x}$ because we need to align $c$ with itself to exploit the assumption.
The idea is to use a variable $r$ that can be seen as predicting the final value of $x$ on the left side.  
For this exposition we will consider $r$ to be a logical variable that does not occur in code.  
We introduce $r$ in a judgment that says $y$ can match the predicted value:
\begin{equation}\label{eq:proph0}
\skipc \sep \havc{y} : \aespec{\Agr w}{\Agr w\land r=\rightex{y}}
\end{equation}
This is proved using \rn{eSkipHav} and \rn{eConseq} with validity of 
$\Agr w \imp (\some{|y}{\Agr w \land r = \rightex{y}})$.
Using a frame rule (\rn{eFrame} below), the assumption 
that $c$ does not write $y$ yields
\begin{equation}\label{eq:proph1}
c \sep c : \aespec{\Agr w\land r=\rightex{y}}{\Agr w\land r=\rightex{y}}
\end{equation}
Next, by \rn{eHavSkip} and \rn{eConseq} we get
\begin{equation}\label{eq:proph2}
\havc{x} \sep \skipc : \aespec{\Agr w\land r=\rightex{y}}{\Agr w\land (r=\leftex{x} \imp r=\rightex{y})}
\end{equation}
using that this is valid:
$\Agr w\land r=\rightex{y} \imp (\all{x|}{\Agr w\land (r=\leftex{x} \imp r=\rightex{y})})$.
From (\ref{eq:proph0}--\ref{eq:proph2}) by \rn{eSeq}, and \rn{eRewrite} to eliminate skips, we get this key judgment that embodies the prophecy variable pattern: 
\[ c; \havc{x} \sep \havc{y}; c : \aespec{\Agr w}{\Agr w \land (r=\leftex{x} \imp r=\rightex{y})} \]
Now $r$ occurs in neither the code nor the precondition,
so by a rule \rn{eForall} below we get
\[ c; \havc{x} \sep \havc{y}; c : \aespec{\Agr w}{(\all{r}{\Agr w \land (r=\leftex{x} \imp r=\rightex{y})})} \]
This yields (\ref{eq:prophSpec}) by \rn{eConseq}.

For both $\forall\forall$ and $\forall\exists$ judgments there is a sound frame rule like that for unary logic.\footnote{Historically, variants had names including Invariance and Constancy,
with Frame used where independence for heap locations is expressed
using separating conjunction.}
We used this one:
\[
  \inferrule[eFrame]{
    c\sep c' : \aespec{\P}{\Q} \\
    \indep(\mathconst{mods}(c) | \mathconst{mods}(c'), \R)
  }{c \sep c' : \aespec{\P \land \R}{\Q \land \R}}
\]
where $\mathconst{mods}(c)$ are the variables assigned or havoc'd in $c$.  
Finally, just as there are disjunction and conjunction rules, for pre- and post-conditions respectively, 
logics often include a rule for introducing existential in preconditions and less commonly a rule like the following one that we used. Here $v$ is a logical variable that does not occur in code.\footnote{Strictly speaking, in accord with our shallow embedding of store relations, one should consider that $\Q$ is a family of store relations $\Q_v$ indexed over $v\in\Z$ and define
$\all{v}{\Q}$ as $\{(s,s')\mid \all{v\in\Z}{(s,s')\in\Q_v}\}$.
}
\[
  \inferrule[eForall]{
    c\sep c' : \aespec{\P}{\Q} \\
    \indep(v|v, \P)
  }{c \sep c' : \aespec{\P}{\all{v}{\Q}}}
\]

\section{Conclusion}\label{sec:concl}

In this paper we augment a collection of relational Hoare logic rules with a straightforward
rule for elimination of ghost variables, and a rule for deriving one correctness
judgment from another by rewriting the commands involved to equivalent ones.
The chosen notion of equivalence is that of KAT,
allowing for the use of hypotheses to axiomatize the meaning of primitive
commands and expressions when reasoning with KAT. 
The rewrites needed to derive a number of frequently proposed RHL rules require no hypotheses at all. 
Using a small set of
hypotheses, we prove that any command is equivalent to one in automaton normal
form, once it is instrumented with assignments to an explicit program counter variable. 
On this basis, we show that any correctness judgment proved in IAM style 
using an alignment automaton can be turned into a deductive
proof in RHL+ using essentially the same assertions.
One practical consequence is that automata-based alignment is not better than deductive in terms of strength of assertions needed.
If some decidable fragment like linear arithmetic suffices for an automaton-based proof then that same fragment suffices for a deductive proof of the same judgment.

We also introduce a new notion, filtered alignment automata, for $\forall\exists$ properties.  We introduce a new logic ERHL+ for $\forall\exists$ properties and show its alignment completeness.  For both kinds of automata we show semantic completeness with respect
to the relevant properties.  Together with alignment completeness, this entails 
that RHL+ and ERHL+ are Cook complete with respect to $\forall\forall$ and $\forall\exists$ properties respectively.

Some rules which embody natural reasoning principles, such as the rule of conjunction (for $\forall\forall$) and transitive composition rules like \rn{rTrans},  are not included in our logics because they are not needed for alignment completeness.  We conjecture these are needed for entailment completeness which we posed as an interesting open problem which may lead to discovery of additional rules and reasoning principles.

\paragraph{Acknowledgments}

We are grateful to the anonymous LMCS reviewers for insightful feedback which in particular helped improve the positioning of alignment completeness with respect to other criteria for program logics.

\bibliographystyle{alphaurl}
\bibliography{extracted}

\newpage

\appendix

\section{Additional definitions}\label{sec:additional}

\autoref{fig:lab} defines the label of a command, recursing only in the sequence case.
\autoref{fig:denot} presents the pre-post relation denoted by a command, defined inductively in big-step style.

\begin{figure}[h]
\begin{small}
\[
\begin{array}{l}
\lab(\lskipc{n}) = 
\lab(\lassg{n}{x}{e}) = 
\lab(\lhavc{n}{x}) =
\lab(\lifgc{n}{gcs}) =
\lab(\ldogc{n}{gcs}) = n \\
\lab(c;d) =  \lab(c)
\end{array}
\]
\end{small}
\vspace{-3ex}
\caption{Definition of $\lab(c)$.}\label{fig:lab}
\end{figure}

\begin{figure}[h]
\begin{small}
\begin{mathpar}
\inferrule{}{
\means{\lskipc{n}}\, s\, s
}

\inferrule{t = \update{s}{x}{\means{e}(s)} }{
\means{\lassg{}{x}{e}}\, s\, t
}

\inferrule{v\in\Z}{
\means{\havc{x}}\, s\, \update{s}{x}{v}
}

\inferrule{\means{c}\,s\, u \\ \means{d}\,u\, t}{
\means{c;d}\,s\,t
}

\inferrule{e\gcto b \mbox{ is in } gcs\\
s\in\means{e} \\
\means{b}\,s\,t
}{
\means{\ifgc{gcs}}\,s\,t
}

\inferrule{e\gcto b \mbox{ is in } gcs\\
s\in\means{e} \\
\means{b}\,s\,u \\
\means{\dogc{gcs}}\,u\,t
}{
\means{\dogc{gcs}}\,s\,t
}

\inferrule{ s\in\means{\neg\enab(gcs)}}{
\means{\dogc{gcs}}\,s\,s
}

\end{mathpar}
\end{small}
\vspace{-2ex}
\caption{Denotational semantics $\means{c}$ (omitting labels which are unconstrained).}\label{fig:denot}
\end{figure}

\section{Axioms for normal form equivalence}\label{sec:KATequiv}

For the sake of straightforward presentation, command equivalence has been 
formulated in terms of a single fixed set of hypotheses that axiomatize simple
assignments and boolean expressions (Defs.~\ref{def:Hyp} and~\ref{def:kateq}).
However, many useful equivalences require no hypotheses.
Only a few specific hypotheses are needed to prove the normal form theorem,
and in this section we spell those out.

Observe that  if $\norm{c}{f}{gcs}$ then there is a finite set of instances of this relation that supports the fact, namely the normal forms of subprograms of $c$.
This enables us to define a set of axioms that are useful for reasoning about 
a given program $c$ and its normal form.  The definition is parameterized on $c$ and on the choice of a $pc$ variable and final label.

\begin{defi} 
The \dt{normal form axioms}, $\nfax(pc,c,f)$, 
comprises the following set of equations.
\begin{description}
\item[(diffTest)] 
$\mkt{ \tpc i } \kdot \mkt{ \tpc j } = \kzero$  for 
$i$ and $j$ in $\labs(c)\union\{f\}$ such that $i\neq j$
\item[(setTest)]
$\mkt{ \spc  i }\kdot\mkt{ \tpc  i } = \mkt{ \spc  i }$ for $i$ in $\labs(c)\union\{f\}$ 
 
\item[(totIf)]
$\kneg\mkt{\enab(gcs)} = 0$ for every $\lifgc{}{gcs}$ that occurs in $c$

\item[(testCommuteAsgn)] 
$\mkt{\tpc i} ; \mkt{\lassg{}{x}{e}} = \mkt{\lassg{}{x}{e}} ; \mkt{\tpc i}$
for every $\lassg{}{x}{e}$ in $c$ such that  $x\not\equiv pc$,
and every $i$ in $\labs(c)\union\{f\}$ 

\item[(testCommuteHav)]
$\mkt{\tpc i} ; \mkt{\havc{x}} = \mkt{\havc{x}} ; \mkt{\tpc i}$
for every $\havc{x}$ in $c$ such that  $x\not\equiv pc$,
and every $i$ in $\labs(c)\union\{f\}$ 

\end{description}
\end{defi}

\begin{lem}\label{lem:nfaxHyp}
\upshape
For any $pc,c,f$, the equations $\nfax(pc,c,f)$ 
follow by KAT reasoning from the equations of $\Hyp$.
\end{lem}
\begin{proof}
Observe that, in light of Lemma~\ref{lem:correctInterp},
the set $\Hyp$ can characterized in terms of $\relKAT$ as follows:
\begin{itemize}
\item $\mkt{e}=0$ for every boolean expression $e$
such that $\mkt{e}^\relKAT = \emptyset$
\item  $\mkt{e_0};\mkt{x:=e};\kneg\mkt{e_1}=\kzero$
for every $x,e,e_0,e_1$ such that 
$\mkt{e_0}^\relKAT;\mkt{x:=e}^\relKAT;\kneg\mkt{e_1}^\relKAT=\emptyset$.
\item  $\mkt{e_0};\mkt{\havc{x}};\kneg\mkt{e_1}=\kzero$
for every $x,e,e_0,e_1$ such that 
$\mkt{e_0}^\relKAT;\mkt{\havc{x}}^\relKAT;\kneg\mkt{e_1}^\relKAT=\emptyset$.
\end{itemize}
Now we proceed.

(diffTest) 
If $i\neq j$ then $\mkt{ \tpc i \land \tpc j }^\relKAT = \emptyset$, 
so $\Hyp$ contains $\mkt{ \tpc i \land \tpc j } = \kzero$.
So we have $\mkt{ \tpc i }\kdot \mkt{ \tpc j } = \mkt{ \tpc i \land \tpc j } =\kzero$
using the definition of $\mkt{-}$.

(setTest)
$\Hyp$ contains $\mkt{\skipc};\mkt{\spc i}\kdot\kneg\mkt{\tpc i}=\kzero$,
and $\mkt{\skipc}$ is $\kone$.
Now observe that 
\(\mkt{\spc i}
= 
\mkt{\spc i}\kdot(\mkt{\tpc i}+\kneg\mkt{\tpc i}) 
= 
\mkt{\spc i}\kdot\mkt{\tpc i} + \kone;\mkt{\spc i}\kdot\kneg\mkt{\tpc i}
= 
\mkt{\spc i}\kdot\mkt{\tpc i}\)
using KAT laws and $\kone;\mkt{\spc i}\kdot\kneg\mkt{\tpc i}=\kzero$.

(totIf)
If $\lifgc{}{gcs}$ that occurs in $c$ then 
we have $(\kneg\mkt{\enab(gcs)})^\relKAT = \emptyset$ 
as a consequence of the $\totalIf$ condition (Def.~\ref{def:lang}).
So the equation $\kneg\mkt{\enab(gcs)} = 0$ is in $\Hyp$.

(testCommuteAsgn)
The equation 
$\mkt{\tpc i} ; \mkt{\lassg{}{x}{e}} = \mkt{\lassg{}{x}{e}} ; \mkt{\tpc i}$,
is equivalent to the conjunction of 
$\mkt{\tpc i} ; \mkt{\lassg{}{x}{e}} ; \kneg\mkt{\tpc i} = 0$
and $\kneg\mkt{\tpc i} ; \mkt{\lassg{}{x}{e}} ; \mkt{\tpc i} = 0$
using KAT laws.  
By definition of $\mkt{-}$ and boolean algebra, 
$\kneg\mkt{\tpc i} ; \mkt{\lassg{}{x}{e}} ; \mkt{\tpc i} = 0$
is equivalent to 
$\mkt{\neg \tpc i} ; \mkt{\lassg{}{x}{e}} ; \kneg\mkt{\neg\tpc i} = 0$.
Both $\tpc i\imp \subst{(\tpc i)}{x}{e}$ and 
$\neg \tpc i\imp \subst{(\neg \tpc i)}{x}{e}$ 
are valid because $pc\not\equiv x$,
so $\Hyp$ contains both equations 
$\mkt{\tpc i} ; \mkt{\lassg{}{x}{e}} ; \kneg\mkt{\tpc i} = 0$
and 
$\mkt{\neg \tpc i} ; \mkt{\lassg{}{x}{e}} ; \kneg\mkt{\neg\tpc i} = 0$.

(testCommuteHav) 
The equation 
$\mkt{\tpc i} ; \mkt{\havc{x}} = \mkt{\havc{x}} ; \mkt{\tpc i}$,
is equivalent to the conjunction of 
$\mkt{\tpc i} ; \mkt{\havc{x}} ; \kneg\mkt{\tpc i} = 0$
and $\kneg\mkt{\tpc i} ; \mkt{\havc{x}} ; \mkt{\tpc i} = 0$
using KAT laws.  
By definition of $\mkt{-}$ and boolean algebra, 
$\kneg\mkt{\tpc i} ; \mkt{\havc{x}} ; \mkt{\tpc i} = 0$
is equivalent to 
$\mkt{\neg \tpc i} ; \mkt{\havc{x}} ; \kneg\mkt{\neg\tpc i} = 0$.
Both $\mkt{\tpc i} ; \mkt{\havc{x}} ; \kneg\mkt{\tpc i} = 0$ and 
$\mkt{\neg \tpc i} ; \mkt{\havc{x}} ; \kneg\mkt{\neg\tpc i} = 0$
are instances of the form $\mkt{e};\mkt{\havc{x}};\kneg\mkt{e}=0$,
and we are assuming $x\not\equiv pc$ so they are both in $\Hyp$ and we are done.  
\end{proof}

\begin{rem} 
\upshape
$\nfax(pc,c,f)$ is finite, and every equation is equivalent to one of the form $\KE=0$.  
So entailments of the form $\nfax(pc,c,f) \proves \KE_0 = \KE_1$
are decidable in PSPACE~\cite{KozenKATcomplex}.
\qed\end{rem}

\begin{lem}\label{lem:nfax}
\upshape
All equations in $\nfax(pc,c,f)$ are true in $\relKAT$, for any $pc,c,f$.
\end{lem}
This holds because the equations follow from $\Hyp$ (Lemma~\ref{lem:nfaxHyp})
and the equations in $\Hyp$ are true in $\relKAT$ (Lemma~\ref{lem:Hyp}).
Note that $c$ need not be $\ok$ for this result.
However, truth of (totIf) does depend on the $\totalIf$ condition (Def.~\ref{def:lang}).

\begin{lem}\label{lem:nf-conseq}
\upshape 
The following hold for any $c,pc,f$ such that $pc$ does not occur in $c$.
\begin{list}{}{}
\item[(diffTestNeg)] $\nfax(pc,c,f)\proves\mkt{\tpc i} = \mkt{\tpc i};\neg\mkt{\tpc j}$ 
and $\mkt{\tpc i} \leq \neg\mkt{\tpc j}$ 
\\
for $i\neq j$ with $i,j$ in $\labs(c)\union\{f\}$.

\item[(nf-enab-labs)]
\( \nfax(pc,c,f)\proves
\mkt{\enab(gcs)} = \mkt{ \quant{\lor}{i}{i\in\labs(c)}{ \tpc  i }} \)
\\
if $\okf(c,f)$ and $\norm{c}{f}{gcs}$.

\item[(nf-lab-enab)] $\nfax(pc,c,f)\proves \mkt{\tpc \lab(c)}\leq\mkt{\enab(gcs)}$
\\
if $\okf(c,f)$ and $\norm{c}{f}{gcs}$.

\item[(nf-enab-disj)] $\nfax(pc,c,f)\proves \mkt{\enab(gcs_0)};\mkt{\enab(gcs_1)}= 0$
\\
if $\okf(c,f)$ and $gcs_0$ and $gcs_1$ are the normal form bodies of disjoint subprograms\linebreak of $c$.

\item[(nf-enab-corr)]
\( \nfax(pc,c,f)\proves\\
\mkt{\enab(gcs)};\mkt{gcs} = 
\mkt{\enab(gcs)};\mkt{gcs};(\mkt{\enab(gcs)}\kplus\mkt{\tpc  f}) \)    
\\
if $\okf(c,f)$ and $\norm{c}{f}{gcs}$ and $pc\notin vars(c)$.
\end{list}
(nf-enab-corr) can be abbreviated 
$\nfax(pc,c,f)\proves \mkt{gcs} : \spec{ \mkt{\enab(gcs)} }{ \mkt{\enab(gcs)}\kplus\mkt{\tpc  f}}$.
\end{lem}
\begin{proof}
(diffTestNeg) follows from (diffTest) using boolean algebra.

(nf-enab-labs) 
By rule induction on the normal form judgment.
\\
For $\ldogc{n}{gcs_0}$, the normal form body is explicitly defined to include, 
not only a command with guard $\tpc n\land e$ for each $e\gcto d$ in $gcs_0$
but also a command with guard $\tpc n\land\neg\enab(\ldots)$ for their complement.
The disjunction of their enabling conditions simplifies to $\tpc n$.
By the normal form  rule and induction hypothesis, the enabling condition for the collected normal form bodies is the disjunction over their label tests, hence the result.
\\
For $\lifgc{n}{gcs_0}$, the normal form body includes disjuncts
$\tpc n\land e$ for each of the guard expressions $e$ in $gcs_0$.
Consider the case of two branches, i.e., $gcs_0$ is $e_0\gcto d_0\gcsep e_1\gcto d_1$,
the disjunction is $(\tpc n\land e_0) \lor (\tpc n\land e_1)$.
Observe
\[\begin{array}{lll}
  & \mkt{(\tpc n\land e_0) \lor( \tpc n\land e_1)}\\
= & \mkt{\tpc n};\mkt{e_0} + \mkt{\tpc n};\mkt{e_1} & \mbox{def $\mkt{-}$} \\
= & \mkt{\tpc n};(\mkt{e_0}+\mkt{e_1}) &\mbox{KAT law} \\ 
= & \mkt{\tpc n};\mkt{e_0\lor e_1} & \mbox{def $\mkt{-}$} \\
= & \mkt{\tpc n} & \mbox{axiom (totIf), unit law} 
\end{array}\]
The rest of the argument is like for \keyw{do}.
The general case (multiple branches) is similar.

(nf-lab-enab) follows from (nf-enab-labs), using boolean algebra and 
definition of $\mkt{-}$ (distributing over $\lor$). 

(nf-enab-disj) is proved by induction on normal form, 
using that by $\ok$ subprograms have distinct labels,
and (nf-enab-labs) and (diffTest).

(nf-enab-corr)
The proof is by induction on the normal form relation and by calculation
using KAT laws and the axioms.
It uses (testCommuteAsgn) so $pc$ must not occur in $c$. 
\end{proof}

\lemnfEnabLabs*
\begin{proof}
Follows by definitions from provability lemma (nf-enab-labs) of Lemma~\ref{lem:nf-conseq},
using Lemmas~\ref{lem:nfax} and~\ref{lem:correctInterp}.
\end{proof}

\section{Normal form bodies}\label{sec:app:nfbodies}

\newcommand{\bodies}{\mathconst{bodies}}
\newcommand{\concat}{\mathconst{concat}}

\begin{figure}[t]
\begin{mathpar}

\inferrule{
\norm{c}{\lab(d)}{gcs_0} \\
\norm{d}{f}{gcs_1}
}{
\norm{c;d}{f}{gcs_0\gcsep gcs_1}
}

\inferrule{
\norm{\bodies(gcs)}{f}{\mathit{nfs}}
}{ 
\norm{ \lifgc{n}{gcs} }{f}{ 
  map((\lambda(e\gcto c)\,.\, \tpc n\land e\gcto \spc\lab(c)),\, gcs)
  \gcsep \concat(\mathit{nfs})
}}

\scalebox{0.97}{%
\inferrule{
\norm{\bodies(gcs)}{f}{\mathit{nfs}}
}{ 
\norm{ \ldogc{n}{gcs} }{f}{ 
  map((\lambda(e\gcto c)\,.\, \tpc n\land e\gcto \spc\lab(c)),\, gcs)
  \gcsep \tpc n\land\neg\enab(gcs) \gcto \spc f
  \gcsep \concat(\mathit{nfs})
}}}

\inferrule{ \norm{c}{f}{gcs} }{ \norm{[c]}{f}{[gcs]} }

\inferrule{ \norm{c}{f}{gcs} \\ \norm{cs}{f}{\mathit{nfs}} }
{\norm{c::cs}{f}{gcs::\mathit{nfs} }}

\end{mathpar}
\vspace*{-2ex}
\caption{Normal form bodies for if and do.}\label{fig:norm:general}
\end{figure}

\autoref{fig:norm:general} gives the general cases of normal forms for if- and do-commands, which subsume the special cases given in \autoref{fig:norm}.
The definition of normal form bodies for commands is mutually inductive 
with the definition of normal form body list for nonempty command lists,
which is given by the last two rules in \autoref{fig:norm:general}.
The rules use square brackets for singleton list and $::$ for list cons.
They use $\bodies(gcs)$ and $\concat(nfs)$ defined by
\[\begin{array}{lll}
\bodies(e\gcto c) &=& [c] \\
\bodies(e\gcto c\gcsep gcs) &=& c::\bodies(gcs)\\
\concat([gcs]) &=& gcs\\
\concat(gcs:nfs) &=& gcs \gcsep \concat(nfs)
\end{array}\]

\section{Additional proofs}\label{sec:app:proofs}

\thmFiltSoundCompl*
\begin{proof}  
We prove the main statement by mutual implication, and afterwards consider the issue of finite support.

(right implies left)
If $an$ is a valid annotation of $\aprod(A,A',L,R,J,K)$ for $\aespec{\P}{\Q}$
and $K$ is an adequate filtering for $an$,
then by Lemma~\ref{lem:FiltAdequate} the product is $\P$-adequate.
Since $an$ is valid for $\spec{\P}{\Q}$ we have
$\aprod(A,A',L,R,J,K)\models\spec{\P}{\Q}$ by Lemma~\ref{prop:IAM}. 
So by Lemma~\ref{lem:aeAdequate} we have $A,A'\models\aespec{\P}{\Q}$.

\newcommand{\Traces}{\mathconst{Traces}}
\newcommand{\termQtraces}{\mathit{termQtr}}   
\newcommand{\shortQtraces}{\mathit{shortQtr}} 
\newcommand{\len}{\mathconst{len}}

(left implies right)
Suppose $A,A'\models\aespec{\P}{\Q}$. 
The idea is to use a left-first sequential alignment automaton such that,
once the left-only execution has finished, with final store $t$,
the filter condition only keeps states that are in some terminating run of 
$A'$ that ends in a store $t'$ such that $(t,t')\in\Q$.

To make this precise we need a few definitions.
Let $\Traces(A)$ be the set of all nonempty and finite sequences of states of
automata $A$ that are consecutive under $A$'s transition relation.  In what
follows, we use $\alpha, \beta$ to range over traces, and write
$\mathit{start}(\alpha)$ and $\mathit{end}(\alpha)$ for the first and last
states of $\alpha$.
For $s\in Sto$, $n'\in Ctrl'$, and $s'\in Sto'$, 
define $\termQtraces(s,n',s')$ to be the set of terminated traces of $A'$
  that begin at $(n',s')$ and end in a state $t'$ for which $(s,t')\in\Q$.
  That is,
  \[
  \small
    \termQtraces(s,n',s') \eqdef
    \{ \alpha \in\Traces(A') \mid 
      \mathit{start}(\alpha) = (n',s') \land
      \some{t'}{\mathit{end}(\alpha) = (\fin',t') \land (s,t')\in\Q} \}.
  \]
Define the restriction to traces of minimum length, as follows.\footnote{One 
may wonder how $\min$ is defined, in case $\termQtraces(s,n',s')$ is empty
(and hence so is $\map(\len, \termQtraces(s,n',s'))$),
but that does not matter because in this case $\shortQtraces(s,n',s')$ is empty.
}
\[
\shortQtraces(s,n',s') \eqdef 
  \{\alpha \in \termQtraces(s,n',s') \mid \len(\alpha) = \min(\map(\len, \termQtraces(s,n',s'))) \}
\]
We need the following facts:
\begin{list}{}{}
\item[(i)]
Suppose $\alpha$ is in $\termQtraces(t, m', s')$
and $(n',t')$ is in $\alpha$.
Then $\termQtraces(t, n', t')$ is nonempty.
\item[(ii)]
$\termQtraces(t, n', s')$ is nonempty 
iff $\shortQtraces(t, n', s')$ is nonempty.
\item[(iii)]
\begin{sloppypar}
If $(s,s')$ is in $\P$ and there is an $A$-trace from $(\init,s)$ to $(\fin,t)$,
then $\termQtraces(t, init', s') \neq\emptyset$. 
\end{sloppypar}
\end{list}
Both (i) and (ii) follow directly by definitions.
Fact (iii) is a consequence of the theorem's assumption 
that $A,A'\models\aespec{\P}{\Q}$.
If $(s,s')\in\P$ and there is an $A$-trace from $(\init,s)$ to $(\fin,t)$
then by $A,A'\models\aespec{\P}{\Q}$ there is some 
$t'$ and $A'$-trace $\alpha$ from $(\init',s')$ to $(\fin',t')$ 
with $(t,t') \in \Q$. 
By definition, $\alpha$ is in $\termQtraces(t,\init',s')$.  

Next we define $L,R,J$ as needed for left-first sequential alignment product.
So we let $J\eqdef\emptyset$.  
In case the transition relations $\trans,\trans'$ of $A,A'$ are total (as in the case of program automata) it suffices to define 
$L \eqdef [*|\init']\land\neg[\fin|*]$ and  
$R \eqdef [\fin|*]\land\neg[\fin|\fin']$. 
In general, to ensure liveness in the sense required by Def.~\ref{def:alignProd},
we need 
\[ \begin{array}{l}
L \eqdef [*|\init']\land\neg[\fin|*] \land (\dom(\trans)\times states') \\
R \eqdef [\fin|*]\land\neg[\fin|\fin'] \land (states\times\dom(\trans'))
\end{array}\]
Define $K$ by\footnote{It also works to define $K$ using shortest traces only.}
\[ ((n,n'),(s,s')) \in K \quad\eqdef\quad  
   n'=\init' \lor (n=\fin \land \termQtraces(s,n',s') \neq\emptyset) \]

Let $an$ be the canonical semantic annotation for $\P$,
i.e., $an(n,n')$ is the strongest invariant of 
$\aprod(A,A',L,R,J,K)$ with respect to $\P$.
To be precise, for any $n,n'$:
\[ an(n,n') \eqdef \{ (t,t') \mid
        \some{s,s'}{(s,s')\in \P \land 
       ((\init,\init'),(s,s')) \biTrans_K^* ((n,n'),(t,t')) } \} \]
By construction, this annotation satisfies $\P\imp an(\init,\init')$
and moreover it satisfies the verification conditions, i.e., it is valid.
It remains to show that $an(\fin,\fin')\imp \Q$ (i.e., it is an annotation for $\spec{\P}{\Q}$).  
By definition of $K$, any stores $t,t'$ for which $((\fin,\fin'),(t,t'))$
is reachable via $\biTrans_K$ satisfy $\Q$.  
Because $an(\fin,\fin')$ is exactly these stores,
we conclude that $an(\fin,\fin')\imp \Q$.

It remains to show $K$ is an adequate filtering for $an$.  

For the enabled condition in the Def.~\ref{def:adequateFiltering}, we need 
for any $n,n'$, $\breve{an}(n,n') \imp L\lor R\lor J\lor[\fin|\fin']$.
By construction, an invariant of the product is
$[*|\init'] \lor [\fin|*]$.  
So by definition of $an$ we get 
$\breve{an}(n,n') \imp L\lor R\lor J\lor[\fin|\fin']$ for any $n,n'$.

The left-permissiveness condition in Def.~\ref{def:adequateFiltering}
holds because $K$ allows all left steps.
Joint-productivity holds because $J$ is empty.
To show right-productivity we need a measure $V$.

To define $V$, first observe that all elements of $\shortQtraces(t,n',s')$ have the same length,
if there are any elements.
So we can write $\len(\shortQtraces(t,n',s'))$ for that number, provided the set is not empty.
Now define $V$, a function from product states to naturals,  by 
\[\begin{array}{lcll}
V ((n,n'),(s,s'))    & = & 0 &\mbox{if $n \neq \fin$} \\
V ((\fin,n'),(t,s')) & = & \len(\shortQtraces(t,n',s')) &\mbox{if $\shortQtraces(t,n',s')\neq\emptyset$}\\
V ((\fin,n'),(t,s')) & = & 0 &\mbox{otherwise}
\end{array}\]
For right productivity, consider any product state $((n,n'),(t,t'))$
and assume $(t,t')\in an(n,n')$ and
$((n,n'),(t,t'))\in R$.
We must show there is some $m'$ and $u'$, and a transition
from $(n',t') \trans' (m',u')$
such that 
\begin{itemize}
\item $((n,m'),(t,u')) \in K(n,m')$, and 
\item $V((n,m'),(t,u')) < V((n,n'),(t,t'))$ 
\end{itemize}
Observe that in general, from $(t,t')\in an(n,n')$ we have that $((n,n'),(t,t'))$ is reachable from some $((\init,\init'),(s,s'))$ with $(s,s')\in \P$.  

From $((n,n'),(t,t'))\in R$ we have that $n=\fin$ and $n'\neq\fin'$.
Now, in general $K\lor[\init|\init']$ is an invariant of an LRJK-automaton, 
so because $\fin\neq\init$ we have that $\breve{an}(\fin,n')$ implies $K(\fin,n')$.
In particular we have  $((\fin,n'),(t,t'))$ is in $K(\fin,n')$.
Thus by definition of $K$ we have that $\termQtraces(t,n',t')\neq\emptyset$,
unless $n'=\init'$.  
In case $n'=\init'$ we get $\termQtraces(t,\init',t')\neq\emptyset$ as follows.
By the general observation above,
from $(t,t')\in an(\fin,\init')$ we have that $((\fin,\init'),(t,t'))$ is reachable from some $((\init,\init'),(s,s'))$ with $(s,s')\in \P$.  
This gives a terminated $A$-trace. Now we can appeal to 
fact (iii) above to get $\termQtraces(t,\init',t')\neq\emptyset$.

Now by fact (ii) we have $\shortQtraces(t,n',t')$ is nonempty.
Choose any $\alpha$ in $\shortQtraces(t,n',t')$.
The first state of $\alpha$ is $(n',t')$, by definitions,
and $\len(\alpha)>1$ because $n'\neq\fin'$.
Let $(m',u')$ be the second state of $\alpha$,
so we have $(n',t')\trans'(m',u')$. 
Using facts (i) and (ii), the set $\termQtraces(t,m',u')$ is nonempty,
so by definition of $K$ we have $((\fin,m'),(t,u')) \in K(\fin,m')$, proving the first bullet above.  

\begin{sloppypar}
Since $\shortQtraces(t,n',t')$ is nonempty (and traces have nonzero length) 
we have $V((\fin,n'),(t,t')) \neq 0$.
Since $(m',u')$ is the second state in $\alpha$ which is in $\shortQtraces(t,n',t')$,
we have $V((\fin,m'),(t,u')) = V((\fin,n'),(t,t')) - 1$
whence 
$V((\fin,m'),(t,u')) < V((\fin,n'),(t,t'))$, proving the second bullet above and completing the proof of the main statement of the theorem.
\end{sloppypar}

Concerning finite support, suppose the stores of $A$ and $A'$ are 
variable stores,  and suppose $A,A',\P,\Q$ are finitely supported. 
We must show that $L,R,J,K,an,V$ are too.
Let $X$ be all the variables on which $\P$ or $\Q$ depend (on the left or right), together with all the variables in the support of the transition relation of $A$ or $A'$.
Note that in any trace $\alpha$ of either automaton, any variable $y\notin X$ is unchanged through the trace.  Moreover if $\beta$ is obtained from $\alpha$ by setting $y$ to another fixed value, then $\beta$ is a trace of the automaton.  Because $X$ supports $\P$, we have that strongest invariants, 
are supported by $X$; in brief, $an$ is supported by $X$.    
Because $X$ supports $\Q$, each set $\termQtraces(s,n',s')$ is determined by the projections of $s$ and $s'$ on $X$, and the same for $\shortQtraces(s,n',s')$. 
Thus the condition
$\termQtraces(s,n',s') \neq\emptyset$ in the definition of $K$ is supported by $X$,
so $K$ is as well.
We have $L,R,J$ supported by $X$ because $\dom(\trans)$ and $\dom(\trans')$ are by assumption.
Finally, $V$ is supported by $X$ because the condition $\shortQtraces(s,n',s') \neq\emptyset$ in its definition is.
\end{proof}

\begin{rem}
A natural question is whether the $L,R,J,K$ conditions, annotation $an$,
and measure $V$ used in the proof of \autoref{thm:FiltSoundCompl} are expressible in a first order assertion language.  
Prior works on Hoare logic showed expressivity of weakest conditions in Peano arithmetic, based on G\"{o}del encodings of stores, execution traces, etc.~\cite{AptOld3,deBakkerBook} as needed for $an,K,V$.  
Thus Peano arithmetic suffices for our result.
\qed\end{rem}

\thmaeAlignComp*
\begingroup
\newcommand{\bothnn}{\bothF{\tpc n\sep\tpc n'}}
\begin{proof}
  The proof of this theorem proceeds in the same way as the proof of
  \autoref{thm:acomplete}.  As in \autoref{thm:acomplete}, the proof uses
  only relational assertions derived from $an(i,j)$, $L$, $R$, and $K$.  

  We start by choosing a variable $pc$ that is
  fresh with respect to $\S$, $\T$, $c$, $c'$, $an$, $L$, $R$, $J$, $K$, and the $V$ that
  witnesses adequacy assumption (c).
  Existence of such a variable is ensured by the assumption (d) of finite support.
  By Lemma~\ref{lem:normExists} there are $gcs$ and $gcs'$ such that
  $\normSmall{c}{f}{gcs}$ and $\normSmall{c'}{f'}{gcs'}$.  Let $n = \lab(c)$
  and $n' = \lab(c')$.

  By Theorem~\ref{thm:normEquiv} we have
  \[
      \spc n; \ldogc{}{gcs} \: \kateq \: \addPC(c); \spc f 
\qquad\mbox{and}\qquad
      \spc n'; \ldogc{}{gcs'} \: \kateq \: \addPC(c'); \spc f' 
  \]
  Define store relation $\Q$ to be $\Q_{an}\land\Q_{pc}$ where 
  \[ \begin{array}{l}
       \Q_{an}: \qquad
       \quant{\land}{i,j}{i\in \labs(c)\union\{f\} \land j\in \labs(c')\union\{f'\}
       }{\bothF{\tpc i\sep \tpc j} \imp an(i,j)} 
       \\
       \Q_{pc}: \qquad 
       \quant{\lor}{i,j}{i\in \labs(c)\union\{f\} \land j\in \labs(c')\union\{f'\}
       }{\bothF{\tpc i\sep \tpc j}}
     \end{array}
   \]
   We will derive 
   \begin{equation}\label{eq:Cx}
     \ldogc{}{gcs} \sep \ldogc{}{gcs'} : \aespec{\Q}{\Q\land\neg\leftF{\enab(gcs)}\land\neg\rightF{\enab(gcs')}}  
   \end{equation}
To do so, we start by noting $K$ being an adequate filtering implies there
   is a variant function $V$ from 
   $((\labs(c)\cup\{f\})\times (\labs(c')\cup\{f'\}))
   \times ((\Var\to\Z)\times (\Var\to\Z))$
   to $D$, where $(D,\prec)$ is some well-ordered set.  Define the
   \dt{$pc$-encoded variant} \graybox{$\encode{V}$} as
   $\encode{V}(s,s') \eqdef V((s(pc),s'(pc)),(s,s'))$.  Note that $\encode{V}$ is a function
   on variable stores.  In what follows, we will also use the curried form 
   \graybox{$\proj{V}(n,n')$} defined for any $n,n',s,s'$ by 
   $\proj{V}(n,n') (s,s') \eqdef V((n,n'),(s,s'))$.

   Now to show~(\ref{eq:Cx}), we instantiate rule \rn{eDo} with $\Q:=\Q$,
   $\Lrel:=\encode{L}$, $\R:=\encode{R}$ and $V := \encode{V}$.  The side
   condition of \rn{eDo} is the same as the side condition of \rn{rDo}.
   Further, the instantiation of $\Q$, $\Lrel$, and $\R$ is the same as in the
   proof of Theorem~\ref{thm:acomplete}.  Thus, the proof that the side
   condition is valid is identical.  We now turn to proofs of the premises of
   \rn{eDo}.  For each, we list a few representative cases.

  There are three sets of premises of \rn{eDo} for the two loops, with these
  forms:
  \\
  \begin{tabular}{ll}
    (left-only) & $b\sep \skipc : \aespec{\Q\land \leftF{e}\land\encode{L} }{\Q}$
                  for all $e\gcto b$ in $gcs$ 
    \\
    (right-only) & \begin{tabular}[t]{@{}r@{}}$\skipc\sep b' : \aespec{\Q\land \rightF{e'}\land\encode{R} \land (\encode{V}=k)}{\Q \land (\encode{V}\prec k)}$  for all $e'\gcto b'$ in $gcs'$,\\ all $k\in D$\end{tabular}
    \\
    (joint) & $b\sep b' : \aespec{\Q\land \bothF{e\sep e'} \land \neg\encode{L} \land \neg\encode{R}}{\Q}$ for all $e\gcto b$ in $gcs$ and $e'\gcto b'$ in $gcs'$
  \end{tabular}
  \\

  \paragraph{Joint cases}

  We start by making note of the following.
  For any $m$ and $m'$ we can prove
  \begin{equation}\label{eqn:aemm}
    \spc m \sep \spc m' : \aespec{an(m,m')}{\Q}.
  \end{equation}
  We arrive at this as follows.  Since $an(m,m')$ is independent of $pc$ on
  both sides, the judgment
  $\spc m\sep\spc m':\aespec{an(m,m')}{an(m,m')\land\bothF{\tpc m\sep\tpc m'}}$ 
  can be proved using \rn{eAsgnAsgn} with \rn{eConseq} to simplify the precondition. 
  Then, we use \rn{eConseq} with the fact that
  $an(m,m') \land \bothF{\tpc m\sep\tpc m'} \imp \Q$ to
  obtain~(\ref{eqn:aemm}).
  Additionally, the following implication is valid.
  \begin{equation}\label{eqn:aejoint}
    \Q \land \bothF{\tpc n \sep \tpc n'} \land \neg\encode{L} \land \neg\encode{R}
    \imp an(n,n') \land \bothF{\tpc n\sep \tpc n'} \land \encode{J}
  \end{equation}
  We arrive at this using adequacy condition (c) of the theorem 
  (in particular, the enabled condition in Def.~\ref{def:adequateFiltering}) 
  and the fact that $Q \land \bothF{\tpc n\sep\tpc n'}\imp an(n,n')$.
  See~(\ref{eq:dliftJ}). 

  Now for the joint premises. We spell out just two cases.

  \begin{itemize}
    \newcommand{\Jpre}{\Q\land\bothnn\land\neg\encode{L}\land\neg\encode{R}}
  \item
    $\lhavc{}{x};\; \spc m \sep \lskipc{};\; \spc m' : \aespec{\Jpre}{\Q}$,
    where $\sub(n,c) = \lhavc{n}{x}$, $m = \fsuc(n,c,f)$,
    $\sub(n',c') = \lskipc{n'}$, and $m' = \fsuc(n',c',f')$.

    We construct a deductive proof using \rn{eSeq} with judgments~(\ref{eqn:aemm})
    and
    \[\lhavc{n}{x} \sep \lskipc{n'} : \aespec{\Jpre}{an(m,m')} \]
To prove the latter, first  by \rn{eHavSkip} we have  \( \lhavc{n}{x}\sep\lskipc{n'} : \aespec{(\allSet{x\smSep}{an(m,m')})}{an(m,m')}\).  Now strengthen the precondition using \rn{eConseq}
    for which we need to show $\Jpre\imp (\allSet{x\smSep}{an(m,m')})$.  In doing so, we'll
    appeal to the corresponding $pc$-encoded VC (Lemma~\ref{lem:liftEVC}) which is
    \[\all{v\in\Z}{an(n,n')\land\bothnn\land\encode{J}\land\subst{\proj{K}(m,m')}{x|}{v|}\imp
        \subst {an(m,m')}{x|}{v|}} \]

    The \rn{eConseq} step is justified as follows.
    \[
    \begin{array}{lll}
      & \Jpre \\
      \imp & an(n,n') \land \bothnn \land \encode{J}
           & \mbox{by~(\ref{eqn:aejoint})} \\
      \imp & an(n,n') \land \bothnn \land \encode{J} \land (\all{v\in\Z}{\subst{\proj{K}(m,m')}{x|}{v|}})
           & \mbox{semantics, adequacy, see below $(\dagger)$} \\
      \iff & \all{v}{an(n,n') \land \bothnn \land \encode{J} \land \subst{\proj{K}(m,m')}{x|}{v|}}
           & \mbox{pred calc} \\
      \imp & \all{v}{\subst{an(m,m')}{x|}{v|}}
           & \mbox{apply VC (Lemma~\ref{lem:liftEVC})} \\
      \iff & \allSet{x\smSep}{an(m,m')}
           & \mbox{by def}
    \end{array}
    \]
To justify the step marked $(\dagger)$ above, we
rely on the theorem's assumption of adequate filtering. 
Specifically, we use use joint-productivity to prove that 
$an(n,n') \land \bothnn \land \encode{J}$
implies $(\all{v\in\Z}{\subst{\proj{K}(m,m')}{x|}{v|}})$.
Because Def.~\ref{def:adequateFiltering} is in terms of automata states,
we need to reason pointwise and take care with our notational abuses.
Observe for any $s,s'$ that 
\[ (s,s')\in (an(n,n') \land \bothnn \land \encode{J}) \]
equivales, by definitions and freshness of $pc$, 
\[ (s,s')\in an(n,n') \land s(pc)=n \land s'(pc)=n' \land ((n,n'),(s,s'))\in J \]
By the assumptions $\sub(n,c) = \lhavc{n}{x}$ and $m = \fsuc(n,c,f)$,
using \autoref{def:aut} of program automata
which in this case is based on semantics of $\lhavc{n}{x}$,

we have $(n,s)\trans (m,\update{s}{x}{v})$ for all $v\in\Z$.
By joint-productivity, for any $v$ there is some $m'',t'$
such that $(n',s')\trans'(m'',t')$ and $((m,m''),(\update{s}{x}{v}))\in K$.
But by semantics and assumptions $\sub(n',c') = \lskipc{n'}$ and $m' = \fsuc(n',c',f')$
we must have $m''=m'$ and $t'=s'$, so we have
$((m,m'),(\update{s}{x}{v},s'))\in K$.
This is equivalent to 
$(\update{s}{x}{v},s')\in \proj{K}(m,m')$
and to 
$(s,s')\in \subst{\proj{K}(m,m')}{x|}{v|}$.
So we have 
$(\all{v\in\Z}{\subst{\proj{K}(m,m')}{x|}{v|}})$,
so $(\dagger)$ is proved.

  \item
    $\spc m\sep \spc m' : \aespec{\Q \land \bothF{\tpc n\sep \tpc n' \land e'} \land\neg\encode{L}\land\neg\encode{R}}{\Q}$
    where
    $\sub(n,c) = \lskipc{n}$, $m = \fsuc(n,c,f)$,
    $\sub(n',c') = \lifgc{n'}{gcs'_0}$ with $e'\gcto d'$ in $gcs'_0$, $m' = \lab(d')$.

The corresponding VC is similar to the assign/if VC in \autoref{fig:EVCjo},
specifically:
\[ J \land \breve{an}(n,n')\land \rightF{e'} \land K(m,m') \imp\hat{an}(m,m') \]
Of course we use the pc-encoded version.  

We prove the judgment using (\ref{eqn:aemm}) and \rn{eConseq} with the following implication.
    \[
    \begin{array}{lll}
      & \Q\land\bothF{\tpc n\sep\tpc n'\land e'}\land\neg\encode{L}\land\neg\encode{R} \\
\imp  & an(n,n') \land \bothF{\tpc n\sep\tpc n'\land e'}\land\encode{J} 
      & \mbox{using (\ref{eqn:aejoint})} \\
\imp  & an(n,n') \land \bothF{\tpc n\sep\tpc n'\land e'}\land\encode{J} \land \proj{K}(m,m')
      & \mbox{adequacy, semantics} \\
\imp  & an(m,m') 
      & \mbox{apply VC} 
      \end{array}
\]
The adequacy step uses in particular joint productivity, which says for any $n,n',s,s',m,t$
that if $(s,s') \in an(n,n')$  and $((n,n'),(s,s')) \in J$ and $(n,s)\trans (m,t)$, then  
there are $m'',t'$ with $(n',s')\trans'(m'',t')$ and $((m,m''),(t,t')) \in K$.
Instantiating this with $m := \fsuc(n,c,f)$ and $t := s$ (since the command at $n$ is skip),
we get that there exist $m'',t'$ 
with $(n',s') \trans' (m'',t')$  and $((m,m''),(t,t')) \in K$.
By semantics and control determinacy (assumption (a) of the theorem,
and Lemma~\ref{lem:controlDet}) and the fact that $e'$ is true in $s'$,
we have that $m''=m'$ since $m'$ is the unique successor,
and by semantics $t'=s'$.
So we have 
$(n',s') \trans' (m',s')$  and $((m,m'),(s,s')) \in K$,
and equivalently $(s,s')\in \proj{K}(m,m')$.

  \end{itemize}

  \paragraph{Right-only cases}

  These cases are proved using the same rules as the joint cases, plus one
  additional rule: \rn{eDisj}, in the form \rn{eDisjN} derived from it.  This
  is needed for the same reason \rn{rDisjN} is needed in the proof of
  alignment completeness for RHL+ (\autoref{thm:acomplete}).  The joint cases
  determine a unique starting and ending pair of control points, which
  determines the VC to appeal to.  The right-only cases do not determine a
  control point on the left.  
  So we consider an arbitrary left control point,
  and then combine all cases using \rn{eDisjN}.
  In passing, we note the following:
  \begin{equation}
    \label{eqn:VCright}
    (s,s') \in \bothF{\tpc m\sep \tpc m'} \imp \proj{V}(m,m')(s,s') = \encode{V}(s,s')
    \quad\mbox{for any $s,s',m,m'$}
  \end{equation}
  \begin{equation}
    \label{eqn:rightQequiv}
    \Q \land \rightF{\tpc n'} \iff \quant{\lor}{n}{n\in\labs(c)\union\{f\}}{ \Q
        \land \bothF{\tpc n\sep\tpc n'}}    
  \end{equation}

  Additionally, to streamline the proofs below, we note the following can be proved.
  \begin{equation}
    \label{eqn:aerightHelper}
    \lskipc{}\sep\spc m' :
    \aespec{an(n,m') \land \leftF{\tpc n} \land \proj{V}(n,m')\prec k}
    {\Q \land \encode{V}\prec k}
  \end{equation}
  We obtain this judgment as follows.  By \rn{eSkipAsgn} and the fact that all
  the terms in the precondition are independent of $pc$ on the right, we have
  \(\lskipc{}\sep\spc m' : \aespec{an(n,m')\land\leftF{\tpc n}\land
    \proj{V}(n,m')\prec k}{an(n,m')\land\bothF{\tpc n\sep\tpc m'}\land
    \proj{V}(n,m')\prec k} \).  We then weaken the postcondition using
  \rn{eConseq}: suppose
  $an(n,m')\land\bothF{\tpc n\sep\tpc m'}\land \proj{V}(n,m')\prec k$.
  By~(\ref{eq:IanRel}), this implies
  $\Q\land\bothF{\tpc n\sep\tpc m'}\land \proj{V}(n,m')\prec k$. 
  We then have $\Q \land \encode{V}\prec k$ by~(\ref{eqn:VCright}).

  We now detail the proofs for a couple of representative right-only cases.  Towards that end, we
  pick an arbitrary control point $n \in \labs(c) \cup \{f\}$ on the left.
  The judgments below conjoin $\leftF{\tpc n}$ to preconditions, and hence, don't
  exactly match the premises of \rn{eDo}.  However, since $n$ is an arbitrary label,
  by~(\ref{eqn:rightQequiv}) and application of \rn{eDisjN}, each proof below
  gives rise to a corresponding right-only premise of rule \rn{eDo}
  (such use of a disjunction rule is spelled out in more detail in the proof of
  \autoref{thm:acomplete}).
  Note there's a premiss for each literal value $k$ in $D$.
  So let $k\in D$ be arbitrary, in the following.

  \begin{itemize}

  \item
    $\lskipc{} \sep \spc m' : \aespec{\Q\land\bothF{\tpc n\sep\tpc n'}\land\encode{R}\land\encode{V}=k}{\Q\land\encode{V}\prec k}$, where
    $\sub(n',c') = \lskipc{n'}$ and $m' = \fsuc(n',c',f')$.

We obtain this from~(\ref{eqn:aerightHelper}) by \rn{eConseq}, strengthening the precondition as follows.
    \[
    \begin{array}{lll}
      & \Q\land\bothF{\tpc n\sep\tpc n'}\land\encode{R}\land\encode{V}=k \\
      \imp & an(n,n') \land \bothF{\tpc n\sep\tpc n'} \land \encode{R} \land \proj{V}(n,n') = k
           & \mbox{(\ref{eq:IanRel}) and~(\ref{eqn:VCright})} \\
      \imp & an(n,n') \land \bothF{\tpc n\sep\tpc n'} \land \encode{R}
             \land \proj{K}(n,m') \land \proj{V}(n,m')\prec k
           & \mbox{adequacy, semantics $(\dagger)$} \\
      \imp & an(n,m') \land \leftF{\tpc n} \land \proj{V}(n,m')\prec k
           & \mbox{apply VC (Lemma~\ref{lem:liftEVC})}
    \end{array}
    \]
In the step marked $(\dagger)$, we use the semantics of $\lskipc{}$
and apply the right-productivity condition,
noting that the only transitions from $n'$ are to $m'$,
to obtain $\proj{K}(n,m') \land \proj{V}(n,m')\prec \proj{V}(n,n')$.
Using $\proj{V}(n,n')=k$,
this yields $\proj{K}(n,m') \land \proj{V}(n,m')\prec k$.  
As in the proofs of the joint-only
premises of \rn{eDo} above, we do not spell this step out in detail.  

\item

    $\lskipc{} \sep \spc m' : \aespec{
         \Q\land\bothF{\tpc n\sep\tpc n' \land e'}\land\encode{R}\land\encode{V}=k
       }{\Q\land\encode{V}\prec k}$, where
    $\sub(n',c') = \lifgc{n'}{gcs'}$ and $e'\gcto d'$ is in $gcs'$ and 
    $m' = \lab(d')$.

As in the preceding case we obtain this from~(\ref{eqn:aerightHelper}) by \rn{eConseq}, strengthening the precondition as follows.
\[ \begin{array}{lll}
     & \Q\land\bothF{\tpc n\sep\tpc n' \land e'}\land\encode{R}\land\encode{V}=k \\
\imp & \hint{ (\ref{eq:IanRel}) and~(\ref{eqn:VCright})} \\
     & an(n,n') \land \bothF{\tpc n\sep\tpc n'\land e'} \land \encode{R}
                \land \proj{V}(n,n') = k \\
\imp & \hint{$(\dagger)$ adequacy, semantics, see below} \\
     & an(n,n') \land \bothF{\tpc n\sep\tpc n'\land e'} \land \encode{R}
                \land \proj{K}(n,m') \land \proj{V}(n,m')\prec k \\
\imp & \hint{apply VC from \autoref{fig:EVCro}, see \autoref{lem:liftEVC}} \\
     & an(n,m') \land \leftF{\tpc n} \land \proj{V}(n,m')\prec k    
\end{array} \]
To prove the step marked $(\dagger)$, consider any $(s,s')$ that satisfies the antecedent.
By right productivity there are $m'',t'$ such that 
$(n',s')\trans'(m'',t')$ and $((n,m''),(s,t'))\in K$ and $V((n,m''),(s,t'))\prec V((n,n'),(s,t'))$.
By definition of automata and semantics of if, the store is unchanged so $t'$ is $s'$.
That is, we have $m''$ such that $(n',s')\trans'(m'',s')$ and $((n,m''),(s,s'))\in K$
and $V((n,m''),(s,s'))\prec V((n,n'),(s,s'))$.
Because $(s,s')$ satisfies $\bothF{\tpc n\sep\tpc n'\land e'}$ and $c'$ is control deterministic, we must have $m''=m'$, hence
$((n,m'),(s,s'))\in K$ and $V((n,m'),(s,s'))\prec V((n,n'),(s,s'))$.
So $(s,s')$ satisfies $\proj{K}(n,m')$ and $\proj{V}(n,m')\prec k$    
as needed for the consequent of $(\dagger)$.
  \end{itemize}

The other right-only cases are similarly proved.

  \paragraph{Left-only cases}

  These are very similar to the right-only cases and are omitted.
  Justifications of \rn{eConseq} are simpler since we don't have to reason
  about $\proj{V}$ decreasing.  Rather than using right-productivity of $K$, these cases use its left-permissivity.

  \paragraph{Finishing the proof}

  Now that we've established all the premises of \rn{eDo}, the rest of the
  proof proceeds exactly like the proof of \autoref{thm:acomplete}
but using \rn{eSeq}, \rn{eConseq}, \rn{eRewrite}, \rn{eGhost}.
\end{proof}
\endgroup

\end{document}